\documentclass[rmp,amsmath,amssymb,twocolumn,floatfix]{revtex4}
\usepackage{graphicx}
\usepackage{bm}
\usepackage{type1cm}
\begin{document}
\title{Nuclear weak-interaction processes in stars}
\author{K. Langanke}
\email{langanke@phys.au.dk}
\affiliation{Institut for Fysik og Astronomi, {\AA}rhus Universitet,
  DK-8000 {\AA}rhus, Denmark} 
\author{G. Mart\'{\i}nez-Pinedo}
\email{martinez@ieec.fcr.es}
\affiliation{Department f\"ur Physik und Astronomie, Universit\"at
  Basel, CH-4056 Basel, Switzerland} 
\altaffiliation[Present address: ]{Institut d'Estudis Espacials de
  Catalunya, Edifici Nexus,  Gran Capit\`a 2, E-08034 Barcelona, Spain}

\date{\today}

\begin{abstract}
  Recent experimental data and progress in nuclear structure modeling
  have lead to improved descriptions of astrophysically important
  weak-interaction processes. The review discusses these advances and
  their applications to hydrostatic solar and stellar burning, to the
  slow and rapid neutron-capture processes, to neutrino
  nucleosynthesis, and to explosive hydrogen burning. Special emphasis
  is given to the weak-interaction processes associated with
  core-collapse supernovae. Despite some significant progress,
  important improvements are still warranted. Such improvements are
  expected to come from future radioactive ion-beam facilities.
\end{abstract}

\maketitle

\tableofcontents

\section{Introduction}
\label{sec:introduction}

The weak interaction is one the four fundamental forces in nature.
Like the other three -- strong, electromagnetic and gravitation -- it
plays a keyrole in many astrophysical processes. This can be nicely
illustrated by the observation that new insights into the nature of
the weak interaction usually were closely followed by the recognition
of their importance in some astrophysical context. Shortly after Pauli
postulated the existence of the neutrino and Fermi developed the first
theory of weak interaction~\cite{Fermi:1934}, Gamow and Schoenberg
speculated about the possible role of neutrinos in stellar evolution
and proposed their production in the star as an important source for
stellar energy losses
\cite{Gamow.Schoenberg:1940,Gamow:1941,Gamow.Schoenberg:1941}.  The
development of the universal $V-A$
theory~\cite{Feynman.Gell-Mann:1958} led Pontecorvo to realize that
the bremsstrahlung radiation of neutrino pairs by electrons would be a
very effective stellar energy loss mechanism \cite{Pontecorvo:1959}.
Just after the discovery of neutral weak current
\citet{Freedman:1974}, \citet{Mazurek:1975}, and \citet{Sato:1975}
recognized that this interaction would result in a sizable elastic
scattering cross section between neutrinos and nucleons, leading to
neutrino trapping during the core collapse of a massive star in a type
II supernova.

The unified model of electroweak
interaction~\cite{Weinberg:1967,Salam:1968,Glashow.Iliopoulos.Maiani:1970}
allows derivation of accurate cross sections for weak processes among
elementary particles (i.e.\ electrons, neutrinos, quarks), but also
for neutron and protons if proper formfactors are taken into account
which describe the composite nature of the nucleons.  However, the
situation is different for weak interaction processes involving
nuclei. Clearly, the smallness of the weak interaction coupling
parameter allows treatment of these processes in perturbation theory,
reducing the calculation basically to a nuclear structure problem.
However, it has been the inability to adequately treat the nuclear
many-body problem, which has -- and in many cases still does -
introduced a substantial uncertainty into some of the key weak
interaction rates used in astrophysical simulations.  However, the
recent few years have witnessed a tremendous progress in nuclear
many-body theory, made possible by new approaches and novel computer
realizations of established models, but also by the availability of
large computational capabilities. This progress allowed calculation of
the rates for many of the stellar weak-interaction processes involving
nuclei with significantly improved accuracy or for the first time. To
actually know that the calculations are more reliable, implies the
availability of experimental data which test, constrain and guide the
theoretical models. Thus, the advances in modelling nuclear
weak-interaction processes in stars also reflects the progress made by
experimentalists in recent years which have succeeded to measure data
which are relevant for the astrophysical applications discussed in
this review either directly, e.g.\ half-lives for some short-lived
nuclei on the r-process path \cite{Pfeiffer.Kratz.ea:2001}, or
indirectly like the Gamow-Teller distributions for nuclei in the iron
mass range \cite{Osterfeld:1992} which decisively constrain the
nuclear models. Another recent experimental first has been the
measurement of charged- and neutral-current neutrino-nucleus cross
sections.

This review will report about progress in modelling nuclear
weak-interaction processes and their possible implications for stellar
evolution and nucleosynthesis. We will restrict ourselves to advances
achieved by improved nuclear models, treating the weak interaction
within the standard model. Of course, it has long been recognized
\citep[e.g.][]{Sato.Sato:1975} that stars can be used as
\emph{laboratories for fundamental physics}~\cite{Raffelt:1996}
searching for new weakly interacting particles or constraining exotic
components of the weak interaction outside the standard model. This
field is rapidly growing \citep[see for
example][]{Raffelt:1996,Raffelt:1999,Raffelt:2000,%
Corsico.Benvenuto.ea:2001,Dominguez.Straniero.Isern:1999}. 

Our review is structured as follows. Following a very brief discussion
of the required ingredients of the weak interaction we introduce the
nuclear many-body models which have been used in the studies of the
weak-interaction processes (section~\ref{sec:theor-descr}). The
remaining sections are devoted to the results of these calculations
and their applications to astrophysics which include the solar nuclear
reaction network and neutrino problem, the core collapse of massive
stars, s- and r-process nucleosynthesis, neutrino nucleosynthesis,
explosive hydrogen burning and type Ia supernovae.

Although generally quite important, weak-interaction processes
constitute only a part of the many nuclear reactions occuring in
stars. For recent reviews about other stellar nuclear reactions
networks and nucleosynthesis the reader is referred to the
comprehensive and competent work by
\citep{Wallerstein.Iben.ea:1997,Arnould.Takahashi:1999,
Boyd:2000,Smith.Rehm:2001}.

\section{Theoretical description}
\label{sec:theor-descr}

\subsection{Weak interactions in nuclei}
\label{sec:weak-inter-nucl}

\begin{figure}[htbp]
  \includegraphics[width=0.9\linewidth]{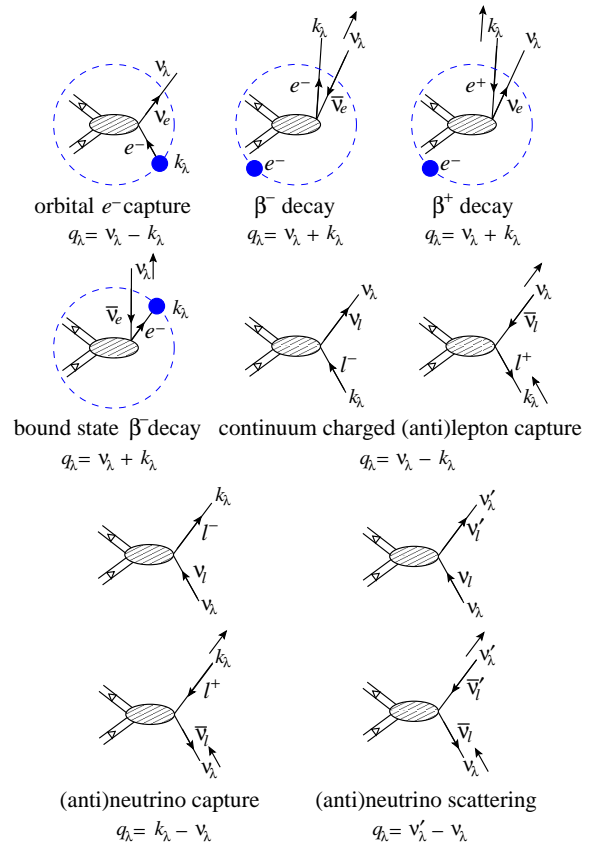}
  \caption{Semileptonic weak processes that occur during the evolution
    of stars. For each process the hadronic current is on the left and
    the leptonic current to the right. The dashed circle indicates a
    bound electron in the initial or final state. The four-momentum
    transfer $q_\lambda = (-\omega,\bm{q})$ for each process is given
    in terms of the charged lepton four-momentum $k_\lambda =
    (\epsilon,\bm{k})$ and the neutrino four-momentum $\nu_\lambda =
    (\nu,\bm{\nu})$. $\omega$, $\epsilon$, and $\nu$ represent the
    energy transfer, lepton energy and neutrino energy, respectively.
    In the case of antiparticles the directions of the momenta are
    show as an arrow close to the four-momentum label. The first row
    shows the usual decay modes in the laboratory. The second and
    third rows show processes that can occur under stellar
    conditions.\label{fig:weakprocess}}
\end{figure}

Processes mediated by the weak interaction in stars can be classified
as leptonic (all interacting particles are leptons) and semileptonic
(leptons interact with hadrons via the weak interaction). Leptonic
processes can be straightforwardly computed using the standard
electroweak model \cite{Grotz.Klapdor:1990}. The calculation of
semileptonic processes (i.e.\ neutrino-nucleus reactions,
charged-lepton capture, and $\beta$-decay) is more complicated due to
the description of the nuclear states involved.  Fortunately the
momenta of the particles turn out to be small compared with the masses
of the $Z,W$ bosons. Thus it is sufficient to consider the
semileptonic processes of interest in the lowest-order approximation
in the weak interaction. Then the interaction can be described by a
current-current Hamiltonian density:

\begin{equation}
  \label{eq:1}
  \mathcal{H}(\bm{x})= -\frac{G}{\sqrt{2}}
  \mathcal{J}_\mu(\bm{x})j_\mu(\bm{x}),
\end{equation}
where $G=G_F V_{ud}$ for charge-current processes and $G=G_F$ for
neutral current processes, with $G_F$ the Fermi coupling constant and
$V_{ud}$ the up-down entry of the Cabibbo-Kobayashi-Maskawa matrix
\cite{PDBook}.  $j_\mu(\bm{x})$ and $\mathcal{J}_\mu(\bm{x})$
are the weak leptonic and hadronic density operators
\cite{Walecka:1975,Walecka:1995,Donnelly.Peccei:1979}. The structure
of the leptonic current $j_\mu(\bm{x})$ for a particular process
is given by the standard electroweak model
\cite{Weinberg:1967,Salam:1968,Glashow.Iliopoulos.Maiani:1970}, and
contains both, vector and axial-vector components.  The standard model
describes the hadronic current in terms of quark degrees of freedom.
Since we are only interested in the matrix elements of
$\mathcal{J}_\mu(\bm{x})$ in nuclei we need only to retain the
pieces which involve $u$ and $d$ quarks. (The contribution from
strange quarks is normally neglected, but see the discussion in
section~\ref{sec:delay-supern-mech}.) As in nuclear physics the
nucleons are treated as elementary spin-1/2 fermions, the Standard
Model current is not immediately applicable.  Moreover, nucleons in
nuclei interact also via the strong interaction.  It is then
convenient to define an effective hadronic current using arguments of
Lorentz covariance and isospin invariance of the strong interaction.
The effective hadronic current can be decomposed into strong isoscalar
($T=0$) and isovector ($T=1$) components and contains both vector
($V$) and axial-vector ($A$) pieces. The weak charge-changing current
is isovector with $M_T = \pm 1$ and can be written in a general form
as:

\begin{equation}
  \label{eq:2}
  \mathcal{J}_\mu = V_\mu^{1M_T} + A_\mu^{1M_T}.
\end{equation}
This current governs processes (see figure~\ref{fig:weakprocess}) such
as $\beta^\pm$-decay, $e$-capture, neutrino $(\nu_l,l^-)$ and
anti-neutrino $(\bar{\nu}_l,l^+)$ reactions ($l=e, \mu $ or $\tau$).
Under the conserved vector current (CVC) hypothesis
\cite{Feynman.Gell-Mann:1958} the current $V_\mu^{1M_T}$ has a
structure identical to the isovector part of the electromagnetic
current. As a consequence of this hypothesis the weak charge-changing
vector current is a conserved quantity.  For the weak neutral current
one has $M_T =0$ and, in general, both $T=0$ and $T=1$ pieces can
occur. The general form of this current is

\begin{equation}
  \label{eq:3}
  \mathcal{J}_\mu = \beta_V^{(0)} V_\mu^{00} + \beta_A^{(0)}
  A_\mu^{00} + \beta_V^{(1)} V_\mu^{10} + \beta_A^{(1)} A_\mu^{10}.
\end{equation}
Assuming that the coupling constants are given by the Standard Model
we have: $\beta_V^{(0)}=-2\sin^2 \theta_W$, $\beta_A^{(0)}= 0$,
$\beta_V^{(1)}= 1-2\sin^2 \theta_W$, $\beta_A^{(1)}= 1$ \citep[p.
25]{Donnelly.Peccei:1979}. $\theta_W$ is the weak mixing angle.  The
neutral current describes weak interactions such as neutrino
$(\nu,\nu^\prime)$ and anti-neutrino $(\bar{\nu},\bar{\nu}^\prime)$
scattering.

The nuclear transitions that are induced by such weak currents
(operators) involve initial and final states that are usually assumed
to be eigenstates of angular momentum, parity, as well as isospin. It
is then convenient to do a multipole expansion of the current
operators.  In that way one obtains the Coulomb, longitudinal,
transverse electric and transverse magnetic multipoles defined in
\citep[p.  136]{Walecka:1975}. The expressions necessary for the
calculation of the processes shown in figure~\ref{fig:weakprocess} can
be obtained from references \cite{Walecka:1975,Donnelly.Peccei:1979}
in terms of the multipole operators. In general the multipole
operators are $A$-body nuclear operators (with $A$ the nucleon
number). In practice, at the energy scales we are interested in, weak
interactions perturb the nucleus only slightly, so that to a good
approximation one-body components dominate most of the transitions.
Two-body meson exchange currents and other many body effects are
neglected \citep[see][for a description of the nuclear current
including two-body
operators]{Marcucci.Schiavilla.ea:2001,Schiavilla.Wiringa:2002}. It is
further assumed that a nucleon in a nucleus undergoing a weak
interaction can be treated as a free nucleon, which for the purpose of
constructing interaction operators satisfies the Dirac equation.  This
latter approximation is known as the impulse approximation. For a
single free nucleon, we have, using Lorentz covariance, conservation
of parity, time-reversal invariance, and isospin invariance, the
following general form for the vector and axial vector currents

\begin{widetext}
\begin{subequations}
  \label{eq:4}
  \begin{eqnarray}
    \langle \bm{k}' \lambda'; 1/2 m_{t'} | V_\mu^{TM_T} | \bm{k}
    \lambda; 1/2 m_t \rangle = i  \bar{u}(\bm{k}' \lambda') \left[
    F_1^{(T)} \gamma_\mu + F_2^{(T)} \sigma_{\mu\nu}q_\nu \right]
    u(\bm{k}\lambda) \langle 1/2 m_{t'} | I_T^{M_T} | 1/2 m_t
    \rangle, \label{eq:4a}\\
    \langle \bm{k}' \lambda'; 1/2 m_{t'} | A_\mu^{TM_T} | \bm{k}
    \lambda; 1/2 m_t \rangle = i \bar{u}(\bm{k}' \lambda') \left[
    F_A^{(T)} \gamma_5 \gamma_\mu - i F_P^{(T)} \gamma_5 q_\mu \right]
    u(\bm{k}\lambda) \langle 1/2 m_{t'} | I_T^{M_T} | 1/2 m_t
    \rangle. \label{eq:4b}
  \end{eqnarray}
\end{subequations}
\end{widetext}
Here, the plane-wave single nucleon states are labelled with the
three-momenta $\bm{k}$ ($\bm{k}'$), helicities $\lambda$ ($\lambda'$),
isospin $1/2$ and isospin projections $m_t$ ($m_{t'}$). The momentum
transfer, $q_\mu^2 = q^2 - \omega^2$, with $q=|\bm{q}|$, is defined in
figure~\ref{fig:weakprocess}. Bold letters denote the three-momentum.
The single-nucleon form factors $F_X^{(T)}=F_X^{(T)}(q_\mu^2)$,
$T=0,1$, $X=1,2,A,P$ (vector Dirac, vector Pauli, axial,
and pseudoscalar) are all functions of $q_\mu^2$
\cite{Donnelly.Peccei:1979,Kuramoto.Fukugita.ea:1990,Beise.McKeown:1991,%
Musolf.Donnelly:1992}.  Second class currents are not included in
equation~\eqref{eq:4}.  The isospin dependence in
equations~\eqref{eq:4} is contained in

\begin{equation}
  \label{eq:5}
  I_T^{M_T} \equiv \frac{1}{2} \times \left\{
    \begin{array}{ll}
      1 & T=0, M_T =0 \\
      \tau_0 & T=1, M_T = 0 \\
      \tau_{\pm 1} = \mp \frac{1}{\sqrt{2}} (\tau_1 \pm i \tau_2) &
      T=1, M_T = \pm 1
    \end{array} \right.
\end{equation}

To evaluate weak-interaction processes in nuclei, one needs
matrix elements of the multipole operators between nuclear
many-body states, labeled $|J_i M_{J_i}; T_i M_{T_i}\rangle$ which are
complicated nuclear configurations of protons and neutrons. Using the
Wigner-Eckart theorem we can write the matrix element of an arbitrary
multipole operator $\hat{T}_{JM_J;TM_T}$ as \cite{Edmonds:1960}

\begin{widetext}
\begin{eqnarray}
  \label{eq:6}
  \langle J_1 M_{J_1}; T_1 M_{T_1}|\hat{T}_{JM_J;TM_T} (q)|J_2 M_{J_2}; T_2
  M_{T_2} \rangle & = & (-1)^{J_1-M_{J_1}} \left(
    \begin{array}{ccc}
      J_1 & J & J_2 \\
      -M_{J_1} & M_J & M_{J_2}
    \end{array} \right) \nonumber \\
  & \times & (-1)^{T_1 - M_{T_1}} 
  \left(\begin{array}{ccc}
    T_1 & T & T_2 \\
    -M_{T_1} & M_T & M_{T_2}
  \end{array}\right) \langle J_1; T_1|||\hat{T}_{J;T}(q)|||J_2; T_2
  \rangle 
\end{eqnarray}
\end{widetext}
where the symbol $|||$ denotes that the matrix element is reduced in
both angular momentum and isospin. If we assume that the multipole
operators are one-body operators, we can write \cite{Heyde:1994}

\begin{widetext}
\begin{equation}
  \label{eq:7}
  \langle J_1; T_1|||\hat{T}_{J;T}(q)|||J_2; T_2
  \rangle= \sum_{\alpha \alpha'} \frac{\langle J_1;
  T_1|||[a^\dagger_{\alpha'} \otimes \tilde{a}_\alpha]_{J;T}|||J_2; T_2
  \rangle}{\sqrt{(2 J+1)(2 T+1)}} \langle \alpha' ||| T_{J;T}(q)
  |||\alpha\rangle 
\end{equation}
\end{widetext}
with the sums extending over complete sets of single-particle
wavefunctions $\alpha={n,l,j}$. The tensor product involves the
single-particle creation operator $a^\dagger_\alpha \equiv
a^\dagger_{\alpha; m_{j_\alpha} m_{t_\alpha}}$ and $\tilde{a}_\alpha
\equiv (-1)^{j_\alpha - m_\alpha} (-1)^{1/2 - m_{t_\alpha}} a_{\alpha;
  -m_{j_\alpha} -m_{t_\alpha}}$, with $a$ the destruction operator.
The phase factor is introduced so that the operator $\tilde{a}$
transforms as a spherical tensor \cite{Edmonds:1960}.

In practice the infinite sums in equation~\eqref{eq:7} are
approximated to include a finite number of (hopefully) dominant terms.
The number of terms to include depends both of the computed observable
and the model used (Shell-Model, Random phase approximation, \ldots).
Typical nuclear models are non-relativistic, requiring a
non-relativistic reduction of the single-particle operators; the
respective expressions are given for example by \citet{Walecka:1975}
and \citet{Donnelly.Peccei:1979}.  \citet{Donnelly.Haxton:1979} give
the expressions for the single-particle matrix elements of these
operators with harmonic oscillator wave functions.
\citet{Donnelly.Haxton:1980} provide expressions for general wave
functions.

The above discussion presents the general theory of semileptonic
processes. However, in many applications the momentum transfers
involved are small compared with the typical nuclear momentum $Q
\approx R^{-1}$, with R the nuclear radius. In that case, the above
formulas can be expanded in powers of $(qR)$ (long-wavelength limit)
and one obtains the standard approximations to allowed (Gamow-Teller
and Fermi) and forbidden transitions \cite{Behrens.Buehring:1982,%
Behrens.Buehring:1971,Bambynek.Behrens.ea:1977}. In these limits the
effect of the electromagnetic interaction on the initial or final
charged lepton, that has been neglected in the above expressions, can be
included \cite{Schopper:1966}.

\subsection{Nuclear models}
\label{sec:nuclear-models}

As discussed in the previous section one of the basic ingredients for
the evaluation of weak-interaction processes involving nuclei is the
description of the nuclear many-body states. Moreover, the calculation
of weak processes in stars have to account for the peculiarities of
the medium (high temperatures and densities) and the presence of an
electron plasma.  When the temperatures and densities are small (for
example during the r- and s-processes) weak transitions could be
determined using the experimentally measured half-lives (in some cases
one has to account for the presence of low lying isomeric states).
However, as many of the very neutron-rich nuclei that participate in
the r-process, are not currently accessible experimentally
\citep[see][for recent experimental advances in the study of r-process
nuclei]{Pfeiffer.Kratz.ea:2001}, the necessary nuclear properties have
to be extracted from theoretical models.

\begin{figure*}[htbp]
  \includegraphics[width=0.9\linewidth]{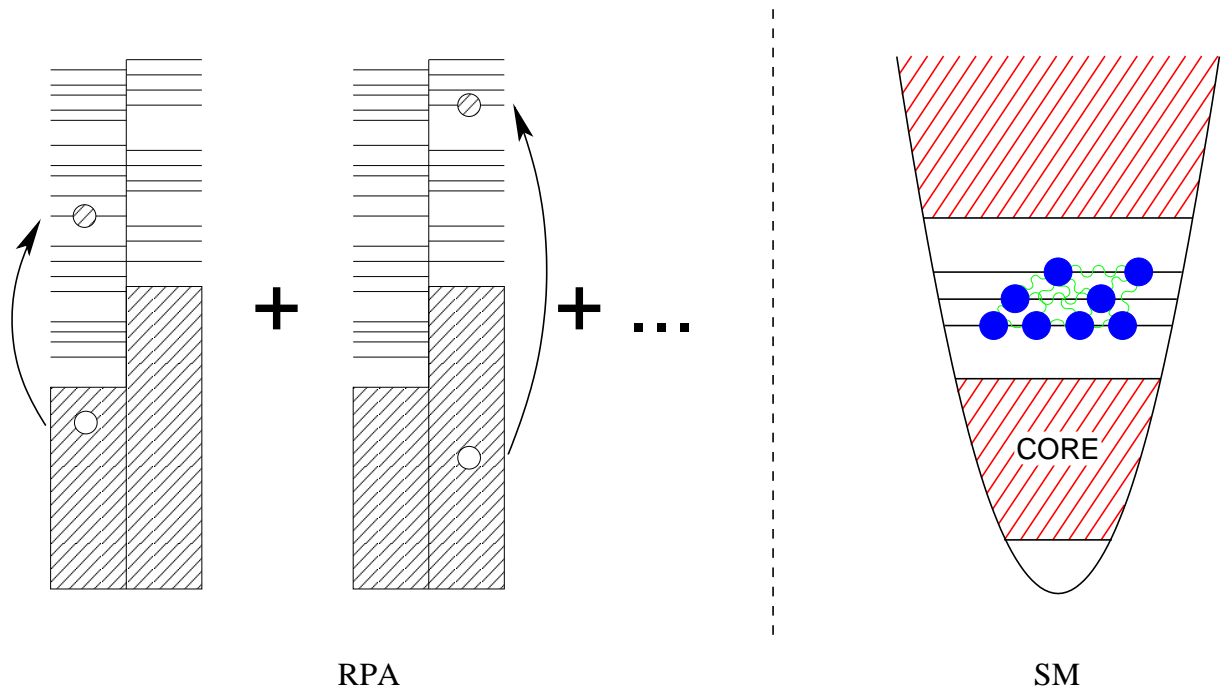}
  \caption{The most commonly used nuclear models for the calculation
    of weak processes in stars are the Random Phase Aproximation (RPA)
    and the Shell-Model (SM). In the RPA the basis states are
    characterized by particle-hole excitations around a given
    configuration (typically a closed-shell nucleus). In the SM, all
    the possible two-body correlations in a given valence space are
    considered. Excitations from the core or outside the model space
    are neglected, but his effect can be included perturbatively
    using effective interactions and operators.\label{fig:rpasm}}
\end{figure*}

As the degrees of freedom increase drastically with the number of
nucleons, models of different sophistication have to be chosen for the
various regions in the nuclear charts.  Exact calculations using
realistic nucleon-nucleon interactions, e.g.\ by Green's Function Monte
Carlo techniques, are restricted to light nuclei with mass $A \le
10$ \cite{Carlson.Schiavilla:1998,Wiringa.Pieper.ea:2000,Pieper:2002}. As an
alternative, methods based on effective field
theory \cite{Kolck:1999,Beane.Bedaque.ea:2001} have recently been
developed for very light
nuclei ($A \le 3$) \cite{Marcucci.Schiavilla.ea:2001,%
Park.Marcucci.ea:2001a,Park.Marcucci.ea:2001b,%
Marcucci.Schiavilla.ea:2000}. For heavier nuclei different
approximations are required. In particular, restricted model spaces
are used so that effective interactions and operators are necessary
\cite{Hjorth-Jensen.Kuo.Osnes:1995}.  For medium-mass nuclei ($A \le
70$) the shell model is the method of choice \cite{Talmi:1993}. This
model explicitly treats all two-body correlations among a set of
valance particles by a residual interaction.  By diagonalizing the
respective Hamiltonian matrix in the model space spanned by the
independent particle states of the valence particles a quite
satisfactory description of the ground state, the spectrum at moderate
excitation energies and the electromagnetic and weak transitions among
these states are obtained
\cite{Caurier.Zuker.ea:1994,Martinez-Pinedo.Poves.ea:1997}.  In recent
times due to progress both in computer technology and programming
techniques shell-model calculations are now possible in model spaces
which seemed impossible only a few years ago; i.e.\ the
diagonalization codes \textsc{antoine} or \textsc{nathan} developed by
Etienne Caurier allow for complete calculations in the $pf$-shell
where the maximum dimension currently attained is $2.3 \times 10^9$
$M=0$ slater determinants for a complete diagonalization of $^{60}$Zn
\cite{Caurier:2002,Mazzocchi.Janas.ea:2001}. To treat even larger
model spaces in the diagonalization shell model, different schemes are
required.  One method is to expand the nuclear many-body wave
functions in terms of a few symmetry-projected Hartree-Fock-Bogoliubov
(HFB) type quasiparticle determinants
\cite{Schmid.Grummer.Faessler:1987,Schmid.Zheng.ea:1989,Schmid:2001}.
A novel approach, introduced by \citet{Honma.Mizusaki.Otsuka:1995}
\citep[see also][]{Otsuka.Honma.ea:2001}, employs stochastical methods
to determine the most important Slater determinants in the chosen
model space.  As an alternative to the diagonalization method, the
Shell Model Monte Carlo (SMMC)
\cite{Johnson.Koonin.ea:1992,Koonin.Dean.Langanke:1997} allows
calculation of nuclear properties as thermal averages, employing the
Hubbard-Stratonovich transformation to rewrite the two-body parts of
the residual interaction by integrals over fluctuating auxiliary
fields.  The integrations are performed by Monte Carlo techniques,
making the SMMC method available for basically unrestricted model
spaces.  While the strength of the SMMC method is the study of nuclear
properties at finite temperature, it does not allow for detailed
nuclear spectroscopy.

The evaluation of nuclear matrix elements for the Fermi operator is
straightforward. The Gamow-Teller operator connects Slater
determinants within a model space spanned by a single harmonic
oscillator shell ($0 \hbar \omega$ space).  The shell model is then
the method of choice to calculate the nuclear states involved in
weak-interaction processes dominated by allowed transitions as
complete or sufficiently converged truncated calculations are nowadays
possible for such $0 \hbar \omega$ model spaces.  The practical
calculation of the Gamow-Teller distribution is achieved by adopting
the Lanczos method \cite{Wilkinson:1965} as proposed
by \citet{Whitehead:1980}; \citep[see
also][]{Poves.Nowacki:2001,Langanke.Poves:2000}.

The calculation of forbidden transitions, however, involves nuclear
transitions between different harmonic oscillator shells and thus
requires multi-$\hbar \omega$ model spaces. These are currently only
feasible for light nuclei where \emph{ab initio} shell model
calculations are possible
\cite{Navratil.Vary.Barrett:2000,Caurier.Navratil.ea:2001}.  Such
multi-$\hbar\omega$ calculations have been used for the calculation of
neutrino scattering from $^{12}$C
\cite{Hayes.Towner:2000,Volpe.Auerbach.ea:2000}.  However, for heavier
nuclei one has to rely on more strongly truncated nuclear models. As
the kinematics of stellar weak-interaction processes are often such
that forbidden transitions are dominated by the collective response of
the nucleus the Random Phase Appoximation \cite{Rowe:1968} is usually
the method of choice (figure~\ref{fig:rpasm}).  Another advantage of
this method is that, in contrast to the shell model, it allows for
global calculations of these processes for the many nuclei often
involved in nuclear networks. An illustrative example is the
evaluation of nuclear half-lives based on the calculation of the GT
strength function within the Quasiparticle RPA model
\cite{Krumlinde.Moeller:1984,Moeller.Randrup:1990}.  The RPA method
considers the residual correlations among nucleons via one particle
one hole (1p-1h) excitations in large multi-$\hbar \omega$ model
spaces.  Compared to the shell model, the neglect of higher-order
correlations renders the RPA method inferior for matrix elements
between individual, non-collective states. A prominent example is the
GT transition from the $^{12}$C ground state to the $T=1$ triad in the
$A=12$ nuclei \citep[e.g.][]{Engel.Kolbe.ea:1996}. While the shell
model is able to reproduce the GT matrix element between these states
\cite{Cohen.Kurath:1965,Warburton.Brown:1992}, RPA calculations miss
an important part of the nucleon correlations and overestimate these
matrix elements by about a factor of 2
\cite{Kolbe.Langanke.Vogel:1994,Engel.Kolbe.ea:1996}.  Recent
developments have extended the RPA method to include the complete set
of 2p-2h excitations in a given model space
\cite{Drozdz.Nishizaki.ea:1990}. Such 2p-2h RPA models have, however,
not yet been applied to semileptonic weak processes in stars.
Moreover, the RPA allows for the proper treatment of the
momentum-dependence in the different multipole operators, as it can be
important in certain stellar neutrino-nucleus processes (see below),
and for the inclusion of the continuum \cite{Buballa.Drozdz.ea:1991}.
Detailed studies indicate that standard and continuum RPA calculations
yield nearly the same results for total semileptonic cross sections
\cite{Kolbe.Langanke.Vogel:2000}.  This is related to the fact that
both RPA versions obey the same sumrules.  The RPA has also been
extended to deal with partial occupation of the orbits so that
configuration mixing in the same shell is included schematically
\cite{Rowe:1968,Kolbe.Langanke.Vogel:1999}.

\section{Hydrogen burning and solar neutrinos}

The tale of the solar neutrinos and their `famous' problem took an
exciting twist from its original goal of measuring the central
temperature of the Sun to providing convincing evidence for neutrino
oscillations, thus opening the door to physics beyond the standard
model of the weak interaction. In 1946, Pontecorvo suggested
\cite{Pontecorvo:1946,Pontecorvo:1991} \citep[later independently
proposed by][]{Alvarez:1949} that chlorine would be a good detector
material for neutrinos and subsequently in the 1950's Davis built a
radiochemical neutrino detector which observed reactor neutrinos via
the $^{37}$Cl($\nu_e,e^-)^{37}$Ar reaction \cite{Davis:1955}. After
the $^{3}$He($\alpha,\gamma){}^7$Be cross section at low energies had
been found to be significantly larger than expected
\cite{Holmgren.Johnston:1958} and, slightly later, the
$^7$Be$(p,\gamma){}^8$B cross section at low energies had been
measured \cite{Kavanagh:1960}, it became clear that the Sun should
also operate by what are now known as the ppII and ppIII chains and in
that way generate neutrinos with energies high enough to be detectable
by a chlorine detector \cite{Fowler:1958,Cameron:1958}.  This idea was
then seriously pursued by Davis, in close collaboration with Bahcall.
The observed solar neutrino flux turned out to be lower than predicted
by the solar models (the original solar neutrino problem)
\cite{Davis.Harmer.Hoffman:1968,Bahcall.Bahcall.Shaviv:1968},
triggering the development of further solar neutrino detectors,
initiating the field of neutrino oscillation experiments and, after
precision helioseismology data \cite{Dalsgaard:2002}
boosted the confidence in the solar
models, finally culminating in the conclusive evidence for neutrino
oscillations in the solar flux.  A detailed recent review of the solar
hydrogen burning and neutrino problem is given in \cite{Kirsten:1999}.

\begin{figure}[htbp]
  \includegraphics[angle=270,width=0.9\linewidth]{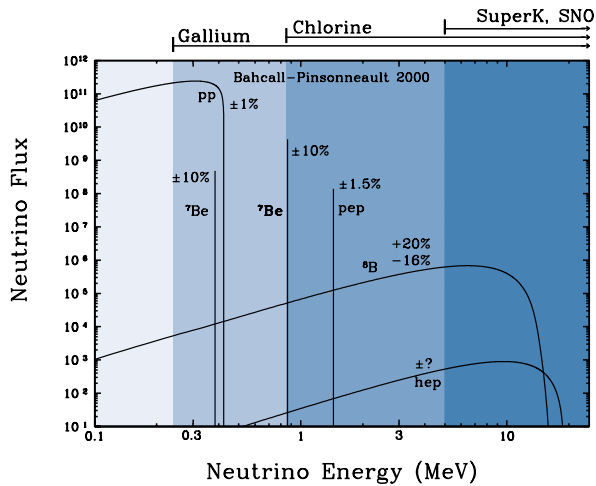}
  \caption{The energy spectrum of neutrinos predicted by the standard
    solar model \protect\cite{Bahcall.Pinsonneault.Basu:2001}. The
    neutrino fluxes from continuum sources (like pp and $^8$B) are
    given in the units of number per cm$^2$ per second per MeV, while
    the line fluxes (e.g.  $^7$Be) are given in number per cm$^2$ per
    second. The ranges of neutrino energies observable in the various
    detectors are indicated by arrows. The uncertainties in the
    various fluxes are given in percent (courtesy of J. N. Bahcall).
     \label{fig:snuspect}}
\end{figure}

\begin{table}[htbp]
  \newlength{\tint}
  \setlength{\tint}{\baselineskip}
  \begin{tabular}{lcc}
    \hline\hline
    \multicolumn{1}{c}{Reaction} & Term & $\nu$ Energy \\
    & (\%) &  (MeV) \\ \hline 
    $p + p \rightarrow {}^2\text{H} + e^+ + \nu_e$  & 99.96 & $\leq
    0.423$\\
    \multicolumn{1}{c}{or} & & \\
    $p + e^- + p \rightarrow {}^2\text{H} + \nu_e$ & 0.44 & 1.445
    \\[1.5\tint] 
    $^2\text{H} + p \rightarrow {}^3\text{H} + \gamma$ & 100 & \\
    $^3\text{He} + {}^3\text{He} \rightarrow \alpha + 2p$ & 85 & \\
    \multicolumn{1}{c}{or} & & \\
    $^3\text{He} + {}^4\text{He} \rightarrow {}^7\text{Be} + \gamma$
    & 15 & \\
    $\qquad {}^7\text{Be} + e^- \rightarrow {}^7\text{Li} + \nu_e$ &
    15 & $\left\{\begin{array}{r} 0.863\ 90\%\\ 0.385\ 10\%
      \end{array}\right.$ \\ 
    $\qquad {}^7\text{Li} + p \rightarrow 2\alpha$ & & \\
    \multicolumn{1}{c}{or} & & \\
    $^7\text{Be} + p \rightarrow {}^8\text{B} + \gamma $ & 0.02 & \\
    $\qquad {}^8\text{B} \rightarrow {}^8\text{Be}^* + e^+ + \nu_e$ &
    & $< 15$\\
    $\qquad {}^8\text{Be}^* \rightarrow 2\alpha$ \\
    \multicolumn{1}{c}{or} & & \\
    $^3\text{He} + p \rightarrow {}^4\text{He} + e^+ + \nu_e$ & 0.00003 &
    $<18.8$ \\ \hline\hline
  \end{tabular}
  \caption{The solar pp chains. The neutrino terminations are from
    the BP2000 solar model \cite{Bahcall.Pinsonneault.Basu:2001}. The neutrino
    energies include the solar corrections
    \cite{Bahcall:1997}. \label{tab:ppchain}} 
\end{table}

The Sun generates its energy from nuclear fusion reactions in the pp
chain (see table~\ref{tab:ppchain}), with a small contribution by the
CNO cycle.  Several of these reactions are mediated by the weak
interaction, and hence create (electron) neutrinos which in the
standard solar model can leave the Sun unhindered. The predicted flux
of solar neutrinos on the surface of the earth is shown in
figure~\ref{fig:snuspect}. These predictions depend on the knowledge
of the relevant nuclear cross sections at solar energies (a few keV)
which, with the notable exception of the $^3$He($^3$He,2p)$^4$He
reaction \cite{Bonetti.Broggini.ea:1999} which has been measured
directly in the underground laboratory in the Gran Sasso, relies on
the extrapolation of data taken at higher energies. As these reactions
are all non-resonant, the extrapolations are quite mild and appear to
be under control \cite{Adelberger.Austin.ea:1998}. The cross section
for the initial $p+p$ fusion reaction is so low that no data exist and
the respective solar reaction rate relies completely on theoretical
modelling.  Nevertheless the underlying theory is thought to be under
control and the uncertainty in this important rate is estimated to be
about $1\%$ \cite{Kamionkowski.Bahcall:1994}, based on potential model
calculations, and it is probably even smaller if effective field
theory is applied
\cite{Park.Marcucci.ea:2001a,Park.Kubodera.ea:1998,Kong.Ravndal:2001}.
In the solar plasma the reaction rates are slightly enhanced due to
screening effects
\cite{Dzitko.Turck-Chieze.ea:1995,Gruzinov.Bahcall:1998}. The most
significant plasma modification is found for the lifetime of $^7$Be
with respect to electron capture where capture of continuum and bound
electrons with the relevant screening corrections have to be accounted
for \cite{Johnson.Kolbe.ea:1992,Gruzinov.Bahcall:1998}.  It is
generally believed that the $^7$Be$(p,\gamma){}^8$B reaction is the
least known nuclear input in nuclear models. Although this reaction
occurs in the weak ppIII chain, the decay of $^8$B is the source of
the high-energy neutrinos observed by the solar neutrino detectors.
Recent direct and indirect experimental methods have improved the
knowledge of the $^7$Be$(p,\gamma)^8$B rate considerably
(\citet{Davids.Anthony.ea:2001} and references therein). While these
data point to an astrophysical S-factor in the range of 18--20~eV~b, a
very recent direct measurement with special emphasis on the control
and determination of the potential errors yielded a slightly larger
value \cite{Junghans.Mohrmann.ea:2001,Junghans.Snover:2002}.  The
neutrino energy distribution arising from the subsequent $^8$B decay
has been measured precisely by \citet{Ortiz.Garcia.ea:2000}.  In
principle, high-energy neutrinos are also produced in the
$^3\text{He}+p$ fusion reaction which, however, occurs only in a weak
branch in the solar pp cycles.  Although the calculation of this cross
section represents a severe theoretical challenge, it appears to be
determined now with the required accuracy using state-of-the-art
few-body
methods \cite{Marcucci.Schiavilla.ea:2000,Marcucci.Schiavilla.ea:2001,%
Park.Marcucci.ea:2001a,Park.Marcucci.ea:2001b,Kong.Ravndal:2001}. 


The solar nuclear cross sections have been reviewed by
\citet{Adelberger.Austin.ea:1998}, including also the reactions
occuring in the CNO cycle. Except for some discrepancies in the
$^{14}$N(p,$\gamma)^{15}$O cross section at low energies
\cite{Adelberger.Austin.ea:1998,Angulo.Descouvemont:2001}, all
relevant solar rates are sufficiently well known.

There are currently 5 solar neutrino detectors operating. Three of
them, the homestake chlorine detector \citep[p. 487]{Bahcall:1989},
GALLEX\footnote{The Gallex detector has been recently upgraded and
  has changed its name to GNO \cite{Altmann.Balata.ea:2000}}
\cite{Anselmann.Hampel.ea:1992}, and SAGE
\cite{Abdurashitov.Faizov.ea:1994}) can only observe charge-current
(electron) neutrino reactions, while the two water Cerenkov detectors
(Super-Kamiokande \cite{Fukuda.Hayakawa.ea:1998a}, SNO
\cite{Boger.Hahn.ea:2000}) also observe neutral-current events, which
can be triggered by all neutrino flavors. All neutrino detectors have
characteristic energy thresholds for neutrino detection, dictated by
the various observation schemes; i.e.\ the detectors are blind for
neutrinos with energies less than the threshold energy $E_{th}$. 
The pioneering
chlorine experiment of Davis uses the $^{37}$Cl($\nu_e,e^-)^{37}$Ar
reaction as detector, with $E_{th}=814 $ keV. Gallex and Sage detect
neutrinos via $^{71}$Ga($\nu_e,e^-)^{71}$Ge with the threshold energy
$E_{th}$=233.2~keV. In Super-Kamiokande (SK) solar neutrinos are
identified by the observation of relativistic electrons produced from
inelastic $\nu+e^-$ scattering. Due to high background at low
energies, the observational threshold is set to $\sim 7$ MeV. SNO has
an inner vessel of heavy water, surrounded by normal water. Like SK,
this detector can also observe neutrinos via inelastic scattering off
electrons.  Additionally, and more importantly, SNO can also detect
neutrinos by the dissociation of the deuteron in heavy water,
with the threshold energy of order 6~MeV.


The threshold energies and the predicted solar neutrino fluxes are
shown in figure~\ref{fig:snuspect}. One notes that SK and SNO are only
sensitive to $^8$B neutrinos (neglecting the weak $hep$ flux), the
chlorine experiment detects mainly $^8$B (76$\%$ of the predicted flux
by \citet{Bahcall.Pinsonneault.Basu:2001}) and $^7$Be neutrinos, while
Gallex and Sage can also observe neutrinos generated in the main solar
energy source, the $p+p$ fusion reaction (54$\%$ of the predicted flux).
It is important to note that the solar neutrino detectors have been
calibrated, using known neutrino sources.

\begin{figure*}[htbp]
  \includegraphics[angle=270,width=0.9\linewidth]{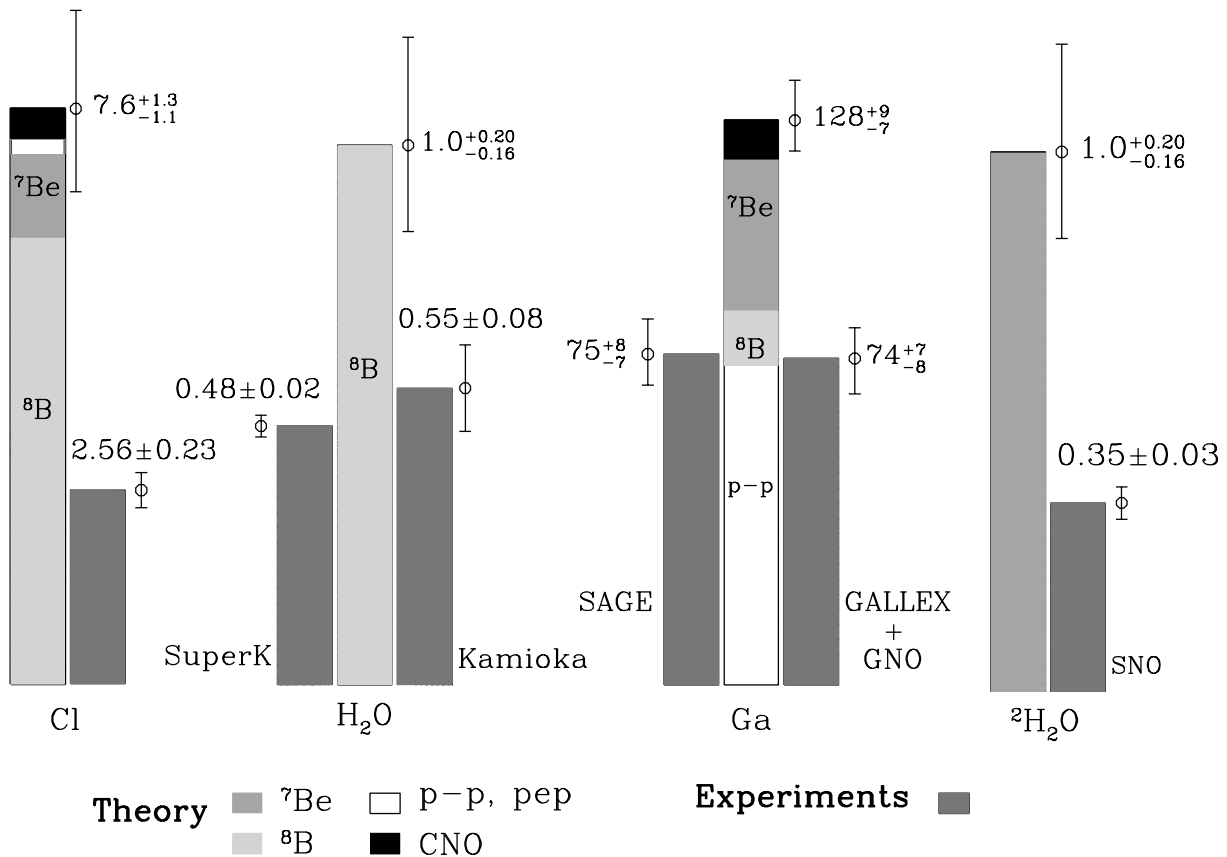}
  \caption{Comparison of the predicted solar neutrino
    fluxes \cite{Bahcall.Pinsonneault.Basu:2001} for the various
    neutrino detectors with the observed fluxes. The unit for the
    chlorine detector (Cl) and the Gallium detectors (Ga) are Solar
    Neutrino Units (SNU) \citep[see][]{Bahcall:1989}, while for SNO
    and SK the comparison is in percentage of the predicted flux
    (courtesy of John N. Bahcall). \label{fig:snudetect}}
\end{figure*}

The original solar neutrino problem constitutes the fact that the
earthbound detectors observe less neutrinos than predicted by the
solar model. The current comparison is depicted in
figure~\ref{fig:snudetect}. Importantly, Sage and Gallex, in close
agreement to each other, observe at least a neutrino flux which is
consistent with the fact that the current solar luminosity is powered
by the $p+p$ fusion
reaction \cite{Hampel.Handt.ea:1999,Abdurashitov.Gavrin.ea:1999}. With
improved input (nuclear reaction rates, opacities, etc.) the solar
models evolved and, as a milestone, passed the stringent test of
detailed comparison to the soundspeed distribution derived from
helioseismology \cite{Christensen-Dalsgaard.Dappen.ea:1996}.  It
became clear that the solution to the solar neutrino problem pointed
to weak-interaction physics beyond the standard model. This line of
reasoning was supported by the
observation \cite{Heeger.Robertson:1996} that any solar model assuming
standard weak-interaction physics leads to contradictions between the
observed fluxes in the various detectors.

It has been speculated already for a long time \cite{Pontecorvo:1968}
that the solution to the deficient observed neutrino flux lies in the
possibility that neutrinos change their flavor on their way from the
center of the Sun to the earthbound detectors. Neutrino oscillations
can occur if the flavor eigenstates (the physical $\nu_e, \nu_\mu,
\nu_\tau$ neutrinos) are not identical with the mass eigenstates
$(\nu_1,\nu_2,\nu_3$) of the weak Hamiltonian, but rather are given by
a unitary transformation of these states defined by a set of mixing
angles. Importantly, oscillations between two flavor states can only
occur if at least one of these states does not propagate with the
speed of light implying that this neutrino has a mass different from
zero; more precisely $\Delta m^2 = m_1^2 - m_2^2 \ne 0$, where
$m_{1,2}$ are the masses of the oscillating neutrinos. As all neutrino
masses are assumed to be zero in the Weinberg-Salam model, the
observation of neutrino oscillations opens the door to new physics
beyond the standard model of weak interaction. Neutrino oscillations
can occur for free-propagating neutrinos (vacuum oscillations).
However, their occurence can also be influenced by the environment. In
particular, it has been pointed out that the high-energy ($\nu_e$)
solar neutrinos can, for a certain range of mixing angles and mass
differences, transform resonantly into other flavors, mediated by the
interaction of the $\nu_e$ neutrinos with the electrons in the solar
plasma, resulting in matter-enhanced oscillations \citep[the so-called
MSW effect][]{Wolfenstein:1978,Mikheyev.Smirnov:1986}.

\begin{figure}[htbp]
  \includegraphics[width=0.9\linewidth]{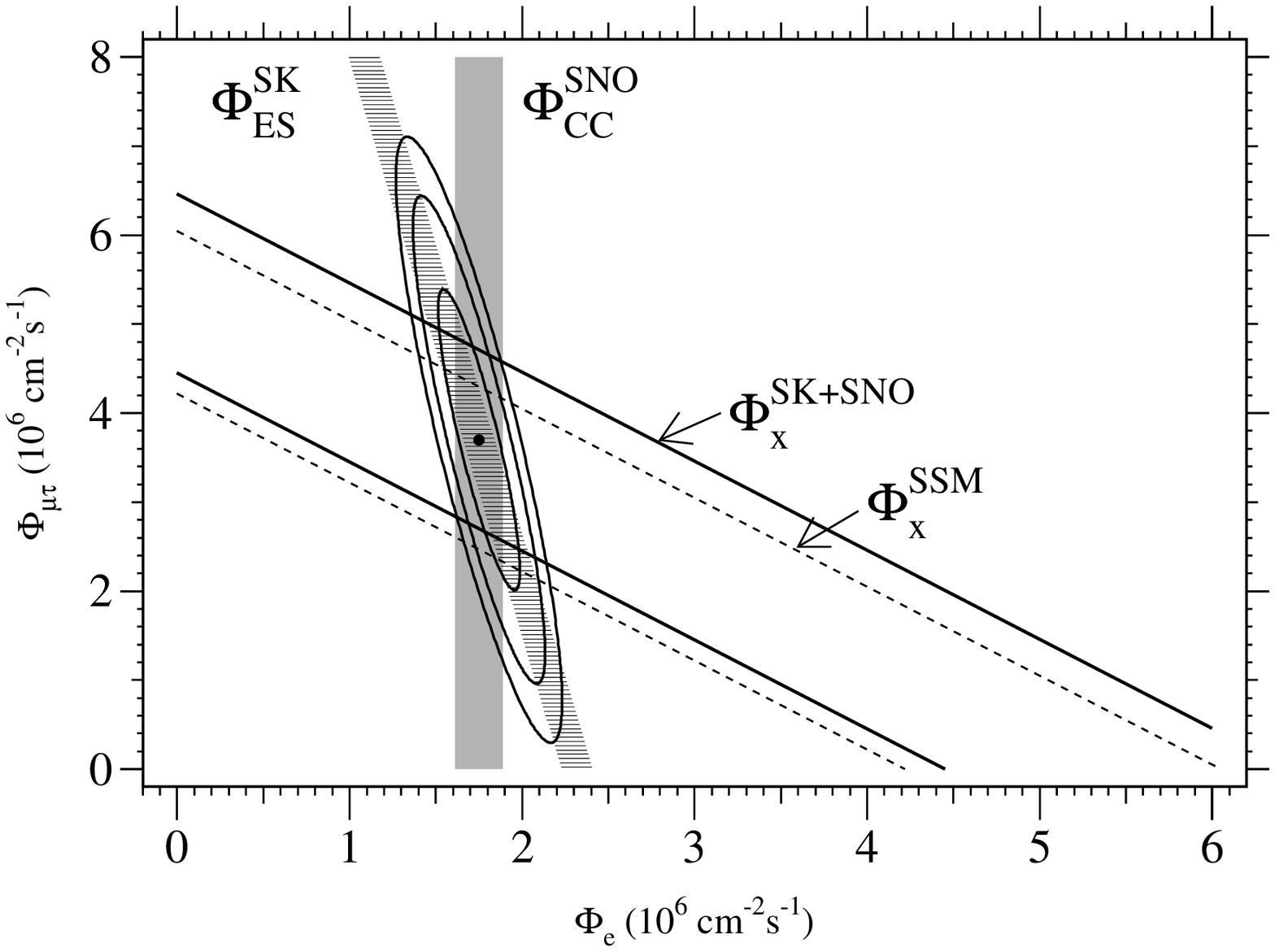}
  \caption{Flux of $\nu_{\mu,\tau}$ neutrinos vs.\ electron neutrinos as
    deduced from the SNO and SK $^8$B neutrino data. The diagonal
    bands show the total $^8$B flux as predicted by the standard solar
    model (SSM, dashed lines) \cite{Bahcall.Pinsonneault.Basu:2001}
    and that derived from the SNO and SK measurements (solid lines).
    The intercepts with the axis represent $1\sigma$ errors
    \citep[from][]{Ahmad.Allen.ea:2001}. \label{fig:SNO}}
\end{figure}

First clear evidence for neutrino oscillations was reported by the SK
collaboration which observed a deficit of $\nu_\mu$-induced events
from atmospheric neutrinos and could link this deficit to $\nu_\mu
\rightleftarrows \nu_\tau$ oscillations
\cite{Fukuda.Hayakawa.ea:1998b,Fukuda.Hayakawa.ea:1999b}. [Further evidence
for neutrino oscillations has been given by the LSND
collaboration \cite{Athanassopoulos.Auerbach.ea:1996,%
  Athanassopoulos.Auerbach.ea:1998}.  This result, however, was, for
most of the allowed parameter space not confirmed by the Karmen
experiment \cite{Armbruster.Blair.ea:1998a,Armbruster.Blair.ea:1998b}.
The complete LSND result will be tested by the
MiniBoone\footnote{http://www-boone.fnal.gov/} experiment which is
currently under construction.]  A clear link between neutrino
oscillations and the solar neutrino problem has been presented
recently by a combined analysis \cite{Ahmad.Allen.ea:2001} of the
first SNO data with the precise SK data \cite{Fukuda.Fukuda.ea:2001b}.
SNO measured the integrated event rate above the kinetic energy
threshold $T_{\text{eff}}=6.75 $ MeV (the electron energy threshold
then is $E_{\text{th}}=T_{\text{eff}}+0.511$ MeV) for charged-current
(CC) reactions on the deuteron and inelastic electron scattering (ES).
As no evidence for a deviation of the spectral shape from the
predicted shape under the no-oscillation hypothesis has been observed,
the integrated rate could be converted into the measured $^8$B
neutrino flux, resulting in $\Phi^{\text{CC}}_{\text{SNO}} (\nu_e) =
1.75 \pm 0.07 (\text{stat}) \pm 0.12 (\text{syst}) \pm 0.05
(\text{theor}) \times 10^6$ cm$^{-2}$~s$^{-1}$,
$\Phi^{\text{ES}}_{\text{SNO}} (\nu) = 2.39 \pm 0.34 (\text{stat}) \pm
0.15 (\text{syst}) \times 10^6$ cm$^{-2}$~s$^{-1}$.  The SNO electron
scattering flux result agrees with the more precise measurement from
SK which yields $\Phi^{\text{ES}}_{\text{SK}} (\nu) = 2.32 \pm 0.03
(\text{stat}) \pm 0.08 (\text{syst}) \times 10^6$~cm$^{-2}$~s$^{-1}$.
We note that the charged-current reaction can only be triggered by
$\nu_e$ neutrinos at the energies of the solar neutrinos.  Thus, from
the measurement of the $\nu_e+D$ event rate, SNO has determined the
solar $\nu_e$-flux arriving on earth stemming from the decay of $^8$B.
On the other hand, neutrino-electron scattering can occur for all
neutrino types, whereby the $\nu_e+e^-$ cross section is about seven
times larger than the $\nu_{\mu,\tau} + e^-$ cross section.  If no
oscillations involving solar $\nu_e$ neutrinos occur, the SNO
charged-current flux $\Phi^{\text{CC}}_{\text{SNO}}$ and the SK
inelastic electron scattering flux $\Phi^{\text{ES}}_{\text{SK}}$
should be the same; that is excluded by 3.3 $\sigma$.  [The exclusion
is even slightly more severe if the recent revision of the $\nu_e+D$
cross section including radiative corrections is considered
\cite{Kurylov.Ramsey-Musolf.Vogel:2002}.]  If $\nu_e \rightleftarrows
\nu_{\mu,\tau}$ oscillations occur, $\Phi^{\text{ES}}_{\text{SK}}$
should be larger than $\Phi^{\text{CC}}_{\text{SNO}}$ as it then
contains additional neutral-current contributions from
$\nu_{\mu,\tau}$ neutrinos. From a best fit to the SNO and SK data
(see figure~\ref{fig:SNO}), this contribution has been determined as
(with 1$\sigma$ uncertainty) $\Phi_{\mu,\tau} = 3.69 \pm 1.13 \times
10^6$~cm$^{-2}$~s$^{-1}$ \cite{Ahmad.Allen.ea:2001}, implying that the
total solar flux is $\Phi_{\text{SNO}}^{\text{CC}} (\nu_e) +\Phi
(\nu_{\mu,\tau}) =5.44 \pm 0.99 \times 10^6$~cm$^{-2}$~s$^{-1}$. This
result agrees very nicely with the $^8$B neutrino flux predicted by
the solar model \cite{Bahcall.Pinsonneault.Basu:2001} ($5.05 \times
10^6$~cm$^{-2}$~s$^{-1}$).

For years the measurement of the neutral-current $\nu$+D reaction at
SNO has been anticipated as the `smoking gun' for solar neutrino
oscillations.  After finishing this review, the first results of this
milestone experiment have been published \cite{Ahmad.Allen.ea:2002a}.
They lead to the same conclusions as the earlier SNO results
\cite{Ahmad.Allen.ea:2001} showing a clear excess of neutral-current
over charged-current events, as expected if neutrino oscillations are
the origin of the solar neutrino problem. Furthermore, the observed
neutral-current event rate is again consistent with the prediction of
the solar model \cite{Bahcall.Pinsonneault.Basu:2001}.

A global analysis of the latest solar neutrino data including the SNO
charged-current rate favors matter-enhanced neutrino oscillations with
large mixing angles \cite{Krastev.Smirnov:2001}. Considering the
recent constraints on the $^7$Be$(p,\gamma){}^8$B cross section and the
respectively predicted $^8$B solar neutrino flux, vacuum oscillations
are essentially excluded. A similar result is obtained by
\citet{Bahcall.Gonzalez-Garcia.Pena-Garay:2002} including the recent
day-night asymmetry measured at SNO \cite{Ahmad.Allen.ea:2002b}

For more than 30 years the solar neutrino problem has been a demanding
challenge for experimentalists and theorists, for nuclear, particle and
astrophysicists alike. The challenge appears to be mastered, leading to
new physics and without the need of the many desperate solution
attempts put forward over the years. 

\section{Late-stage stellar evolution}

\subsection{General remarks}

Weak interactions play an essential role already during hydrostatic
burning. Its importance lies in the fact that the neutrinos generated
by these processes can leave the star unhindered, thus carrying away
energy and hence cooling the star. While the consideration of energy
losses by neutrinos is already required during hydrogen burning (see
above), the heat flux in the early stages of stellar burning is
predominantly by radiation. This changes, following helium burning,
when the stellar temperatures reach $\sim 5 \times 10^8$ K and
neutrino-antineutrino pair production and emission becomes the leading
energy loss mechanism. The respective cooling rate is a local property
of the star depending on density $\rho$ and, very sensitively, on
temperature $T$; i.e.\ the energy loss rate for $\nu{\bar{\nu}}$
emission scales approximately like $T^{11}$, implying that the hot
inner regions of the star cool most effectively. However, the dominant
nuclear reactions, occuring after helium burning, have even stronger
temperature dependences. For example, the heat production $\epsilon$
in the $^{12}$C+$^{12}$C or $^{16}$O+$^{16}$O fusion reactions, which
dominate hydrostatic carbon and oxygen burning, scales like $\epsilon
\approx T^{22}$ and $\approx T^{35}$ around $T= 10^9$ K. As a
consequence of the temperature gradient in the stellar interior and
the vast difference in the temperature sensitivity, nuclear reaction
heating overcomes the neutrino energy loss in the center.  However, in
the cooler mantle region surrounding the core neutrino cooling
dominates. The resulting entropy difference leads to convective
instabilities \citep[see][]{Arnett:1996}. First attempts of modelling
late-stage stellar burning and nucleosynthesis including a
two-dimensional treatment of convection is reported in
\cite{Baleisis.Arnett:2001}.

\begin{figure*}[htbp]
  \includegraphics[width=0.90\linewidth]{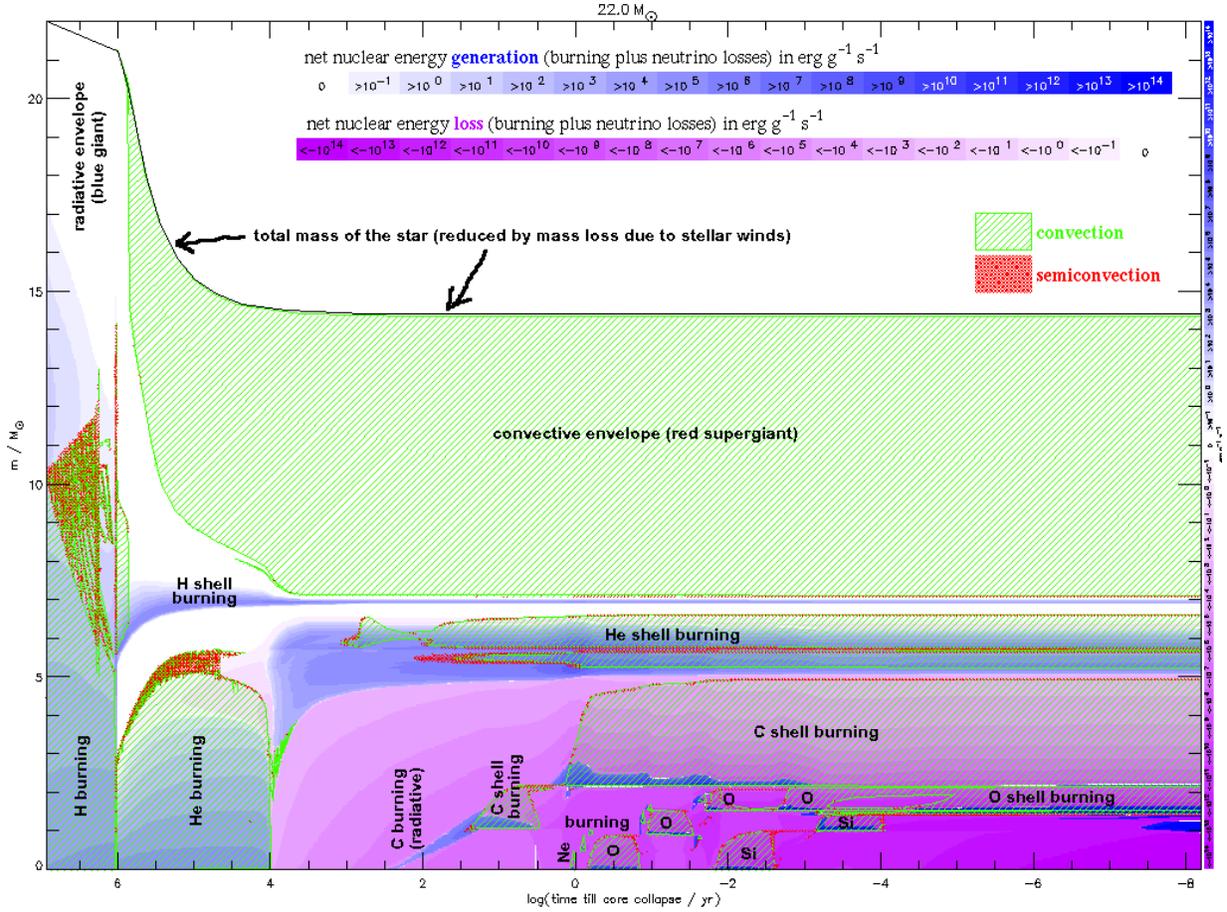}
  \caption{Energy history of a 22~$M_\odot$ star as a function of time
    until core collapse. The y-axis defines the included mass from the
    center. Hydrogen and helium core and shell burning are major
    energy sources. In the later burning stages, following oxygen
    core burning, neutrino losses related to weak processes in the
    stellar interior become increasingly important and can dominate
    over the nuclear energy production.  Convection plays an important
    role in the envelope outside the helium burning shell, but also in
    shells during oxygen and silicon burning
    \citep[from][]{Heger.Woosley.pvt}. \label{fig:Kippenhahn}} 
\end{figure*}

The importance of convection has, of course, already been noticed
before and is accounted for in one-dimensional models within the
so-called mixing-length theory \cite{Clayton:1968}. It has been found
that this convective transport is far more efficient at carrying
energy and mixing the matter composition than radiation transport. For
example, convection dominates the envelope region in massive stars
during helium shell burning, as can be seen in
figure~\ref{fig:Kippenhahn} which shows the energy history of a
22~$M_\odot$ star \cite{Heger.Woosley.pvt}. The figure also identifies
the various subsequent energy reservoirs of the star: hydrogen,
helium, carbon, neon, oxygen, and silicon core and shell burning.
However, the figure also demonstrates the importance of neutrino
losses which, following oxygen core burning, can overcome the nuclear
energy generation, except at the high temperatures in the very inner
core. Obviously weak-interaction processes are crucial in this late
epoch of massive stars. This is not only true for the star's energy
budget, but these processes can also alter the matter composition
and entropy which, in turn, can affect the location and extension of
convective shells, e.g.\ during oxygen and silicon burning, with
subsequent changes in the stellar structure. Such effects have
recently been observed after the improved shell-model weak-interaction
rates (subsection III.2) have been incorporated into stellar models.
(figure~\ref{fig:Kippenhahn} is already based on these rates.)  

\subsection{Shell-model electron capture and $\beta$ decay rates}

The late evolution stages of massive stars are strongly influenced
by weak interactions which act to determine the core entropy and
electron to baryon ratio, $Y_e$, of the presupernova star, hence its
Chandrasekhar mass which is proportional to $Y_e^2$.  Electron capture
reduces the number of electrons available for pressure support, while
beta-decay acts in the opposite direction.  Both processes generate
neutrinos which, for densities $\rho\lesssim 10^{11}$~g~cm$^{-3}$,
escape the star carrying away energy and entropy from the core.

\begin{figure}[htbp]
  \includegraphics[width=0.90\linewidth]{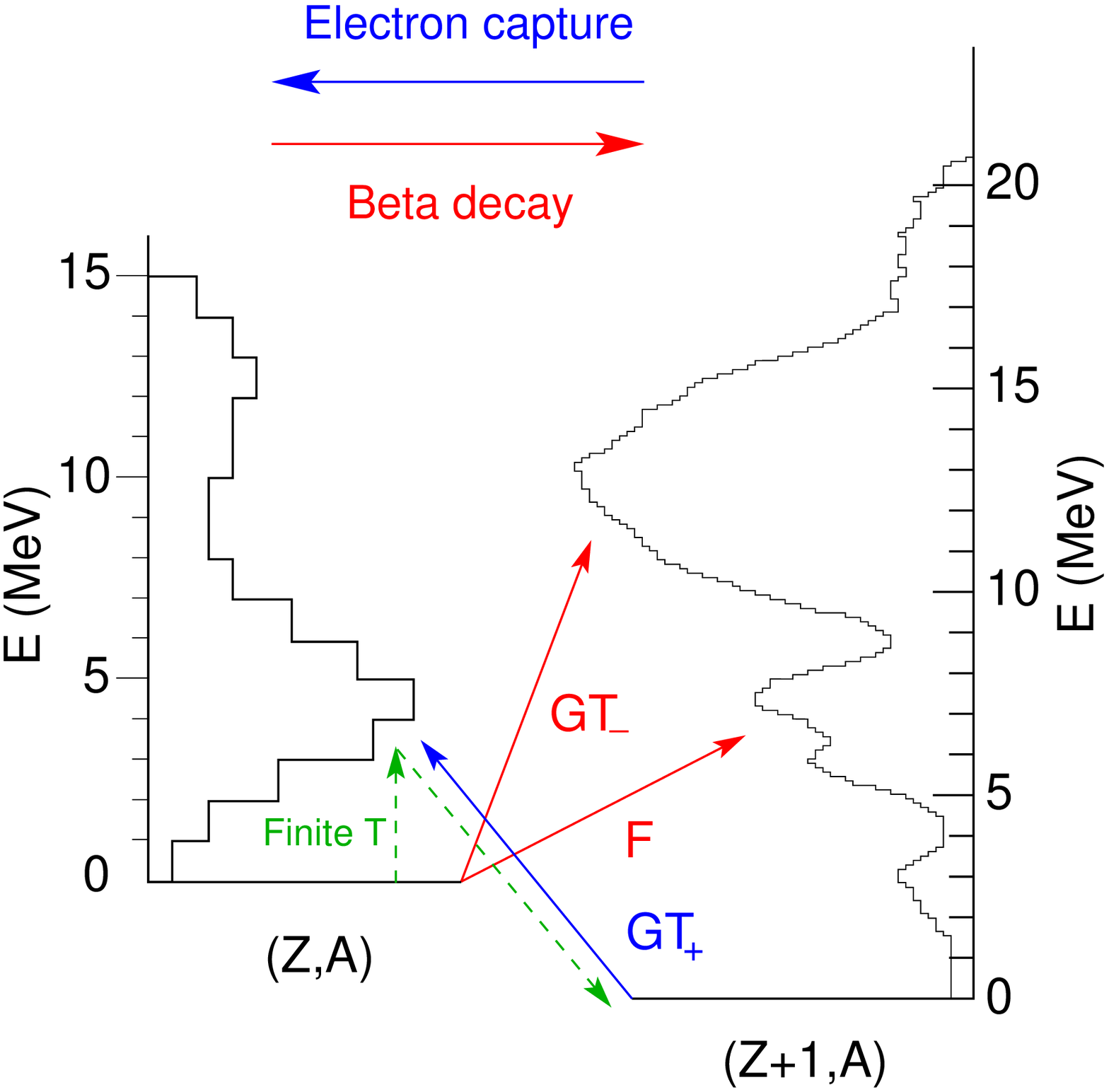}
  \caption{The figure shows schematically the electron capture and
    beta decay processes in the stellar environment. Electron capture
    proceeds by Gamow-Teller transitions to the GT$_+$ resonance. In
    the case of beta decay both the Fermi and Gamow-Teller resonances
    are typically outside of the $Q_\beta$ window,
    and hence are not populated in laboratory decays. Due to the
    finite temperature in stars excited states in the decaying nucleus can be
    thermally populated. Some of these states have strong GT
    transitions to low-lying states in the daughter nucleus. These
    states in the decaying nucleus are called
    ``backresonances''.\label{fig:ecbeta}} 
\end{figure}

Electron capture and beta decay during the final evolution of a
massive star are dominated by Fermi and Gamow-Teller (GT) transitions.
While the treatment of Fermi transitions (important only in beta
decays) is straightforward, a correct description of the GT
transitions is a difficult problem in nuclear structure. In the
astrophysical environment nuclei are fully ionized so one has
continuum electron capture from the degenerate electron plasma. The
energies of the electrons are high enough to induce transitions to the
Gamow-Teller resonance. Shortly after the discovery of this collective
excitation \citet{Bethe.Brown.ea:1979} recognized its importance for
stellar electron capture. The presence of the degenerate electron gas
blocks the phase space for the produced electron in beta decay. Then
the decay rate of a given nuclear state is greatly reduced or even
completely blocked at high densities. However, due to the finite
temperature excited states in the decaying nucleus can be thermally
populated. Some of these states are connected by strong GT transitions
to low-lying states in the daughter nucleus that with increased phase
space can significantly contribute to the stellar beta decay rates.
The importance of these states in the parent nucleus for beta-decay
was first recognized by Fuller, Fowler and Newman 
(commonly abbreviated as FFN) who coined the term
``backresonances'' (see figure~\ref{fig:ecbeta}).

Over the years, many calculations of weak interaction rates for
astrophysical applications have become
available \cite{Hansen:1966,Hansen:1968,Mazurek:1973,%
Mazurek.Truran.Cameron:1974,Takahashi.Yamada.Kondoh:1973,%
Aufderheide.Fushiki.ea:1994a}. For approximately 15 years though, the
standard in the field has been the tabulations
of \citet*{Fuller.Fowler.Newman:1980,%
Fuller.Fowler.Newman:1982a,Fuller.Fowler.Newman:1982b,%
Fuller.Fowler.Newman:1985}. These authors calculated rates for
electron capture, positron capture, beta-decay, and positron emission
plus the associated neutrino losses for all the astrophysically
relevant nuclei ranging in mass number from 21 to 60. Their
calculations were based upon an examination of all available
experimental information in the mid 1980s for individual transitions
between ground states and low-lying excited states in the nuclei of
interest.  Recognizing that this only saturated a small part of the
Gamow-Teller distribution, they added the collective strength
via a single-state representation. Both, energy position and strength
collected in this single state were
determined using an independent particle model (IPM). 

\begin{figure}[htbp]
  \includegraphics[width=0.90\linewidth]{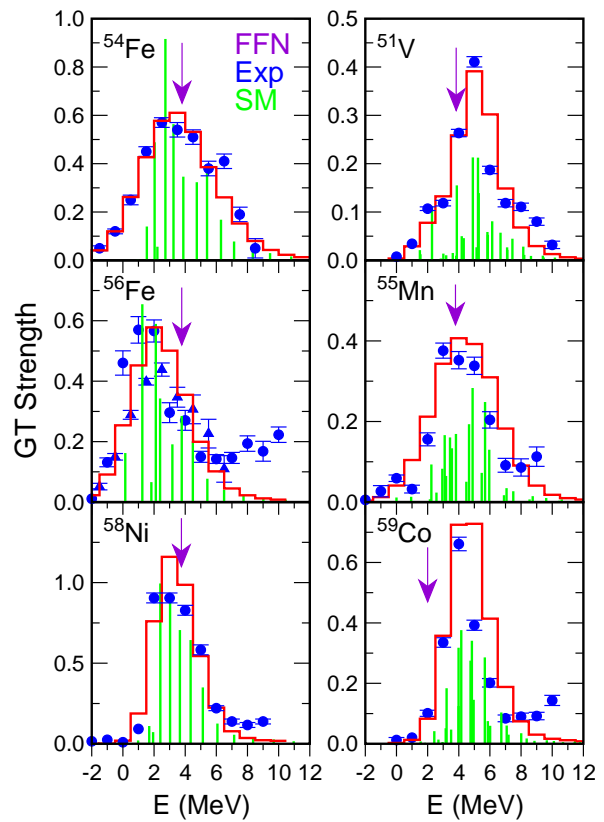}
  \caption{Comparison of shell model GT$_+$ distributions with
    experimental data \cite{Roennqvist.Conde.ea:1993,%
      El-Kateb.Jackson.ea:1994,Alford.Brown.ea:1993} for selected
    nuclei. The shell model results (discrete lines) have been folded
    with the experimental resolution (histograms). The arrows indicate
    the positions where \citet{Fuller.Fowler.Newman:1982a} placed the GT
    resonance in their calculations of the stellar weak-interaction
    rates  \citep[adapted from][]{Caurier.Langanke.ea:1999}.
     \label{fig:GTstrength}}
\end{figure}

Recent experimental data on  GT distributions in iron group
nuclei \cite{Alford.Helmer.ea:1990,Alford.Brown.ea:1993,%
Williams.Alford.ea:1995,Vetterli.Haeusser.ea:1990,%
El-Kateb.Jackson.ea:1994,Rapaport.Taddeucci.ea:1983,%
Anderson.Chittrakarn.ea:1985,Anderson.Lebo.ea:1990,Roennqvist.Conde.ea:1993}
measured in charge exchange reactions
\cite{Goodman.Goulding.ea:1980,Osterfeld:1992}, show that the GT
strength is strongly quenched, compared with the independent particle
model value, and fragmented over many states in the daughter nucleus.
Both effects are caused by the residual interaction among the valence
nucleons and an accurate description of these correlations is
essential for a reliable evaluation of the stellar weak-interaction
rates due to the strong phase space energy dependence, particularly of
the stellar electron-capture rates. The shell model is the only known
tool to reliably describe GT distributions in nuclei
\cite{Brown.Wildenthal:1988}. Indeed, \citet{Caurier.Langanke.ea:1999}
demonstrated that the shell model reproduces 
all measured GT$_+$ distributions (in
this direction a proton is changed into a neutron, like in electron
capture) for nuclei in the iron mass range very well and gives a very
reasonable account of the experimentally known GT$_-$ distributions
(in this direction, a neutron is changed into a proton, like in
$\beta$ decay). Further, the lifetimes of the $pf$-shell nuclei and
their spectroscopy at low energies are simultaneously also described
well.  Figure~\ref{fig:GTstrength} compares the shell model GT$_+$
distributions to the pioneering measurement performed at TRIUMF\@.
These measurements had a typical energy resolution of $\sim 1$ MeV.
Recently developed techniques, involving for example ($^3\text{He},t$)
\cite{Fujita.Akimune.ea:1996} and ($d,{}^2$He) \cite{Woertche:2001}
charge-exchange reactions at intermediate energies, demonstrated in
pilot experiments an improvement in the energy resolution by an order
of magnitude or more. Again, the shell model calculations agree quite
favorably with the improved data.

Several years ago, \citet{Aufderheide:1991} and
\citet{Aufderheide.Bloom.ea:1996,Aufderheide.Fushiki.ea:1994a,%
  Aufderheide.Bloom.ea:1993a,Aufderheide.Bloom.ea:1993b} pointed out
that the interacting shell model is the method of choice for the
calculation of stellar weak-interaction rates.  Following the work by
\citet{Brown.Wildenthal:1988}, \citet{Oda.Hino.ea:1994} calculated
shell-model rates for all the relevant weak processes for $sd$-shell
nuclei ($A=17$--39). This work was then extended to heavier nuclei
($A=45$--65) by \citet{Langanke.Martinez-Pinedo:2001} based on
shell-model calculations in the complete $pf$ shell.  Following the
spirit of FFN, the shell model results have been replaced by
experimental data (energy positions, transition strengths) wherever
available.

Weak interaction rates have also been computed using the proton-neutron
quasiparticle RPA model
\cite{Nabi.Klapdor-Kleingrothaus:1999a,Nabi.Klapdor-Kleingrothaus:1999b}
and the spectral distribution theory
\cite{Kar.Ray.Sarkar:1994,Sutaria.Ray:1995} 

\begin{figure}[htbp]
  \includegraphics[width=0.9\linewidth]{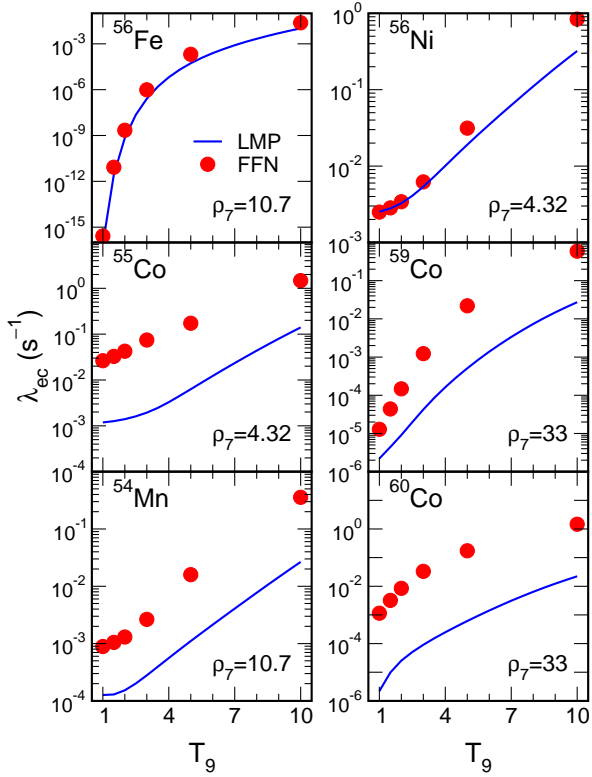}
  \caption{Shell model electron-capture rates as a function of
    temperature ($T_9$ measures the temperature in $10^9$ K) and for
    selected densities ($\rho_7$ defines the density in
    $10^7$~g~cm$^{-3}$) and nuclei.  For comparison, the FFN rates are
    given by the full points.\label{fig:ecrates}}
\end{figure}

\begin{figure}[htbp]
  \includegraphics[width=0.9\linewidth]{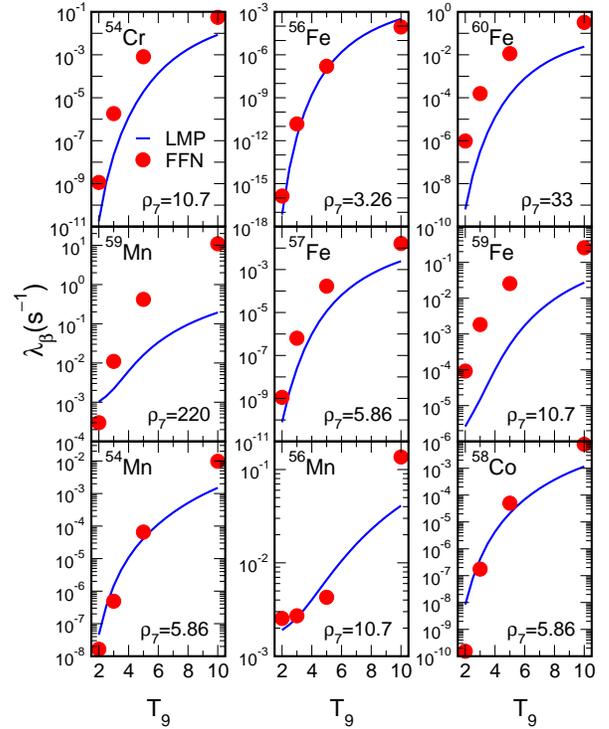}
  \caption{Shell model beta-decay rates as a function of temperature
    ($T_9$ measures the temperature in $10^9$ K) and for selected
    densities ($\rho_7$ defines the density in $10^7$~g~cm$^{-3}$) and
    nuclei.  For comparison, the FFN rates are given by the full
    points \citep[from][]{Martinez-Pinedo.Langanke.Dean:2000}.
      \label{fig:brates}}
\end{figure}

After oxygen burning, the important weak processes are electron
captures and beta decays on nuclei in the iron mass range ($A \sim
45$--65).  Conventional stellar models described these weak processes
using the rates estimated by \citet*{Fuller.Fowler.Newman:1982a}.
These rates are compared to the shell model electron capture rates in
figure~\ref{fig:ecrates} at relevant temperatures and densities.
Importantly the shell model rates are nearly always lower than the FFN
rates. Thus this difference represents a systematic trend, which is
not expected to be washed out if the many nuclei in the stellar
composition are considered. The difference is caused, for example, by
the reduction of the Gamow-Teller strength (quenching) compared to the
IPM value and a systematic misplacement of the Gamow-Teller centroid
in nuclei with certain pairing structure
\cite{Langanke.Martinez-Pinedo:2000}. In some cases, experimental
data, which were not available to FFN, but could be used by
\citet{Langanke.Martinez-Pinedo:2001}, led to significant changes. The
FFN and shell-model beta decay rates are compared in
figure~\ref{fig:brates}, \citet{Martinez-Pinedo.Langanke.Dean:2000}
discuss the differences between the two rate sets.

\subsection{Consequences of the shell model rates in stellar models}

\citet{Heger.Langanke.ea:2001,Heger.Woosley.ea:2001} have investigated
the influence of the shell model rates on the late-stage evolution of
massive stars by repeating the calculations of
\citet{Woosley.Weaver:1995} keeping the stellar physics, except for
the weak rates, as close to the original studies as possible.  The new
calculations have incorporated the shell-model weak interaction rates
for nuclei with mass numbers $A=45$--65, supplemented by rates from
\citet{Oda.Hino.ea:1994} for lighter nuclei.  The earlier calculations
of Weaver and Woosley (WW) used the FFN rates for electron capture and
an older set of beta decay rates
\cite{Mazurek:1973,Mazurek.Truran.Cameron:1974}. As a side-remark we
note that late-stage evolution of massive stars is quite sensitive to
the still not sufficiently well known $^{12}$C($\alpha,\gamma)^{16}$O
rate. The value adopted in the standard WW and in the
\citeauthor{Heger.Langanke.ea:2001} models [$S(E=300\ \text{keV}) =
170$~keV~b] agrees, however, rather nicely with the recent data
analysis [$S(300)= 165\pm50$~keV~b \cite{Kunz.Jaeger.ea:2001}] and the
value derived from nucleosynthesis arguments by
\citet{Weaver.Woosley:1993}, [$S(300) = 170\pm20$~keV~b].

\begin{figure*}[htbp]
  \includegraphics[width=0.3\linewidth]{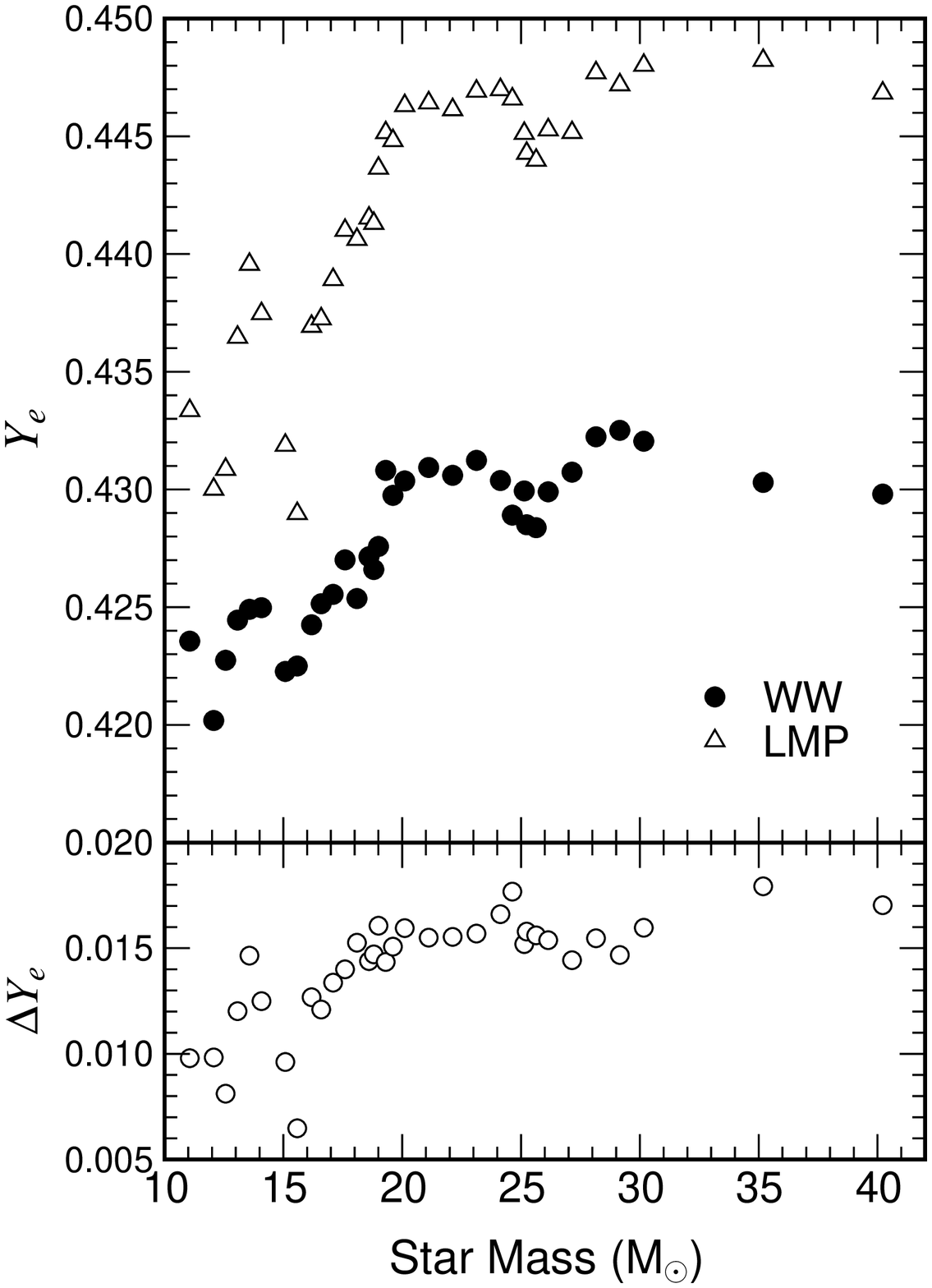}%
  \hspace{0.025\linewidth}%
  \includegraphics[width=0.3\linewidth]{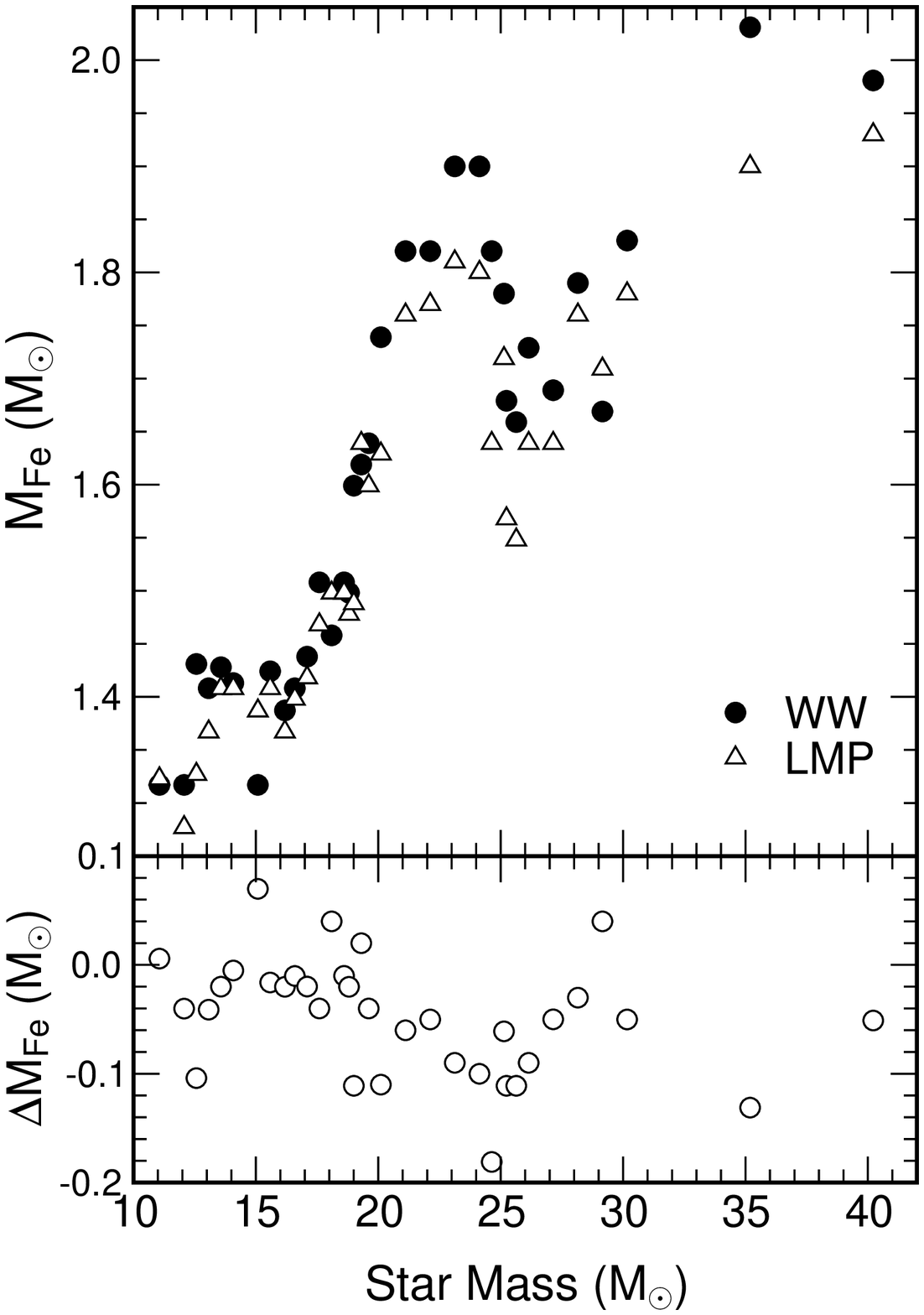}%
  \hspace{0.025\linewidth}%
  \includegraphics[width=0.3\linewidth]{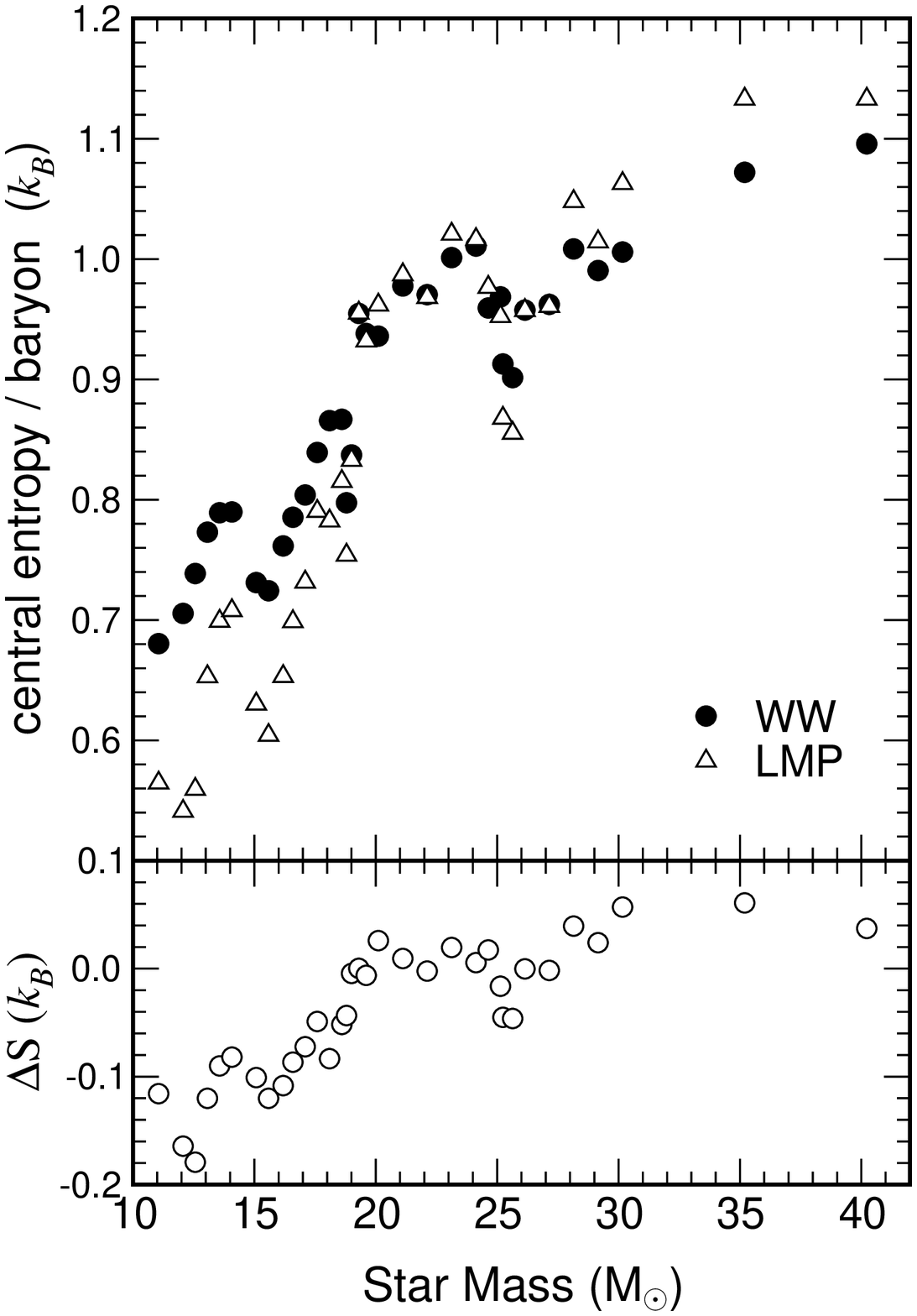}
  \caption{Comparison of the center values of $Y_e$ (left), the iron
    core sizes (middle) and the central entropy (right) for $11\text{--}40\ 
    M_\odot$ stars between the WW models and the ones using the shell
    model weak interaction rates (LMP)
    \citep[from][]{Heger.Langanke.ea:2001}. 
    The lower parts define the changes in the 3 quantities between the
    LMP and WW models.
\label{fig:presn}} 
\end{figure*}

Figure~\ref{fig:presn} illustrates the consequences of the shell model weak
interaction rates for presupernova models in terms of the three
decisive quantities: the central electron-to-baryon ratio $Y_e$, the
entropy, and the iron core mass.  The central values of $Y_e$ at the
onset of core collapse increased by 0.01-0.015 for the new rates. This
is a significant effect.  
For example, a change from $Y_e=0.43$ in the WW model for a
20 $M_\odot$ star to $Y_e=0.445$ in the new models increases the respective
Chandrasekhar mass by about 0.075 $M_\odot$.
We note that the new models also result in
lower core entropies for stars with $M \leq 20\ M_\odot$, while for $M
\geq 20\ M_\odot$, the new models actually have a slightly larger
entropy.  The iron core masses are generally smaller in the new models
where the effect is larger for more massive stars ($M \ge 20\
M_\odot$), while for the most common supernovae ($M \le 20\ M_\odot$)
the reduction is by about 0.05~$M_\odot$.  [We define the iron core as
the mass interior to the point where the composition becomes at least
$50 \%$ of iron group elements $(A \geq 48)$].  This reduction of the
iron core mass appears to be counterintuitive at first glance with
respect to the slower electron capture rates in the new models. It is,
however, related to changes in the entropy profile during silicon
shell burning which reduces the growth of the iron core just prior to
collapse \cite{Heger.Langanke.ea:2001}.

It is intriguing to speculate what effects these changes might have
for the subsequent core collapse and supernova explosion. At first we
note that in the current supernova picture
\cite{Bethe:1990,Burrows:2000,Langanke.Wiescher:2002,%
  Woosley.Heger.Weaver:2002} gravitation overcomes the resisting
electron degeneracy pressure in the core, leading to increasing
densities and temperatures. Shortly after neutrino trapping at
densities of a few $10^{11}$~g~cm$^{-3}$, an homologous core, which
stays in sonic communication, forms in the center. Once the core
reaches densities somewhat in excess of nuclear matter density (a few
$10^{14}$~g~cm$^{-3}$) the nuclear equation of state stiffens and a
spring-like bounce is created triggering the formation of a shock wave
at the surface of the homologous core \cite{Bethe:1990}.  This shock
wave tries to traverse the rest of the infalling matter in the iron
core.  However, the shock loses its energy by dissociation of the
infalling matter and by neutrino emission, and it is generally
believed now that supernovae do not explode promptly due to the bounce
shock. Probably, this happens in the `delayed mechanism' 
\cite{Wilson:1985} where the
shock is revived by energy deposition from the neutrinos generated by
the cooling of the proto-neutron star, the remnant in the center of
the explosion.

With the larger $Y_e$ values, obtained in the calculations with the
improved weak rates, the core contains more electrons whose pressure
acts against the collapse. It is also expected that the size of the
homologous core, which scales like $\sim Y_e^2$ with the $Y_e$ value
at neutrino trapping, should be larger.  This, combined with the
smaller iron cores, yields less material which the shock has to
traverse.  Furthermore, the change in entropy will affect the mass
fraction of free protons, which in the later stage of the collapse
contribute significantly to the electron capture. For presupernova
models with masses $M<20\ M_\odot$, however, the number fraction of
protons is very low \citep[$\lesssim
10^{-6}$,][]{Heger.Langanke.ea:2001} so that for these stars electron
capture should still be dominated by nuclei, even at densities in
access of $10^{10}$~g~cm$^{-3}$. We will return to this problem below.

\begin{figure}[htbp]
  \includegraphics[width=0.9\linewidth]{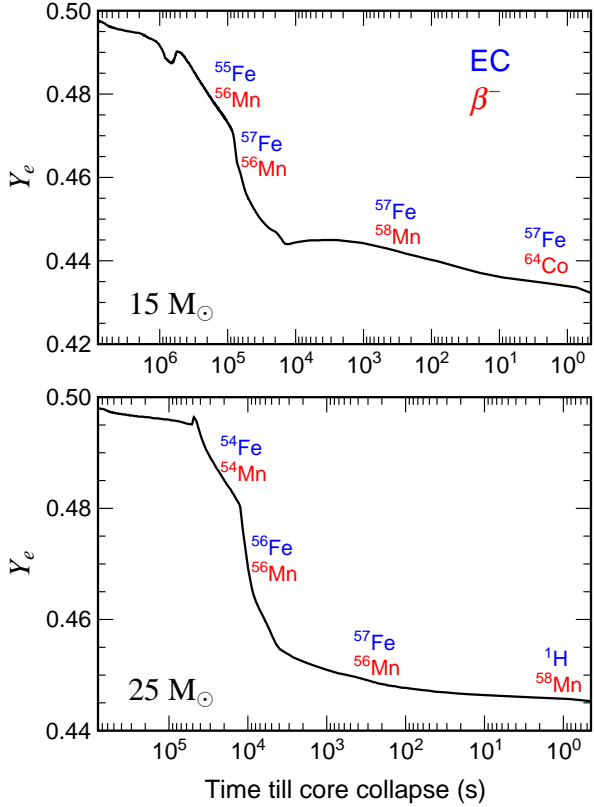}
  \caption{Evolution of the $Y_e$ value in the center of a
    15~$M_\odot$ (upper part) and 25~$M_\odot$ (lower part) as a
    function of time until bounce.  The most important $Y_e$-changing
    nuclei for the calculations adopting the shell model rates are
    indicated at different times, where the upper nucleus refers to
    electron capture and the lower to $\beta$-decay.\label{fig:Ye}}
\end{figure}

\begin{figure}[htbp]
  \includegraphics[width=0.9\linewidth]{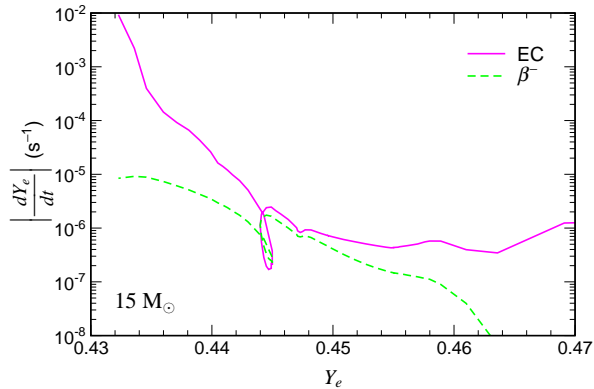}
  \caption{Comparison of the change of the $Y_e$ value with time,
    $|dY_e/dt|$ due to electron capture and beta decay in a
    15~$M_\odot$ star. In general, the $Y_e$ value decreases with time
    during the collapse, caused by electron captures. The loops
    indicate that during this period beta-decay, which increases
    $Y_e$, dominates over electron capture.\label{fig:dyeye}}
\end{figure}

To understand the origin of these differences it is illustrative to
investigate the role of the weak-interaction rates in greater details.
The evolution of $Y_e$ during the collapse phase is plotted in
figure~\ref{fig:Ye}.  Weak processes become particularly important in
reducing $Y_e$ below 0.5 after oxygen depletion ($\sim 10^7$ s and
$10^6$ s before core collapse for the 15~$M_\odot$ and 25~$M_\odot$
stars, respectively) and $Y_e$ begins a decline which becomes
precipitous during silicon burning. Initially electron capture occurs
much more rapidly than beta decay. As the shell model rates are
generally smaller than the FFN electron capture rates, the initial
reduction of $Y_e$ is smaller in the new models; the temperature in
these models is correspondingly larger as less energy is radiated away
by neutrino emission.

An important feature of the new models is demonstrated in
figure~\ref{fig:dyeye}. Beta decay becomes temporarily competitive with
electron capture after silicon depletion in the core and during
silicon shell burning \citep[this had been foreseen
in][]{Aufderheide.Fushiki.ea:1994b}.  The presence of an important
beta decay contribution has two effects.  Obviously it counteracts the
reduction of $Y_e$ in the core, but equally important, beta decays are
an additional neutrino source and thus they add to the cooling of the
core and a reduction in entropy.  This cooling can be quite efficient
as often the average neutrino energy in the involved beta decays is
larger than for the competing electron captures. As a consequence the
new models have significantly lower core temperatures than the WW
models after silicon burning. At later stages of the collapse, beta
decay becomes unimportant again as an increased electron chemical
potential drastically reduces the phase space.

We note that the shell model weak interaction rates predict the
presupernova evolution to proceed along a temperature-density-$Y_e$
trajectory where the weak processes are dominated by nuclei rather
close to stability. Thus it will be possible, after next generation
radioactive ion-beam facilities become operational, to further
constrain the shell model calculations by measuring relevant GT
distributions for unstable nuclei by charge-exchange reaction, where
we like to point out again that the GT$_+$ distribution is also
crucial for stellar $\beta$-decays. Figure~\ref{fig:Ye} identifies
those nuclei which dominate (defined by the product of abundance times
rate) the electron capture and beta decay during various stages of the
final evolution of 15~$M_\odot$ and 25~$M_\odot$ stars.
\citet{Heger.Woosley.ea:2001} give an exhaustive list of the most
important nuclei for both electron capture and beta decay during the
final stages of stellar evolution for stars of different masses.

In total, the weak interaction processes shift the matter composition
to smaller $Y_e$ values (see Fig.~\ref{fig:Ye}) and hence more
neutron-rich nuclei, subsequently affecting the nucleosynthesis. Its
importance for the elemental abundance distribution, however, strongly
depends on the location of the mass cut in the supernova explosion. It
is currently assumed that the remnant will have a baryonic mass
between the iron core and oxygen shell masses
\cite{Woosley.Heger.Weaver:2002}. As the reduction of $Y_e$ occurs
mainly during silicon burning, it is essential to determine how much
of this material will be ejected.  Another important issue is the
possible long-term mixing of material during the explosion
\citep[e.g.][]{Kifonidis.Plewa.ea:2000}.  Changes of the elemental
abundances due to the improved weak-interaction rates are rather small
as the differences, compared to FFN, occur in regions of the star
which are probably not ejected (however, for type Ia supernovae, see
below). The weak interaction also determines the decay of the newly
synthesized nuclei in supernova explosions. Some of them are
proton-rich nuclei that decay by orbital electron capture, leaving
atomic K-shell electron vacancies. The X-rays emitted can escape the
supernova ejecta for sufficiently long-lived isotopes and can possibly
be detected by the current generation of X-ray telescopes
\cite{Leising:2001}.

In dense stellar environment the electron capture rates have to be
corrected for screening effects caused by the relativistically
degenerate electron liquid. Such studies have been recently
performed within the linear response theory \cite{Itoh.Tomizawa.ea:2002} 
who find typical screening corrections of order a few percent.

\section{Collapse and post-bounce stage}

The models, as we have described them above, constitute the so-called
presupernova models. They follow the late-stage stellar evolution
until core densities just below $10^{10}$~g~cm$^{-3}$ and temperatures
between 5 and 10 GK. (More precisely, \citet{Woosley.Weaver:1995}
define the final presupernova models as the time when the collapse
velocity near the edge of the iron core first reached
1000~km~s$^{-1}$.) As we have stressed above, stellar evolution until
this time requires the consideration of an extensive nuclear network,
but is simplified by the fact that neutrinos need only be treated as a
source for energy losses. This is no longer valid at later stages of
the collapse. As neutrinos will eventually be trapped in the
collapsing core and their interaction with the surrounding matter is
believed to be crucial for the supernova explosion, computer
simulations of the collapse, bounce and explosion necessitate a
detailed time- and space-dependent bookkeeping of the various neutrino
($\nu_e, \nu_\mu, \nu_\tau$ neutrinos and their antiparticles)
distributions in the core.  Built on the pioneering work by
\citet{Bruenn:1985}, this is done by multi-group (neutrinos of
different flavor and energy) Boltzmann neutrino
transport \cite{Mezzacappa.Liebendoerfer.ea:2001,%
  Liebendoerfer.Mezzacappa.ea:2001,Rampp.Janka:2000}.  Advantageously,
the temperature during the collapse and explosion are high enough that
the matter composition is given by nuclear statistical equilibrium
without the need of reaction networks for the strong and
electromagnetic interaction. The transition from a rather complex
global nuclear network, involving many neutron, proton and $\alpha$
fusion reactions and their inverse, to a quasi-statistical
equilibrium, in which reactions within mini-cycles are fast enough to
bring constrained regions of the nuclear chart into equilibrium, to
finally global nuclear statistical equilibrium is extensively
discussed by \cite{Woosley:1986}.

Presupernova models are the input for collapse and explosion
simulations. Currently, one-dimensional models with sophisticated
neutrino transport do not
explode \cite{Mezzacappa.Liebendoerfer.ea:2001,%
  Liebendoerfer.Mezzacappa.ea:2001,Rampp.Janka:2000}, including first
attempts with the presupernova models derived with the improved
weak-interaction rates discussed above \cite{Messer.Others:2002}.
Explosions can, however, be achieved if the shock revival in the
delayed mechanism is modelled by two-dimensional hydrodynamics
allowing for more efficient neutrino energy transfer
\cite{Herant.Benz.ea:1994,Burrows.Hayes.Fryxell:1995,Janka.Mueller:1996}.
Thus the intriguing question arises: Are supernova explosions
three-dimensional phenomena requiring convective motion and perhaps
rotation and magnetic fields?  Or do one-dimensional models fail due
to incorrect or insufficient nuclear physics input? Although first
steps have been taken in modelling the multi-dimensional effects
\citep[for reviews and references
see][]{Janka.Kifonidis.Rampp:2001,Woosley.Heger.Weaver:2002}, these
require extremely demanding and computationally challenging
simulations. In the following we will briefly discuss some nuclear
physics ingredients in the collapse models and their possible
improvements.

The crucial weak processes during the collapse are
\cite{Bruenn:1985,Rampp.Janka:2002,Burrows:2001}:

\begin{subequations}
  \label{eq:weakp}
  \begin{eqnarray}
    p + e^- & \rightleftarrows & n + \nu_e \label{eq:epnnu}\\
    n + e^+ & \rightleftarrows & p + \bar{\nu}_e \label{eq:enpnu}\\
    (A,Z) + e^- & \rightleftarrows & (A,Z-1) + \nu_e \label{eq:eAnuA}\\
    (A,Z) + e^+ & \rightleftarrows & (A,Z+1) + \bar{\nu}_e
    \label{eq:posA}\\ 
    \nu + N & \rightleftarrows & \nu + N \label{eq:nuNnuN} \\
    N + N & \rightleftarrows & N + N + \nu + \bar{\nu}
    \label{eq:NNNNnunu}\\ 
    \nu + (A,Z) & \rightleftarrows & \nu + (A,Z) \label{eq:nuAnuA}\\
    \nu + e^\pm & \rightleftarrows & \nu + e^\pm \label{eq:nuenue}\\
    \nu + (A,Z) & \rightleftarrows & \nu + (A,Z)^* \label{eq:nuAnuAin}\\
    e^+ + e^- & \rightleftarrows & \nu + \bar{\nu} \label{eq:eenunu}\\
    (A,Z)^* & \rightleftarrows & (A,Z) + \nu + \bar{\nu}
    \label{eq:AAnunu} 
\end{eqnarray}
\end{subequations}

Here, a nucleus is symbolized by its mass number $A$ and charge $Z$,
$N$ denotes either a neutron or a proton and $\nu$ represents any
neutrino or antineutrino.  In the early collapse stage, before
trapping, these reactions proceed dominantly to the right. We note
that, due to the generally accepted collapse picture
\citep[e.g.][]{Bethe:1990}, elastic scattering of neutrinos on
nuclei~\eqref{eq:nuAnuA} is mainly responsible for the trapping, as it
determines the diffusion time scale of the outwards streaming
neutrinos. Shortly after trapping, the neutrinos are thermalized by
the energy downscattering, experienced mainly in inelastic scattering
off electrons~\eqref{eq:nuenue}. The relevant cross sections for these
processes are readily derived \cite{Bruenn:1985}.  For elastic
neutrino-nucleus scattering one usually makes the simplifying
assumption that the nucleus has a $J=0^+$ spin/parity assignment, as
appropriate for the ground state of even-even nuclei.  The scattering
process is then restricted to the Fermi part of the neutral current
(pure vector coupling) \cite{Freedman:1974,Tubbs.Schramm:1975} and
gives rise to coherent scattering; i.e.\ the cross section scales with
$A^2$, except from a correction $\sim (N-Z)/A$ arising from the
neutron excess. This assumption is, in principle, not correct for the
ground states of odd-$A$ and odd-odd nuclei and for all nuclei at
finite temperature, as then $J \ge 0$ and the cross section will also
have an axial-vector Gamow-Teller contribution.  However, the relevant
GT$_0$ strength is not concentrated in one state, but rather
fragmented over many nuclear levels. Thus, one can expect that the GT
contributions to the elastic neutrino-nucleus cross sections are in
general small enough to be neglected. 

Reactions \eqref{eq:epnnu} and \eqref{eq:eAnuA} are equally important, as
they control the neutronization of the matter and, in a large portion,
also the star's energy losses. Due to their strong phase space
sensitivity ($\sim E_e^5$), the electron capture cross sections
increase rapidly during the collapse as the density (the electron
chemical potential scales like $\sim \rho^{1/3}$) and the temperature
increase.  We already observed above that beta-decay is rather
unimportant during the collapse due to Pauli-blocking of the electron
phase space in the final state. We also noted how sensitively the
electron capture rate on nuclei depends on a proper description of
nuclear structure. As we will discuss now, this is also expected for
this stage of the collapse, although the relevant nuclear structure
issues are somewhat different.

\subsection{Electron capture on nuclei}

The new presupernova models indicate that electron capture on nuclei
will still be important, at least in the early stage of the collapse.
Although capture on free protons, compared to nuclei, is favored by
the significantly lower Q-value, the number fraction of free protons
$Y_p$, i.e.\ the number of free protons divided by the total number of
nucleons, is quite low \citep[$Y_p \sim 10^{-6}$ in the 15~$M_\odot$
presupernova model of][]{Heger.Woosley.ea:2001}. This tendency had
already been observed before, but has been strengthened in the new
presupernova models, where the $Y_e$ values are significantly larger
and thus the nuclei present in the matter composition are less
neutron-rich, implying lower Q-values for electron capture.
Furthermore, the entropy is smaller in stars with $\lesssim 20\ 
M_\odot$, yielding a smaller fraction of free protons.

As the entropy is rather low \cite{Bethe.Brown.ea:1979}, most of the
collapsing matter survives in heavy nuclei. However, $Y_e$ decreases
during the collapse making the matter composition more neutron-rich,
hence energetically favoring increasingly heavy nuclei.  In computer
studies of the collapse, the ensemble of heavy nuclei is described by
one representative which is generally chosen to be the most abundant
in the nuclear statistical equilibrium composition.
Due to a simulation of the infall phase \cite{Mezzacappa.Bruenn:1993a,%
Mezzacappa.Bruenn:1993b}, such representative nuclei are $^{70}$Zn
and $^{88}$Kr at different stages of the
collapse \cite{Mezzacappa.pvt}.

In current collapse simulations the treatment of electron capture on
nuclei is schematic and rather simplistic.  The nuclear structure
required to derive the capture rate is then described solely on the
basis of an independent particle model for iron range nuclei, i.e., 
considering only Gamow-Teller transitions from $f_{7/2}$
protons to $f_{5/2}$ neutrons \cite{Bethe.Brown.ea:1979,Bruenn:1985,%
Mezzacappa.Bruenn:1993a,Mezzacappa.Bruenn:1993b}. In particular,
this model predicts that electron capture vanishes for nuclei with
charge number $Z<40$ and neutron number $N\ge40$, arguing that
Gamow-Teller transitions are blocked by the Pauli principle, as all
possible final neutron orbitals are already occupied in nuclei with
$N\ge40$ (closed $pf$ shell) \cite{Fuller:1982}.  Such a situation
would, for example, occur for the two nuclei $^{70}$Zn and $^{88}$Kr
with $(Z=30,N=40)$ and $(Z=36,N=52)$, respectively.  It has been pointed
out \cite{Cooperstein.Wambach:1984} that this picture is too simple
and that the blocking of the GT transitions will be overcome by
thermal excitation which either moves protons into $g_{9/2}$ orbitals
or removes neutrons from the $pf$ shell, in both ways reallowing GT
transitions.  In fact, due to this `thermal unblocking', GT
transitions again dominate the electron capture on nuclei for
temperatures of order 1.5~MeV \cite{Cooperstein.Wambach:1984}.  An
even more important unblocking effect, which is already relevant at
lower temperatures is, however, expected by the residual interaction
which will mix the $g_{9/2}$ (and higher) orbitals with those in the
$pf$ shell.

\begin{figure}[htbp]
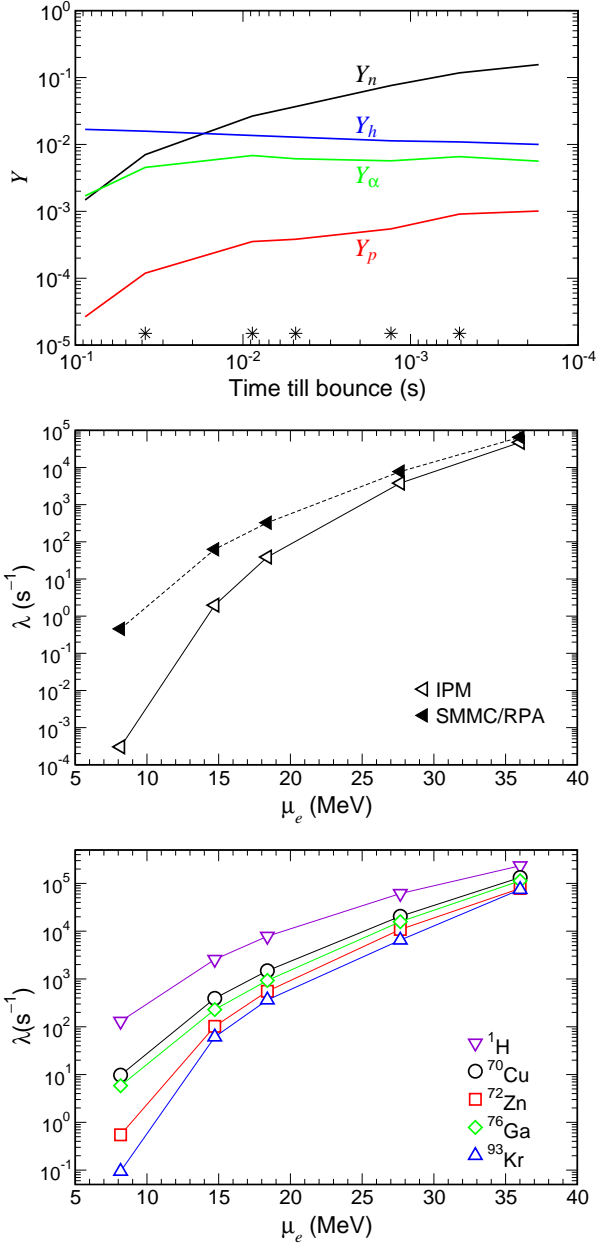

    \includegraphics[bb=80 58 704
    487,width=0.9\linewidth]{abundevol.eps}\\[2mm]  
    \includegraphics[bb=80 58 704
    487,width=0.9\linewidth]{ge78-ec.eps}\\[2mm] 
    \includegraphics[bb=80 58 704
    487,width=0.9\linewidth]{ecngt40.eps} 
    \caption{Electron capture rates on free proton and several
      representative nuclei as a function of electron chemical
      potential $\mu_e$ (lower panel).  The rates have been determined
      for conditions during the infall phase \cite{Liebendoerfer.pvt}:
      ($T,Y_e,\rho_{11}$)=(0.846,0.429,0.102), (1.133,0.410,0.601),
      (1.259,0.400,1.186), (1.617,0.372,4.287), (2.000,0.349,10.063),
      where the temperature $T$ is given in MeV and $\rho_{11}$
      measures the density in $10^{11}$~g~cm$^{-3}$. The electron
      chemical potential increases during the infall, $\mu_e \sim
      \rho^{1/3}$. The middle panel compares the capture rates on
      $^{78}$Ge in the independent particle model (IPM), in which
      Gamow-Teller transitions are Pauli-blocked, with the results
      obtained in the hybrid model (SMMC/RPA) which unblocks these
      transitions due to correlations and finite temperature effects.
      The rates have been calculated without neutrino final-state
      Pauli blocking which will become important at trapping
      densities.  The upper panel shows the time evolution of the
      number abundances for neutrons $Y_n$, protons $Y_p$,
      $\alpha$-particles $Y_\alpha$ and heavy nuclei $Y_h$ calculated
      along the same stellar trajectory for which the rates have been
      calculated at selected points. These points are marked by
      asterisks \citep[from][]{Sampaio.Langanke.ea:2002b}.
      \label{fig:gecapt}}
\end{figure}

A consistent calculation of the electron capture rates for nuclei with
neutron numbers $N>40$ and proton numbers $20<Z<40$, including
configuration mixing and finite temperature, is not yet feasible by
direct shell model diagonalization due to the large model spaces and
many states involved.  It can, however, be performed in a reasonable
way adopting a hybrid model: The capture rates are calculated within
the RPA approach with partial occupation formalism, including allowed
and forbidden transitions. The partial occupation numbers represent an
`average' state of the parent nucleus and depend on temperature. They
are calculated within the Shell Model Monte Carlo approach at finite
temperature \cite{Koonin.Dean.Langanke:1997} and include an
appropriate residual interaction. Exploratory studies, performed for a
chain of germanium isotopes ($Z=32$), confirm that the GT transition is
not blocked for $N\ge40$ and still dominates the electron capture
process for such nuclei at stellar conditions
\cite{Langanke.Kolbe.Dean:2001}.  This is demonstrated in
figure~\ref{fig:gecapt}, which compares electron capture rates for
$^{78}$Ge calculated within the hybrid model with the results in the
independent particle model (IPM).  For this nucleus ($N=46$) the rate
in the IPM is given solely by forbidden transitions (mainly induced by
$1^-$ and $2^-$ multipoles). However, correlations and finite
temperature unblock the GT transitions in the hybrid model which
increases the rate significantly. The differences are particularly
important at lower densities (a few $10^{10}$~g~cm$^{-3}$) where the
electron chemical potential does not suffice to induce forbidden
transitions.

We note again that many nuclei are present with similar mass
abundances during the supernova collapse phase and that their relative
abundances are approximately described by nuclear statistical
equilibrium.  Figure~\ref{fig:gecapt} shows the capture rates for
several representative nuclei during the collapse phase, identified by
the average charge and mass number of the matter composition following
the time evolution of a certain ($M=0.6\ M_\odot$) mass trajectory
\cite{Liebendoerfer.Mezzacappa.ea:2001}.  (For the conditions shown in
figure~\ref{fig:gecapt} $^{93}$Kr and $^{72}$Zn are examples of
representative nuclei at $\rho_{11}=10.063$ and 0.601, respectively).
The general trend of the rates reflects the competition of the two
main energy scales of the capture process: the electron chemical
potential $\mu_e$, which grows like $\rho^{1/3}$ during infall, and
the reaction Q-value. As the Q-value is smaller for free protons
($Q=1.29$ MeV) than for neutronrich nuclei ($Q \sim$ few MeV), the
capture rate on free protons is larger than for the heavy nuclei.
However, this difference diminishes with increasing density.  This is
expected because the electron energies involved (for example, the
electron chemical potential is $\mu_e \sim 18$~MeV at
$\rho=10^{11}$~g~cm$^{-3}$) are then significantly higher than the
$Q$-values for the capture reactions on the abundant nuclei, (i.e.\ 
$^{93}$Kr has a $Q$-value of about 11 MeV).  As also nuclear structure
effects at the relatively high temperatures involved are rather
unimportant, the capture rates on the abundant nuclei at the later
stage of the collapse are rather similar.  However, the capture rate
is quite sensitive to the reaction $Q$-value for the lower electron
chemical potentials.  To quantify this argument we take the point of
the stellar trajectory from figure~\ref{fig:gecapt} with the lowest
electron chemical potential ($\mu_e \sim 8$ MeV) as an example.  Under
these conditions the capture rates on $^{70}$Cu and $^{76}$Ga (both
nuclei have $Q$-values around 4 MeV) are noticeably larger than for
$^{78}$Ge and $^{72}$Zn with $Q$-values around 8 MeV. However, in
nuclear statistical equilibrium the relative mass fraction of
$^{72}$Zn (about $1.2 \times 10^{-2}$) is larger than for $^{70}$Cu
($4.0 \times 10^{-3}$) or $^{76}$Ga ($1.8 \times 10^{-3}$). The most
abundant nucleus, $^{66}$Ni, has a mass fraction of $4.3 \times
10^{-2}$ and a capture rate comparable to $^{72}$Zn. $^{93}$Kr is too
neutron-rich to have a significant abundance at this stage of the
collapse.  This discussion indicates that the most abundant nuclei are
not necessarily the nuclei which dominate the electron capture in the
infall phase.  Thus, a single-nucleus approximation can be quite
inaccurate and should be replaced by an ensemble average.

What matters for the competition of capture on nuclei compared to that
on free protons is the product of number abundance times capture rate.
Figure~\ref{fig:gecapt} shows the time evolution of the number
abundances for free neutrons, protons, $\alpha$-particles and heavy
nuclei, calculated for the same stellar trajectory \citep[obtained
from][]{Liebendoerfer.pvt} for which the capture rates have been
evaluated under the assumption of 
nuclear statistical equilibrium (NSE). (We note that the commonly
adopted equations of states \cite{Lattimer.Swesty:1991,Shen.Toki.ea:1998a}
yield somewhat larger $Y_p$ fractions than obtained in NSE.)
Importantly the number abundance of heavy nuclei is significantly
larger than that of free protons (by more than two orders of
magnitude at the example point discussed above) to compensate for the
smaller capture rates on heavy nuclei.  It appears thus that electron
capture on nuclei cannot be neglected during the collapse. We note
that the average energies of the neutrinos produced by
capture on nuclei are significantly smaller than for capture on free
protons making this process a potentially important source for
low-energy neutrinos.

The neutron number $N=40$ is not magic in nuclear structure, nor
for stellar electron capture rates. Thus the anticipated strong
reduction of the capture rate on nuclei will not occur and we expect
capture on nuclei to be an important neutronization process probably
until neutrino trapping. The magic neutron number $N=50$ is also no
barrier as for nuclei like $^{93}$Kr ($N=57$), the neutron $pf$-shell
is nearly completely occupied, but due to correlations protons
occupy, for example, the $g_{9/2}$ orbital, and thus unblock GT
transitions by allowing transformations into $g_{9/2},g_{7/2}$
neutrons.  The description of electron capture on nuclei in the
collapse simulations needs to be improved.

In the current simulations the inverse reaction rates of the weak
processes listed above are derived by detailed balance. Thus an
improved description of electron capture will then also affect the
neutrino absorption on nuclei, although this process is strongly
suppressed by Pauli blocking in the final state.

\subsection{Neutrino rates}

\begin{figure}[htbp]
  \includegraphics[width=0.9\linewidth]{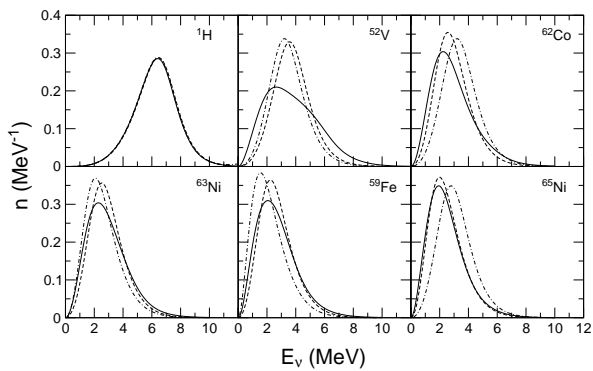}
  \caption{Normalized neutrino spectra for stellar electron capture
    on the six most important `electron-capturing nuclei' in the
    presupernova model of a 15~$M_\odot$ star, as identified in
    \cite{Heger.Langanke.ea:2001}. The stellar parameters are $T=7.2
    \times 10^9$ K, $\rho = 9.1 \times 10^9$~g~cm$^{-3}$ and
    $Y_e=0.43$. The solid lines represent the spectra derived from the
    shell model electron capture rates. The dashed and dashed-dotted
    lines correspond to parametrizations recommended by
    \citet{Langanke.Martinez-Pinedo.Sampaio:2001} and
    \citet{Bruenn:1985}, respectively.\label{fig:nusp}}
\end{figure}

\begin{figure}[htb]
  \includegraphics[width=0.9\linewidth]{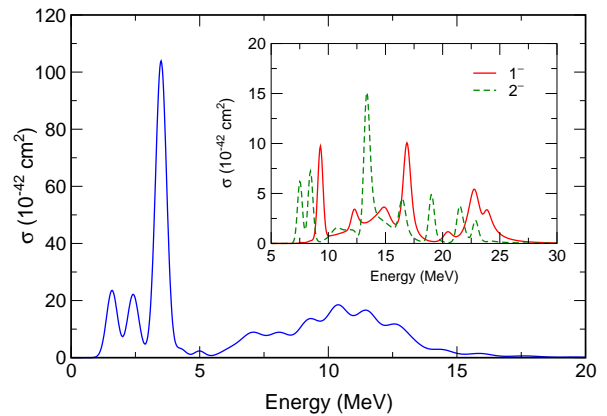}
  \caption{Differential $^{56}$Fe$(\nu_e,e^-){}^{56}$Co cross section
    for the KARMEN neutrino spectrum, coming from the decay at rest of
    the muon, as a function of the excitation energy in $^{56}$Co. The
    figure shows the allowed contributions, while the inset gives the
    contribution of the 1$^-$ and 2$^-$
    multipolarities.\label{fig:fe56nue}} 
\end{figure}

In the capture process on nuclei, the electron has to overcome the
Q-values of the nuclei and the internal excitation energy of the GT
states in the daughter, so the final neutrino energies are noticeably
smaller than for capture on free protons. Typical neutrino spectra for
a presupernova model are shown in figure~\ref{fig:nusp}. In this stage
of the collapse the neutrino energies are sufficiently small that they
only excite allowed transitions.  Consequently neutrino-nucleus cross
sections for $pf$-shell nuclei can be determined on the basis of GT
distributions determined in the shell model. In the later stage of the
collapse, the increased density results also in higher energy
electrons ($E_e \sim \rho^{1/3}$) which in turn, if captured by
protons or nuclei, produce neutrinos with energies larger than
15--20~MeV. For such neutrinos forbidden (mainly $\lambda=1$ dipole
and spin-dipole) transitions can significantly contribute to the
neutrino-nucleus cross section. Such a situation is shown in
figure~\ref{fig:fe56nue} which shows the differential cross section
for the process $^{56}$Fe$(\nu_e,e^-){}^{56}$Co computed using the
Shell-Model for the Gamow-Teller contribution and the CRPA for the
forbidden contributions \cite{Kolbe.Langanke.Martinez-Pinedo:1999}.
The calculation adopts a neutrino spectrum corresponding to a muon
decaying at rest. The average energy, $\bar{E}_\nu\approx 30$~MeV, and
momentum transfer, $q \approx 50$~MeV, represent the maximum values
for $\nu_e$ neutrinos expected during the supernova collapse phase,
i.e., the maximum contribution expected from forbidden transitions to
the total neutrino-nucleus cross sections. In this particular case the
$1^+$ multipole (Gamow-Teller at the $q=0$ limit) represents 50\% of
the cross section.  The $^{56}$Fe$(\nu_e,e^-){}^{56}$Co cross section
for neutrinos from muon-decay at rest has been measured by the Karmen
collaboration.  The measured cross section ($2.56\pm1.08({\rm
  stat})\pm 0.43({\rm syst}) \times 10^{-40}$ cm$^2$)
\cite{Zeitnitz:1994} agrees with the result calculated in the shell
model (allowed transitions) plus CRPA (forbidden transitions) approach
($2.38 \times 10^{-40}$ cm$^2$)
\cite{Kolbe.Langanke.Martinez-Pinedo:1999}.

Although the most important neutrino reactions during collapse are
coherent elastic scattering on nuclei and inelastic scattering off
electrons, it has been noted \cite{Haxton:1988,Bruenn.Haxton:1991}
that neutrino-induced reactions on nuclei can happen as well. Using
$^{56}$Fe as a representative nucleus, Bruenn and Haxton concluded
that charged-current $(\nu_e,e^-)$ reactions do not have an
appreciable effect on the evolution of the core during infall, due to
the high-threshold for neutrino absorption. Based on shell model
calculations of the GT strength distributions,
\citet{Sampaio.Langanke.Martinez-Pinedo:2001} confirmed this finding
for other, more relevant nuclei in the core composition. The same
authors showed that finite-temperature effects can increase the
$(\nu_e,e^-)$ cross sections for low neutrino energies
drastically \cite{Sampaio.Langanke.Martinez-Pinedo:2001}. But this
increase is found to be significantly smaller than the reduction of
the cross section caused by Pauli blocking of the final phase space,
i.e.\ due to the increasing electron chemical potential. This
environmental effect ensures that neutrino absorption on nuclei is
unimportant during the collapse compared with inelastic
neutrino-electron scattering.

\begin{figure}[htbp]
  \includegraphics[width=0.9\linewidth]{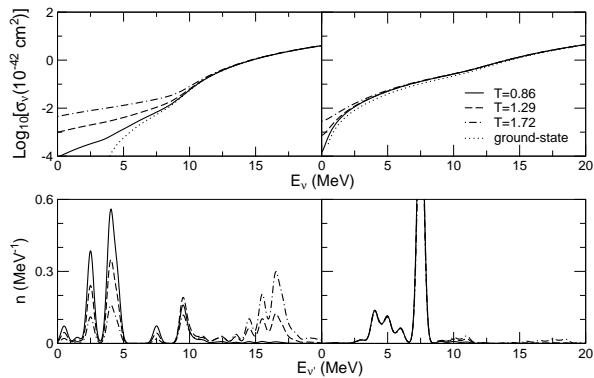}
  \caption{Inelastic neutrino cross sections for $^{56}$Fe (left) and 
    $^{59}$Co (right) as function of initial neutrino energy and for
    selected temperatures (upper part).  Only allowed Gamow-Teller
    transitions have been considered. Temperatures are in MeV.  For
    $T=0$, the cross section is calculated for the ground state only.
    At $T>0$, the cross sections have been evaluated for a thermal
    ensemble of initial states. The corresponding neutrino energy
    distribution in the final state is shown in the lower part,
    assuming an initial neutrino energy of $E_\nu =7.5$ MeV.  Due to
    threshold effects a significant portion of the neutrinos are
    upscattered in energy for even-even nuclei.\label{fig:neut}}
\end{figure}

\citet{Bruenn.Haxton:1991} observed that inelastic neutrino scattering
off nuclei plays the same important role of equilibrating electron
neutrinos with matter during infall as neutrino-electron scattering.
The influence of finite temperature on inelastic neutrino-nucleus
scattering was studied in \cite{Fuller.Meyer:1991}, using an
independent particle model.  While the study in
\cite{Bruenn.Haxton:1991} was restricted to $^{56}$Fe, additional
cross sections have been calculated for inelastic scattering of
neutrinos on other nuclei based on modern shell-model GT strength
distributions \cite{Sampaio.Langanke.ea:2002}. Again, for low neutrino
energies the cross sections are enhanced at finite temperatures
(figure~\ref{fig:neut}).  This is caused by the possibility that, at
finite temperatures, the initial nucleus can reside in excited states
which can be connected with the ground state by sizable GT matrix
elements.  These states can then be deexcited in inelastic neutrino
scattering.  Note that in this case the final neutrino energy is
larger than the initial (see figure~\ref{fig:neut}) so that the
deexcitation occurs additionally with larger phase space.  Until
neutrino trapping there is little phase space blocking in inelastic
neutrino-nucleus scattering.  \citet{Toivanen.Kolbe.ea:2001} presented
the charged- and neutral-current cross sections for neutrino-induced
reactions on the iron isotopes $^{52-60}$Fe, using a combination of
shell model and RPA approach. Other possible neutrino processes, e.g.\ 
nuclear deexcitation by neutrino pair production~\eqref{eq:AAnunu},
have been discussed in \cite{Fuller.Meyer:1991}, but the estimated
rates are probably too small for these processes to be important
during the collapse.

Finally we remark that coherent elastic scattering on nuclei scales
like $\sim E_\nu^2$ so that neutrinos with low energies are the last
to be trapped. In order to fill this important sink for entropy and
energy, processes which affect the production of neutrinos with low
energies can be quite relevant for the collapse. Inelastic neutrino
scattering on nuclei, including finite temperature effects, is one
such process \cite{Bruenn.Haxton:1991}. The significantly lower
energies of the neutrinos generated by electron capture on nuclei than
the ones generated by capture on free protons is another reason to
implement these processes with appropriate care in collapse
simulations.

\subsection{Delayed supernova mechanism}
\label{sec:delay-supern-mech}

\begin{figure}[htbp]
  \includegraphics[width=0.9\linewidth]{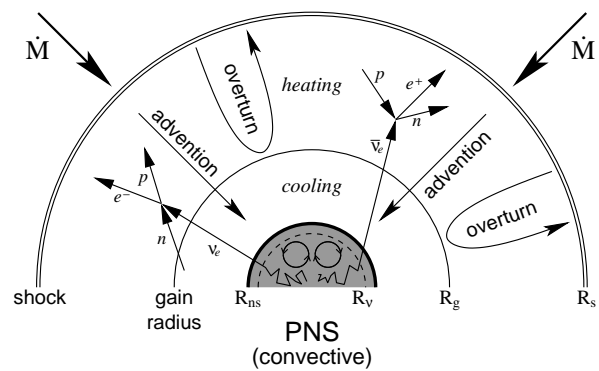}
  \caption{Sketch of the stellar core during the shock revival phase.
    $R_\nu$ is the neutrinosphere radius, from which neutrinos are
    expected to stream out freely, $R_{ns}$ is the radius of the
    protoneutron star, $R_g$ the gain radius (see text) and $R_s$ the
    radius at which the shock is stalled.  The shock expansion is
    impeded by mass infall at a rate ${\dot M}$, but supported by
    convective energy transport from the region of strongest neutrino
    heating to the stalled shock. Convection inside the protoneutron
    star (PNS) as well as correlations in the dense nuclear medium
    increase the neutrino luminosity \citep[adapted
    from][]{Janka.Kifonidis.Rampp:2001}. \label{fig:Janka1}}
\end{figure}

In the delayed supernova mechanism the fate of the explosion is
determined by several distinct neutrino processes. When the shock
reaches the $\nu_e$ neutrinosphere, from which $\nu_e$ are expected to
stream out freely, electron capture on the shock-heated and
shock-dissociated matter increases the $\nu_e$ production rate
significantly. Additionally neutrinos are produced by the
transformation of electron-positron pairs into $\nu \bar{\nu}$ pairs
(equation \ref{eq:eenunu}).  This process is strongly temperature
dependent \citep[e.g.][]{Brown.Soyeur:1979} and occurs most
effectively in the shock-heated regions of the proto-neutron star.
Electron-positron pair annihilation and nucleon-nucleon bremsstrahlung
(equation \ref{eq:NNNNnunu}) generate pairs of all three neutrino
flavors with the same probability and thus are the main mechanisms for
the production of $\nu_\mu$, $\nu_\tau$ neutrinos and antineutrinos
\cite{Raffelt:2001,Hanhart.Phillips.Reddy:2001,Stoica.Horvath:2002,%
Hannestad.Raffelt:1998,Hannestad:2001,Thompson.Burrows:2001,%
Thompson.Burrows.Horvath:2000}.  The emitted $\nu_e$ and $\bar{\nu}_e$
neutrinos, however, can be absorbed again by the free nucleons behind
the shock.  Due to the temperature and density dependences of the
neutrino processes involved, neutrino emission wins over neutrino
absorption in a region inside a certain radius (the \emph{gain
  radius}), while outside the gain radius matter is heated by neutrino
interactions that are dominated by absorption of electron
neutrinos and antineutrinos on free nucleons which have been
previously liberated by dissociation due to the shock (see
figure~\ref{fig:Janka1}). As a net effect, neutrinos transport energy
across the gain radius to the layers behind the shock. Due to the
smaller abundances, neutrino-induced reactions on finite nuclei are
expected to contribute only modestly to the shock revival. It has been
also suggested that the shock revival is supported by `preheating'
\cite{Haxton:1988}.  In this scenario the electron neutrinos, which
have been trapped during the final collapse phase and are liberated in
a very short burst (with luminosities of a few $10^{53}$~erg~s$^{-1}$
lasting for about 10 ms), can partly dissociate the matter (e.g.\ iron
and silicon isotopes) prior to the shock arrival. As reliable
neutrino-induced cross sections on nuclei have not been available
until recently, the neutrino-nucleus reactions have not been included
in collapse and post-bounce simulations.

To describe the important neutrino-nucleon processes, most core
collapse simulations use the same lowest order cross section for both
neutrinos and antineutrinos \cite{Bruenn:1985,Horowitz:2002}, i.e.,
they neglect terms of order $E_\nu/M$, where $E_\nu$ is the neutrino
energy and $M$ the nucleon mass. The most important corrections to the
cross section at this order are the nucleon recoil and the weak
magnetism related to the form factor $F_2$ in Eq.~\eqref{eq:4a}
\cite{Horowitz:2002}. The recoil correction is the same for neutrinos
and antineutrinos and decreases the cross sections. However, the weak
magnetism corrects the cross sections via its parity-violating
interference with the dominant axial-vector component. As the
interference is constructive for neutrinos and destructive for
antineutrinos, inclusion of the weak magnetism correction increases
the neutrino cross section, while it decreases the $\bar{\nu}$-nucleon
cross sections. It is then expected that corrections up to order
$E_\nu/M$ decrease the antineutrino cross section noticeably (by about
$25\%$ for 40 MeV antineutrinos), while the $\nu$-nucleon cross
sections are only affected by a few percents for $E_\nu \le 100$ MeV
\cite{Horowitz:2002}.

\begin{figure}[htbp]
  \includegraphics[width=0.9\linewidth]{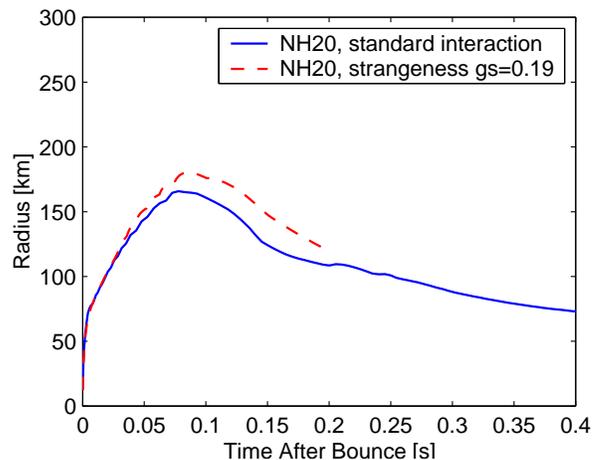}
  \caption{Shock trajectories of a 20 $M_\odot$ star, calculated with
    (dashed) and without (solid) an isoscalar strange axialform factor
    in the neutrino-nucleon elastic cross sections (courtesy of M.
    Liebend\"orfer). \label{fig:strange}}
\end{figure}

Neutral-current processes are sensitive to possible strange quark
contributions in the nucleon which would give rise to an isoscalar
piece $g_A^s$ in the axial-vector form factor besides the standard
isovector form factor $g_A \bm{\tau}$
\cite{Jaffe.Manohar:1990,Beise.McKeown:1991}.  The current knowledge
on $g_A^s$ comes from a $\nu p$ elastic scattering experiment
performed at Brookhaven yielding $g_A^s=-0.15\pm0.08$
\cite{Ahrens.Aronson.ea:1987}, but is considered rather uncertain
\cite{Garvey.Louis.White:1993}. With $g_A=1.26$ and assuming
axial-vector dominance, i.e.\ the cross section scales like $\sigma
\sim |g_A \bm{\tau} - g_A^s |^2$, a non-vanishing strange
axial-vector form factor would reduce the elastic scattering cross
section on neutrons and increase the $\nu p$ elastic cross section
\cite{Garvey.Krewald.ea:1992,Garvey.Kolbe.ea:1993,Horowitz:2002}.  As
the matter behind the shock is neutron-rich, the net effect will be a
reduction of the neutrino-nucleon elastic cross section.  This
increases the energy transfer to the stalled shock, however, a
simulation has shown that this increase is not strong enough for a
successful shock revival \citep[see
figure~\ref{fig:strange}]{Liebendoerfer.Messer.ea:2002}.

The physics involved in the attempt to revive the shock by neutrino
heating is exhaustively reviewed by \citet{Janka.Kifonidis.Rampp:2001}
\citep[see also][]{Burrows.Goshy:1993}. These authors show that the
fate of the stalled shock does not only depend on the neutrino heating
above the gain radius, but is also influenced by the energy loss in
the cooling region below the gain radius \citep[see
also][]{Bethe.Wilson:1985}. \citet{Janka:2001} also demonstrates the
existence of a critical value for the neutrino luminosity from the
neutron star needed to revive the shock. This critical luminosity
depends on the neutron star mass and radius and on the mass infall to
the shock. One expects that the shock expansion is eased for high mass
infall rates, which increase the matter pile-up on the neutron star
and push the shock outwards, and for high $\nu_e$ and $\bar{\nu}_e$
luminosities from the neutron star, which lead to an enhancement of
neutrino absorption relative to neutrino emission in the gain region.

The explosion depends crucially on the effectiveness by which energy
is transported by neutrinos to the region where the shock has stalled.
As stressed before, one-dimensional models including sophisticated
neutrino transport
\citep[e.g.][]{Rampp.Janka:2000,Liebendoerfer.Mezzacappa.ea:2001} fail
to explode. However, the neutrino energy transport is very sensitive
to: i) the effect of nucleon-nucleon correlations on the neutrino
opacities in dense matter and ii) convection both in the
neutrino-heated region and in the proto-neutron star
\cite{Janka.Kifonidis.Rampp:2001}.

In his pioneering work, \citet{Sawyer:1989} calculated the neutrino
mean free path, or equivalently the neutrino opacity, in uniform
nuclear matter and showed that effects due to strong interaction
between nucleons are important.  The same conclusion has been reached
by \citet{Raffelt.Seckel.Sigl:1996} who demonstrated that the average
neutrino-nucleon cross section in the medium is reduced due to spin
fluctuations induced by the spin-dependent interaction among nucleons.
\citep[For earlier calculations of the neutrino mean free path in
uniform nuclear matter,
see][]{Sawyer:1975,Friman.Maxwell:1979,Iwamoto.Pethick:1982}.
\citet{Sawyer:1989} exploited the relation between the equation of
state (EOS) of the matter and the long-wavelength excitations of the
system to calculate the weak interaction rates. However, consistency
between the EOS and the neutrino opacities are more difficult to
achieve for large energy ($q_0$) and momentum ($q$) transfer of the
neutrinos. Here, particle-hole and particle-particle interactions are
examples of effects which might influence the EOS and the neutrino
opacities. For the following discussion it is quite illuminating to
realize the similarity of the neutrino-induced excitations of nuclear
matter with the physics of multipole giant resonances in finite
nuclei.

For muon and tau neutrinos, neutral current reactions are the only
source of opacities. Here, the energy and momentum transfer is limited
by the matter temperature alone. For electron neutrinos the mean free
path is dominated by charged-current reactions, for which the energy
transfer is typically of the order of the difference between neutron
and proton chemical potentials.  During the early deleptonization
epoch of the proto-neutron star the typical neutron momenta are large
($\sim 100$--200~MeV) and the mismatch of proton, neutron and electron
Fermi momenta can be overcome by the neutrino momenta. This is not
longer possible in later stages when the neutrino energies are of
order $k_b T$; then momentum conservation restricts the available
phase space for the absorption reaction.  Pauli blocking of the lepton
in the final state increases the mean free path for charged-current
and neutral-current reactions.

We note an important and quite general consequence of the fact that
muon and tau neutrinos react with the proto-neutron star matter only
by neutral-current reactions: The four neutrino types have similar
spectra. Due to universality, $\nu_\mu$, $\nu_\tau$ and
$\bar{\nu}_\mu$, $\bar{\nu}_\tau$ have identical spectra.  It is
usually even assumed that neutrinos and antineutrinos have the same
spectra (one therefore refers to the 4 neutrino types unifyingly as
$\nu_x$ neutrinos) exploiting axial-vector dominance in the neutrino
cross sections. However, the interference of the axial-vector and the
weak magnetism components makes the $\bar{\nu}_x$ spectra slightly
hotter than the $\nu_x$ spectra.  The $\nu_x$ neutrinos decouple
deepest in the star, i.e., at a higher temperature, than electron
neutrinos and antineutrinos, and hence have higher energies.  As the
matter in the proto-neutron star is neutron-rich, electron neutrinos,
which are absorbed by neutrons, decouple at a larger radius than their
antiparticles, which interact with protons by charged-current
reactions. As a consequence decoupled electron neutrinos have, on
average, smaller energies than electron antineutrinos. Calculated
supernova neutrino spectra can be found in
\cite{Janka.Hillebrandt:1989,Yamada.Janka.ea:1999}, which yield the
average energies of the various supernova neutrinos approximately as:
$\langle E_{\nu_e} \rangle =11$ MeV, $\langle E_{\bar{\nu}_e} \rangle
=16$ MeV, and $\langle E_{\nu_x} \rangle =25$ MeV.
\citet{Burrows.Young.ea:2000} find the same hierarchy, but somewhat
smaller average neutrino energies.

For a much deeper and detailed description of the neutrino mean free
paths in dense matter the reader is refered to
\cite{Reddy.Prakash.ea:1999,Prakash.Lattimer.ea:2001} and the earlier
work
\cite{Reddy.Prakash.Lattimer:1998,Burrows.Sawyer:1998,Burrows.Sawyer:1999}.
We will here only briefly summarize the essence of the work presented
in these references.

\begin{figure}[htbp]
  \includegraphics[angle=270,width=0.9\linewidth]{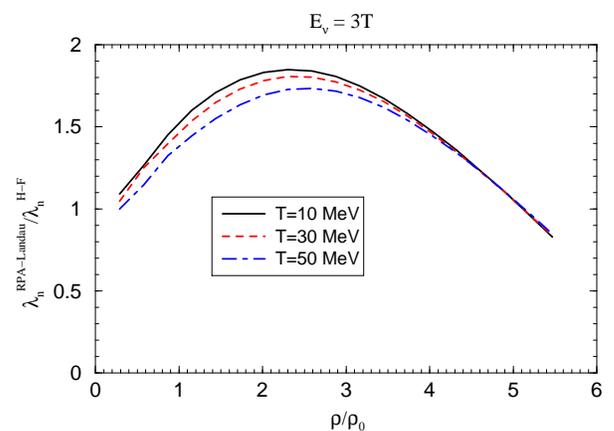}
  \caption{Ratio of neutrino mean free paths in neutron matter
    calculated in RPA and Hartree-Fock approaches at various
    temperatures \protect\cite{Margueron.Navarro.ea:2002}.  The
    interaction is the Gogny force D1P.  The neutrino energy is taken
    as $E_\nu=3T$.\label{fig:nufree}}
\end{figure}

Collapse simulations describe neutrino opacities typically on the
mean-field level or even by a nucleon gas. Then an analytical
expression can be derived for the vector and axial-vector response of
the medium which in turn determines the charged- and neutral-current
cross sections. Effects due to the strong interaction between nucleons
are considered by a medium-dependent effective mass in the dispersion
relation.  Like in finite nuclei, collective excitations in nuclear
matter arise due to nucleon-nucleon correlations beyond the mean-field
approximation.  As it is believed that single-pair excitations
dominate over multi-pair excitations for the kinematics of interest to
neutrino scattering and absorption, it appears to be sufficient to
determine the vector and axial-vector response, in a first step,
within the Random Phase Approximation (RPA). Assuming that the
interaction is short-ranged compared to the wavelength of the
excitations, it is justified to retain only s-wave components in the
interaction which in turn can be related to Fermi-liquid parameters.
It is found that the repulsive nature of the parameter $G_0'$, which
is related to the isovector spin-flip or giant Gamow-Teller resonances
in nuclei, induces a collective state in the region $\omega/q \sim
v_F$ ($v_F$ is the Fermi velocity), while the cross section is reduced
at smaller energies.  However, these smaller energies are important
for the neutrino mean free path at nuclear matter densities ($\rho_0$)
or smaller densities.  Assuming a typical neutrino energy $E_\nu
\approx 3 T$ (corresponding to a Fermi-Dirac distribution with
temperature $T$ and zero chemical potential) RPA correlations increase
the neutral-current neutrino mean-free path (see
figure~\ref{fig:nufree}) at low temperatures and for $\rho = \rho_0$,
compared with the mean-field result.  An enhancement due to RPA
correlations is also found for neutrino absorption mean-free paths for
neutrino-trapped matter.  Like in the case of neutrino-induced
reactions on finite nuclei (see above), finite-temperature effects
allow that nuclear excitation energy is transferred to the neutrino in
inelastic scattering processes.  This contributes to the cooling of
the nuclear matter and increases the neutrino energy in the final
state.

Neutrino heating is maximal in the layer just above the gain radius.
The energy transport from this region to the shock, which is stalled
further out, can be supported by convective overturn and might lead to
successful explosions, as has been demonstrated in several simulations
with two-dimensional hydrodynamics treatment of the region between the
gain radius and the shock \cite{Herant.Benz.ea:1994,%
  Burrows.Hayes.Fryxell:1995,Janka.Mueller:1996}.  The effect of
convection is twofold \cite{Janka.Kifonidis.Rampp:2001}: At first,
heated matter is transported outwards to cooler regions where the
energy loss due to neutrino emission is reduced (the neutrino
production rate for electron and positron captures on nucleons depends
strongly on temperature).  Second, cooler matter is brought down
closer to the gain radius where the neutrino fluxes are larger and
hence the heating is more effective. While this picture is certainly
appealing, it is not yet clear whether multi-dimensional simulations
will indeed lead to explosions as the two-dimensional studies did not
include state-of-the-art Boltzmann neutrino transport, but treated
neutrino transport in an approximate manner. In fact, in a simulation
with an improved treatment of neutrino transport the convective
overturn was found not to be strong enough to revive the stalled shock
\cite{Mezzacappa.Calder.ea:1998}.

The shock revival can also be supported by convection occuring inside
the protoneutron star where it is mainly driven by the negative lepton
gradient which is established by the rapid loss of leptons in the
region around the neutrinosphere
\cite{Burrows:1987,Keil.Janka.Mueller:1996}. By this mode,
lepton-rich matter will be transported from inside the protoneutron
star to the neutrinosphere which increases the neutrino luminosity and
thus is expected to help the explosion. The simulation of protoneutron
star convection is complicated by the fact that neutrinos and matter
are strongly coupled. In fact, two-dimensional simulations found that
neutrino transport can equilibrate otherwise convective fluid elements
\cite{Mezzacappa.Calder.ea:1998}. Such a damping is possible in
regions where neutrinos are still strongly coupled to matter, but
neutrino opacities are not too high to make neutrino transport
insufficient. In the model of \citet{Keil.Janka.Mueller:1996} the
convective mixing occurs very deep inside the core where the neutrino
opacities are high; no damping of the convection by neutrinos is then
found.

Wilson and Mayle attempted to simulate convection in their
spherical model by introducing neutron-fingers and found successful
explosions \cite{Wilson.Mayle:1993}. This idea is based on the
assumption that energy transport (by three neutrino flavors) is more
efficient than lepton number transport (only by electron neutrinos).
However, this assumption is under debate
\cite{Bruenn.Dineva:1996}.

\section{Nucleosynthesis beyond iron}

While the elements lighter than mass number $A\sim60$ are made by
charged-particle fusion reactions, the heavier nuclei are made by
neutron captures, which have to compete with $\beta$ decays.  Already
\cite{Burbidge.Burbidge.ea:1957,Cameron:1957} realized that two
distinct processes are required to make the heavier elements. This is
the slow neutron-capture process (s-process), for which the $\beta$
lifetimes $\tau_\beta$ are shorter than the competing neutron capture
times $\tau_n$. This requirement ensures that the s-process runs
through nuclei in the valley of stability. The rapid neutron-capture
process (r-process) requires $\tau_n \ll \tau_\beta$.  This is achieved
in extremely neutron-rich environment, as $\tau_n$ is inversely
proportional to the neutron density of the environment.  The r-process
runs through very neutron-rich, unstable nuclei, which are far-off
stability and whose physical properties are often experimentally
unknown.

Weak-interaction processes play interesting, but different roles in
these processes. The half-lives of the $\beta$-unstable nuclei along
the s-process path are usually known with good precision. However, the
nuclear half-life in the stellar environment can change due to the
thermal population of excited states in the parent nucleus. This is
particularly interesting if the effective lifetime is then comparable
to $\tau_n$, leading to branchings in the s-process path, from which
the temperature and neutron density of the environment can be
determined. Several recent examples are discussed in the next
subsection.  For the r-process, $\beta$-decays are probably even more
crucial. They regulate the flow to larger charge numbers and determine
the resulting abundance pattern and duration of the process.  Except
for a few key nuclei, $\beta$-decays of r-process nuclei have to be
modelled theoretically; we will briefly summarize the recent progress
below. Although the r-process site is not yet fully determined, it is
conceivable that it occurs in the presence of extreme neutrino fluxes.
As we will discuss, neutrino-nucleus reactions can have interesting
effects during and after the r-process, perhaps allowing for clues to
ultimately identify the r-process site.

\subsection{S-process}

The analysis of the solar abundances have indicated that two
components of the s-process have contributed to the synthesis of
elements heavier than $A\sim60$. The weak component produces the
elements with $A\lesssim 90$. Its site is related with helium core
burning of CNO material in more massive stars
\cite{Couch.Schmiedekamp.Arnett:1974,Kaeppeler.Wiescher.ea:1994}. The
main component, which is responsible for the heavier s-process
nuclides up to Pb and Bi, is associated with helium flashes occurring
during shell burning in low-mass (asymptotic giant branch) stars
\cite{Busso.Gallino.Wasserburg:1999}. The $^{22}$Ne($\alpha,n)^{25}$Mg
and $^{13}$C($\alpha,n)^{16}$O reactions are believed to be the
supplier of neutrons for the weak and main components, respectively.
The s-process abundances $N_s$ are found to be inversely proportional
to the respective (temperature averaged) neutron capture cross
sections $\langle \sigma \rangle$, as expected for a steady-flow
picture \cite{Burbidge.Burbidge.ea:1957}, which, however, breaks down
for the extemely small cross sections at the magic neutron numbers. As
a consequence, the product $N_s \cdot \langle \sigma \rangle$ exhibits
almost constant plateaus between the magic neutron numbers, separated
by pronounced steps, \citep[e.g.][]{Kaeppeler:1999}.

\begin{figure}[htbp]
  \includegraphics[width=0.9\linewidth]{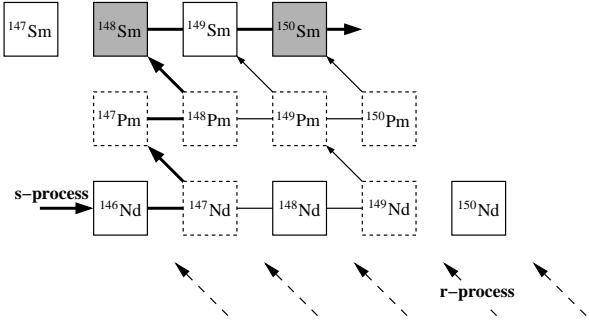}
  \caption{The s-process reaction path in the Nd-Pm-Sm region
    with the branchings at $A=147$, 148, and 149. Note that $^{148}$Sm
    and $^{150}$Sm are shielded against r-process $\beta$-decays
    \citep[adapted from][]{Kaeppeler:1999}.\label{fig:ssm}}
\end{figure}

The neutron density of the stellar environment during the main
s-process component can be determined from branching points occuring,
for example, in the $A=147$--149 mass region (see
figure~\ref{fig:ssm}). Here the relative abundances of the two
s-process-only isotopes $^{148}$Sm and $^{150}$Sm ($Z=62$), which are
shielded against r-process contributions by the two stable Nd ($Z=60$)
isotopes $^{148}$Nd and $^{150}$Nd, is strongly affected by branchings
occuring at $^{147}$Nd and, more importantly, at $^{148}$Pm and at
$^{147}$Pm. As the neutron captures on these branching nuclei will
bypass $^{148}$Sm in the flow pattern, the $^{150}$Sm $N_s \langle
\sigma \rangle$ value will be larger for this nucleus than for
$^{148}$Sm. Furthermore, the neutron capture rate $\lambda_n$ is
proportional to the neutron density $N_n$.  Thus, $N_n$ can be
determined from the relative $^{150}$Sm/$^{148}$Sm abundances,
resulting in $N_n = (4.1 \pm 0.6) \times 10^8$ cm$^{-3}$
\cite{Toukan.Debus.ea:1995}. A similar analysis for the weak s-process
component yields neutron densities of order $(0.5\text{--}1.3) \times
10^8$~cm$^{-3}$ \cite{Walter.Beer.ea:1986a,Walter.Beer.ea:1986b}.  We
stress that a 10\% determination of the neutron density requires the
knowledge of the involved neutron capture cross sections with about
1\% accuracy \cite{Kaeppeler.Thielemann.Wiescher:1998}, which has yet
not been achieved for unstable nuclei.  Improvements are expected from
new time-of-flight facilities like LANSCE at Los Alamos or NTOF at
CERN.

\begin{figure}[htbp]
  \includegraphics[width=0.9\linewidth]{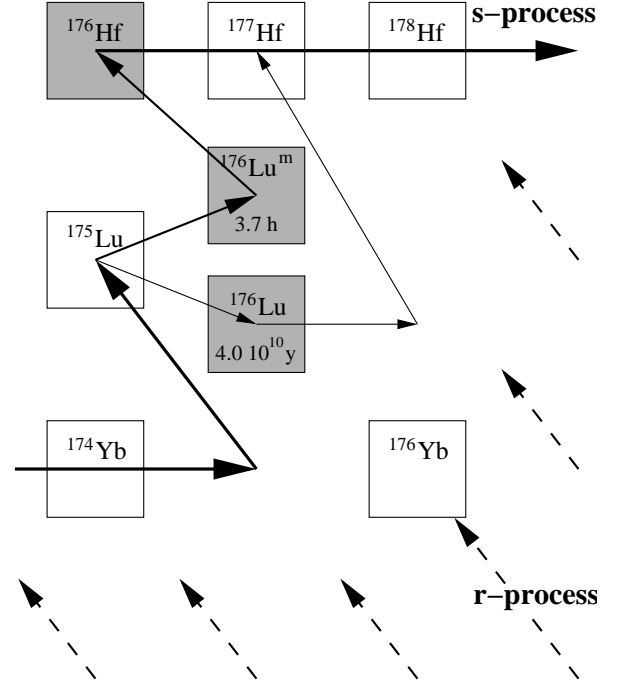}
  \caption{The s-process neutron capture path in the Yb-Lu-Hf region
    (solid lines). For $^{176}$Lu the ground state and isomer are
    shown separately. Note that $^{176}$Lu and $^{176}$Hf are shielded
    against r-process $\beta$-decays \citep[adapted
    from][]{Doll.Boerner.ea:1999}.\label{fig:slu176}}
\end{figure}

\begin{figure}[htbp]
  \includegraphics[width=0.9\linewidth]{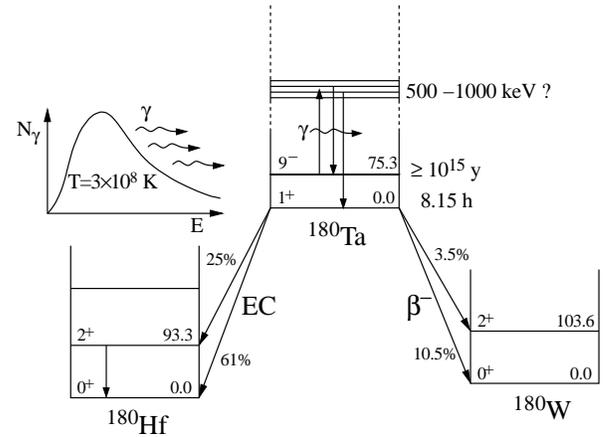}
  \caption{The figure shows an schematic energy-level diagram of
    $^{180}$Ta and its daughters illustrating the possibility for
    thermally enhanced decay of $^{180}$Ta$^m$ in the stellar
    environment of the s-process. The inset shows the photon density
    at s-process temperature \citep[adapted
    from][]{Belic.Arlandini.ea:1999}. \label{fig:mediate}}
\end{figure}

The temperature of the s-process environment can be `measured', if the
$\beta$ half-life of a branching point nucleus is very sensitive to
the thermal population of excited nuclear levels. A prominent example
is $^{176}$Lu \cite{Beer.Kaeppeler.ea:1981}. For mass number $A=176$,
the $\beta$-decays from the r-process terminate at $^{176}$Yb
($Z=70$), making $^{176}$Hf ($Z=72$) and $^{176}$Lu ($Z=71$) s-only
nuclides (see figure~\ref{fig:slu176}). Besides the long-lived ground
state ($t_{1/2}=4.00(22) \times 10^{10}$ y), $^{176}$Lu has an
isomeric state at an excitation energy of 123 keV ($t_{1/2}=3.664(19)$
h). Both states can be populated by $^{175}$Lu$(n,\gamma)$ with known
partial cross sections.  At $^{176}$Lu, the s-process matter flow is
determined by the competition of neutron capture on the ground state
and $\beta$-decay of the isomer.  But importantly, the ground and
isomeric states couple in the stellar photon bath via the excitation
of an intermediate state at 838 keV (figure~\ref{fig:mediate} shows a
similar process taking place in $^{180}$Ta), leading to a matter flow
from the isomer to the ground state, which is very
temperature-dependent. A recent analysis of the $^{176}$Lu s-process
branching yields an environment temperature of $T=(2.5\text{--}3.5)
\times 10^8$ K \cite{Doll.Boerner.ea:1999}.

Similar finite-temperature effects play also an important role in the
s-process production of $^{180}$Ta. This is the rarest isotope
(0.012\%) of nature's rarest element. It only exists in a long-lived
isomer ($J^\pi=9^-$) at an excitation energy of 75.3 keV and with a
half-life of $t_{1/2} \ge 1.2 \times 10^{15}$~y. The $1^+$ ground
state decays with a half-life of 8.152(6)~h, mainly by electron
capture to $^{180}$Hf.  While potential s-process production paths of
the $^{180}$Ta isomer have been pointed out, the survival of this
state in a finite-temperature environment has long been questionable.
While a direct electromagnetic decay to the ground state is strongly
suppressed due to angular momentum mismatch, the isomer can decay via
thermal population of intermediate states with branchings to the
ground state (see figure~\ref{fig:mediate}). By measuring the relevant
electromagnetic coupling strength, the temperature-dependent
half-life, and thus the $^{180}$Ta survival rate, has been determined
by \citet{Belic.Arlandini.ea:1999} under s-process conditions as
$t_{1/2} \lesssim 1$ y, i.e.\ more than 15 orders of magnitude smaller
than the half-life of the isomer!  Accompanied by progress in stellar
modeling of the convective modes during the main s-process component
(which brings freshly produced $^{180}$Ta to cooler zones, where it
can survive more easily, on timescales of days) it appears now likely
that $^{180}$Ta is partly made within s-process
nucleosynthesis~\cite{Wisshak.Voss.ea:2001}. The p-process
\cite{Rayet.Arnould.ea:1995} and neutrino nucleosynthesis
\cite{Woosley.Hartmann.ea:1990} have been proposed as alternative
sites for the $^{180}$Ta production.

The two neighboring isotopes $^{186}$Os and $^{187}$Os are s-process
only nuclides, shielded against the r-process by $^{186}$W and
$^{187}$Rh. The nucleus $^{187}$Rh has a half-life of 42 Gy which is
comparable with the age of the universe. As
the $^{187}$Rh decay has contributed to the observed $^{187}$Os
abundance, the Os/Rh abundance ratio can serve as a sensitive clock
for the age of the universe, once the s-process component is
subtracted from the $^{187}$Os abundance \cite{Clayton:1964}.  To
determine the latter precise measurements of the neutron capture cross
sections on $^{186}$Os and $^{187}$Os are required, which are in
progress at CERN's NTOF facility. A potential complication arises from
the fact that $^{187}$Os has a low-lying state at 9.8 keV which is
populated equally with the ground state at s-process temperatures. The
respective neutron capture cross sections on the excited state can be
indirectly determined from (n,n') measurements.  Furthermore one has
to consider that the half-life of $^{187}$Rh is strongly
temperature-dependent: If stripped of its electrons, the half-life is
reduced by 9 orders of magnitude to $t_{1/2}=32.9\pm2$ y as measured
at the GSI storage ring \cite{Bosch.Faestermann.ea:1996}.  However,
the GSI data can be translated into a $\log ft$ value from which the
$^{187}$Rh half-life can be deduced for every ionization state.

The decay of totally ionized $^{187}$Rh is an example for a bound
state $\beta$-decay, i.e., the decay of bare $^{187}$Rh$^{75+}$ to
continuum states of $^{187}$Os$^{76+}$ is energetically forbidden, but
it is possible if the decay electron is captured in the K-shell
($Q_\beta$=73 keV) or in the L-shell ($Q_\beta$=9.1 keV)
\cite{Johnson.Soff:1985}. We note that the decay of neutral $^{187}$Rh
is energetically allowed for a first-forbidden transition to the
$^{187}$Os ground state; the long half-life results from the small
matrix element and $Q_\beta$ value related to this transition. Another
example of a bound state $\beta$-decay with importance for the
s-process occurs for $^{163}$Dy.  This nucleus is stable as a neutral
atom, but, if fully ionized, can decay to $^{163}$Ho with a half-life
of $47^{+5}_{-4}$~d; the measurement of this half-life at the GSI
storage ring was the first observation of bound state $\beta$-decay
\cite{Jung.Bosch.ea:1992}. The consideration of the $^{163}$Dy decay
at s-process conditions has been found to be essential to explain the
abundance of the s-process only nuclide $^{164}$Er, which is produced
by neutron capture on $^{163}$Ho, the daughter of the bound state
$\beta$-decay of $^{163}$Dy, and the subsequent $\beta$-decay of
$^{164}$Ho to $^{164}$Er.

\subsection{R-process}

\begin{figure}[htbp]
  \includegraphics[width=0.9\linewidth]{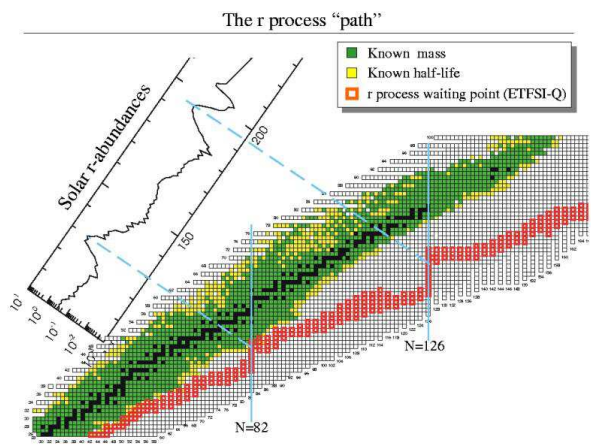}
  \caption{The r-process occurs under dynamically changing astrophysically
    conditions which affect the reaction pathway. The figure shows the
    range of r-process paths, defined by their waiting point nuclei.
    After decay to stability the abundance of the r-process
    progenitors produce the observed solar r-process abundance
    distribution. The r-process paths run generally through
    neutronrich nuclei with experimentally unknown masses and
    halflives. In this calculation a mass formula based on the
    Extended Thomas Fermi model with Strutinski Integral (ETFSI) and
    special treatment of shell quenching (see text) has been adopted.
    (courtesy of Karl-Ludwig Kratz and Hendrik Schatz).
\label{fig:pfad}}
\end{figure}

Phenomenological parameter studies indicate that the r-process occurs
at temperatures around $T \sim 100$ keV and at extreme neutron fluxes
(neutron number densities $n> 10^{20}$~cm$^{-3}$)
\citep[e.g.][]{Cowan.Thielemann.Truran:1991}.  It has also been
demonstrated that not all r-process nuclides can be made
simultaneously at the same astrophysical conditions (constant
temperature, neutron density), i.e.\ the r-process is a dynamical
process with changing (different) conditions and paths
\cite{Kratz.Bitouzet.ea:1993}.  In a good approximation, the neutron
captures proceed in $(n,\gamma) \rightleftarrows (\gamma,n)$
equilibrium, fixing the reaction paths at rather low neutron
separation energies of $S_n \sim 2$--3 MeV
\cite{Cowan.Thielemann.Truran:1991}, implying paths through very
neutron-rich nuclei in the nuclear chart as is shown in
figure~\ref{fig:pfad}. 
While the general picture
of the r-process appears to be well accepted, its astrophysical site
is still open. The extreme neutron fluxes point to explosive scenarios
and, in fact, the neutrino-driven wind above the nascent neutron star
in a core-collapse supernova is the currently favored
model \cite{Witti.Janka.Takahashi:1994a,Witti.Janka.Takahashi:1994b,%
  Woosley.Wilson.ea:1994}. But also shock-processed helium shells in
type II supernovae \cite{Truran.Cowan.Fields:2001} and neutron star
mergers
\cite{Freiburghaus.Rosswog.Thielemann:1999,Rosswog.Davies.ea:2000} are
investigated as possible r-process sites. Recent meteoritic clues
\cite{Wasserburg.Busso.Gallino:1996,Qian.Vogel.Wasserburg:1998a} as
well as observations of r-process abundances in low-metallicity stars
\cite{Sneden.Cowan.ea:2000} point to more than one distinct site for
the solar r-process nuclides. 

In an important astronomical observation \citet{Cayrel.Hill.ea:2001}
have recently detected the r-process nuclides thorium and uranium in
an old galactical halo star. As these two nuclei have halflives, which
are comparable to the expected age of the universe, their measured
abundance ratio serves as a sensitive clock to determine a lower limit
to this age, provided their initial r-process production abundance
ratio can be calculated with sufficient accuracy.
\citet{Cayrel.Hill.ea:2001} have used their observed
$^{238}$U/$^{232}$Th abundance ratio from the star CS31082-001 and the
r-process model predictions from
\cite{Cowan.Pfeiffer.ea:1999,Goriely.Clerbaux:1999} to deduce a value
of $12.5\pm3.3$~Gyr for the age of the star. Recently, a refined
analysis of the CS31082-001 spectra has led to a significant
improvement in the derived abundances which  provides now an age estimate
of $14.0\pm2.4$~Gyr \cite{Hill.Plez.ea:2002}. The effect of different
nuclear physics input on the r-process production of U and Th have
been studied by \citet{Goriely.Arnould:2001} and by
\citet{Schatz.Toenjes.ea:2002}

As relevant nuclear input, r-process simulations require neutron
separation energies (i.e.\ masses), half-lives and neutron capture
cross sections of the very neutron-rich nuclei on the various
dynamical r-process paths. As currently only few experimental data are
known for r-process nuclei, these quantities have to be modelled.
Traditionally this has been done by global models which, by fitting a
certain set of parameters to known experimental data, are then being
used to predict the properties of all nuclei in the nuclear landscape.
Arguably the most important input to r-process simulations are neutron
separation energies as they determine, for given temperature and
neutron density of the astrophysical environment, the r-process paths.
The most commonly used mass tabulations are based on the
microscopic-macroscopic finite-range droplet model (FRDM) approach
\cite{Moeller.Nix.Kratz:1997} or the Extended Thomas-Fermi model with
Strutinski Integral (ETFSI) approach \cite{Aboussir.Pearson.ea:1995}.
In more recent developments mass tabulations have been developed
adopting parametrizations inspired by shell model results
\cite{Duflo.Zuker:1995} or calculated on the basis of nuclear
many-body theories like the Hartree-Fock model with a BCS treatment of
pairing \cite{Goriely.Tondeur.Pearson:2001}.  Special attention has
been paid recently also to the `shell quenching', i.e.\ the
observations made in HFB calculations that the shell gap at magic
neutron numbers is less pronounced in very neutron-rich nuclei than in
nuclei close to stability \cite{Dobaczewski.Hamamoto.ea:1994}. Such a
vanishing of the shell gap has been experimentally verified for the
magic neutron number $N=20$
\cite{Guillemaud-Mueller.Detraz.ea:1984,Motobayashi.Ikeda.ea:1995}.
The confirmation of the predicted quenching at the $N=82$ shell
closure is the aim of considerable current experimental activities
\cite{Kratz.Moeller.ea:2000}.  Recent theoretical studies on this
topic are reported in \cite{Sharma.Farhan:2001,Sharma.Farhan:2002}.
The potential importance of shell quenching for the r-process rests on
the observation
\cite{Chen.Dobaczewski.ea:1995,Pfeiffer.Kratz.Thielemann:1997} that it
can correct the strong trough just below the r-process peaks in the
calculated r-process abundances encountered with conventional mass
models. Neutron capture cross sections become important, if the
r-process flow drops out of $(n,\gamma) \rightleftarrows (\gamma,n)$
equilibrium, which happens close to freeze-out when the neutron source
ceases. They can also be relevant for nuclides with small abundances
for which no flow equilibrium is built up
\cite{Surman.Engel.ea:1997,Surman.Engel:2001}.  In the following we
will summarize recent progress in calculating half-lives for nuclei on
the r-process paths.

\subsubsection{Half-Lives}

The nuclear half-lives determine the relative r-process abundances. In a
simple $\beta$-flow equilibrium picture the elemental abundance is
proportional to the half-life, with some corrections for
$\beta$-delayed neutron emission \cite{Kratz.Thielemann.ea:1988}.
As r-process half-lives are longest for the magic nuclei, these waiting
point nuclei determine the minimal r-process duration time;
i.e.\ the time needed to build up the r-process peak around $A\sim 200$
via matter-flow from the seed nucleus. We note, however, that this time
depends also crucially on the r-process path.

Pioneering experiments to measure half-lives of neutron-rich isotopes
near the r-process path succeeded in determining the half-lives of two
$N=50$ ($^{80}$Zn, $^{79}$Cu) and two $N=82$ ($^{129}$Ag, $^{130}$Cd)
waiting point nuclei
\cite{Gill.Casten.ea:1986,Kratz.Gabelmann.ea:1986,%
Pfeiffer.Kratz.ea:2001}. \citet{Pfeiffer.Kratz.ea:2001} reviewed the
experimental information on r-process nuclei. These data play crucial
roles in constraining and testing nuclear models, which are still
necessary to predict the bulk of half-lives required in r-process
simulations. It is generally assumed that the half-lives are determined
by allowed Gamow-Teller (GT) transitions. 
The calculations of $\beta$-decays require usually two ingredients:
the GT strength distribution in the daughter nucleus and the relative
energy scale between parent and daughter, i.e.\ their mass difference.
However, the $\beta$ decays
only probe the weak low-energy tail of the GT distributions. Only a
few percent of the $3(N-Z)$ Ikeda sum rule
\cite{Ikeda.Fujii.Fujita:1963} lie within the $Q_\beta$ window
(i.e.\ at energies accessible in $\beta$-decay), the
rest being located in the region of the Gamow-Teller resonance at
higher excitation energies. 
Due to the strong $E^5$ energy dependence of the phase space
$\beta$-decay rates are very sensitive to the correct description of the
detailed low-energy GT distribution and its relative energy scale to the
parent nucleus.
This  explains why different calculations
of the $\beta$-decay half-lives present large deviations among them.

\begin{figure*}[htbp]
  \includegraphics[width=0.45\linewidth]{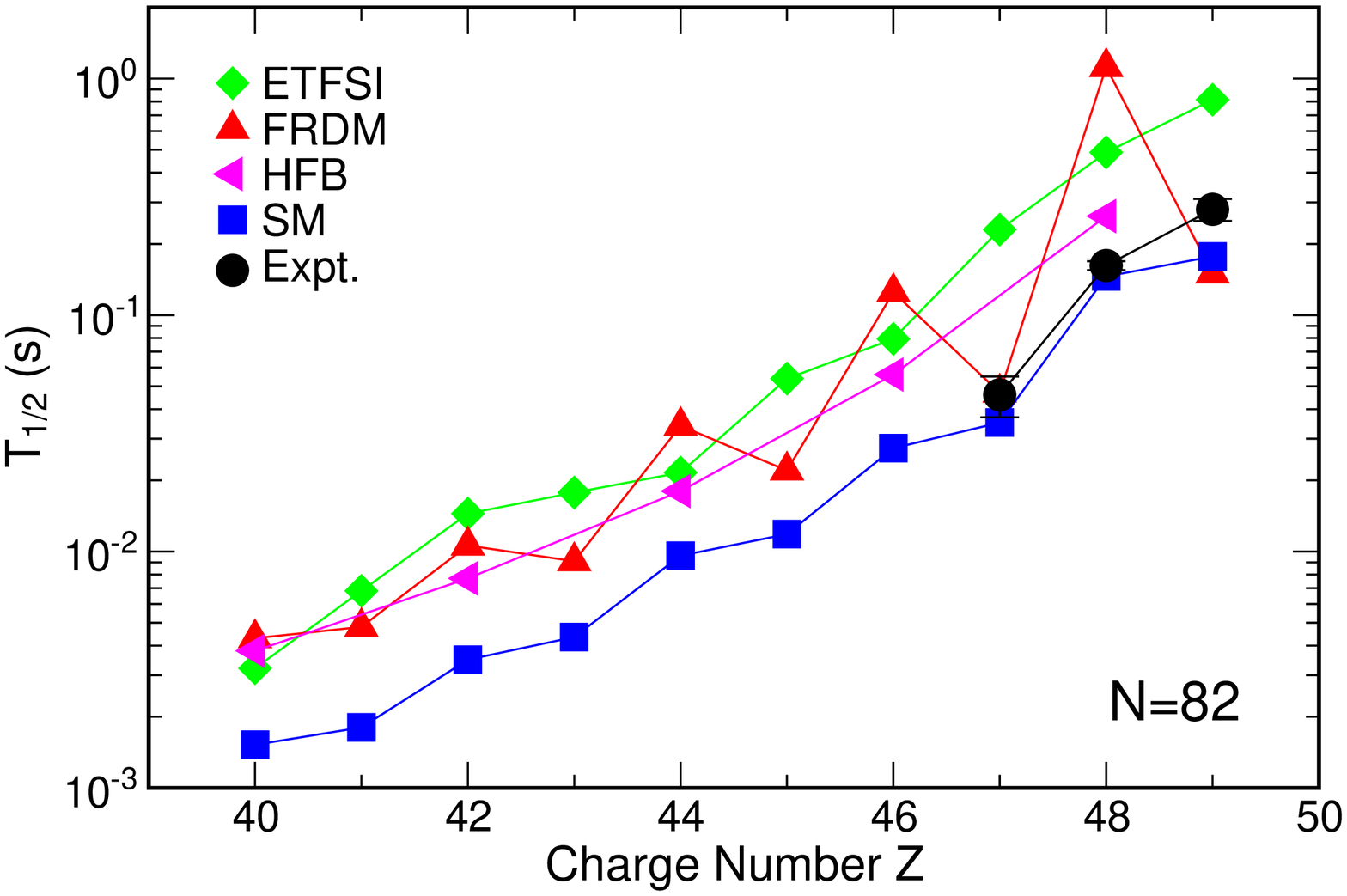}%
  \hspace{0.03\linewidth}%
  \includegraphics[width=0.45\linewidth]{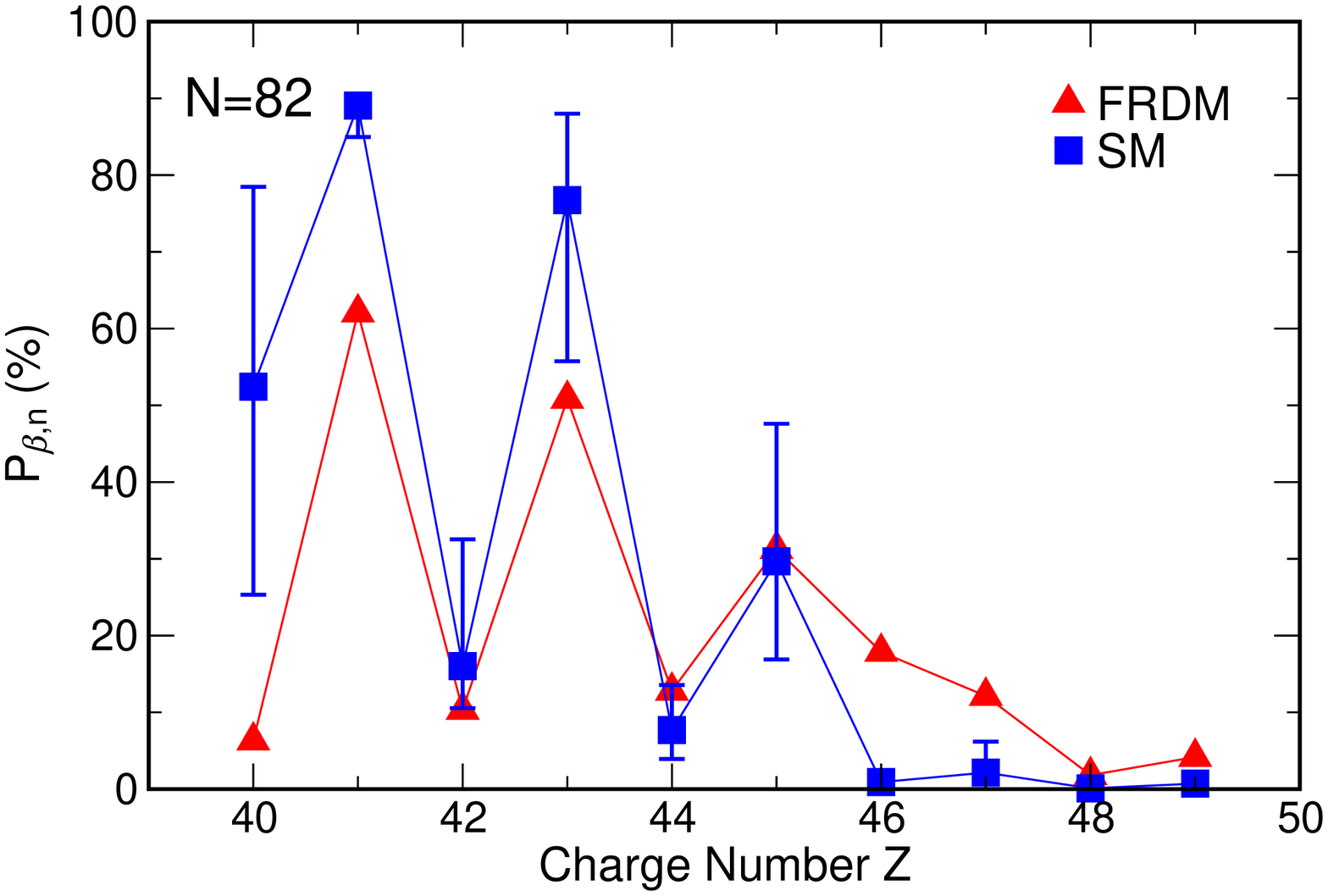}\\[3mm]
  \includegraphics[width=0.45\linewidth]{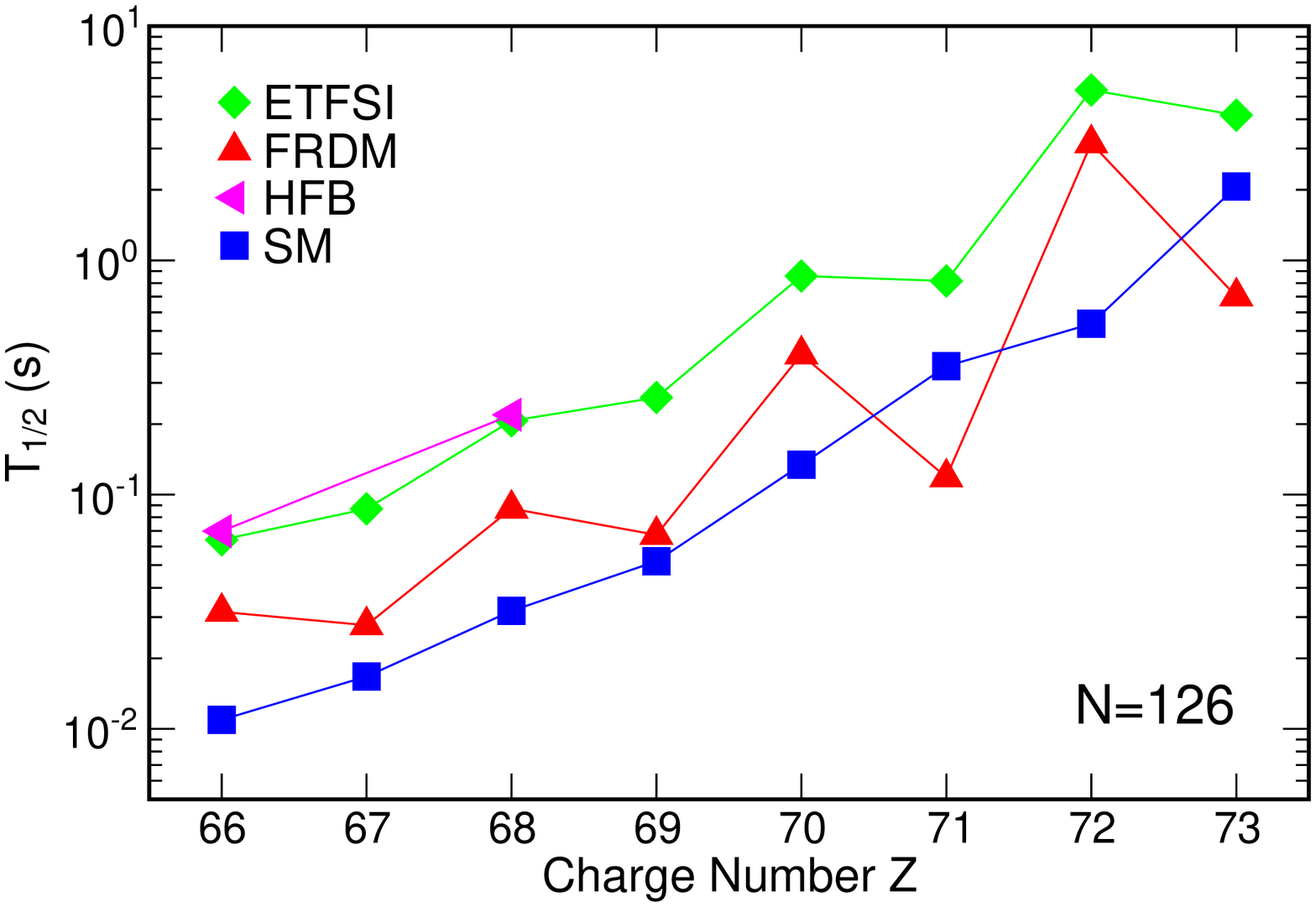}%
  \hspace{0.03\linewidth}%
  \includegraphics[width=0.45\linewidth]{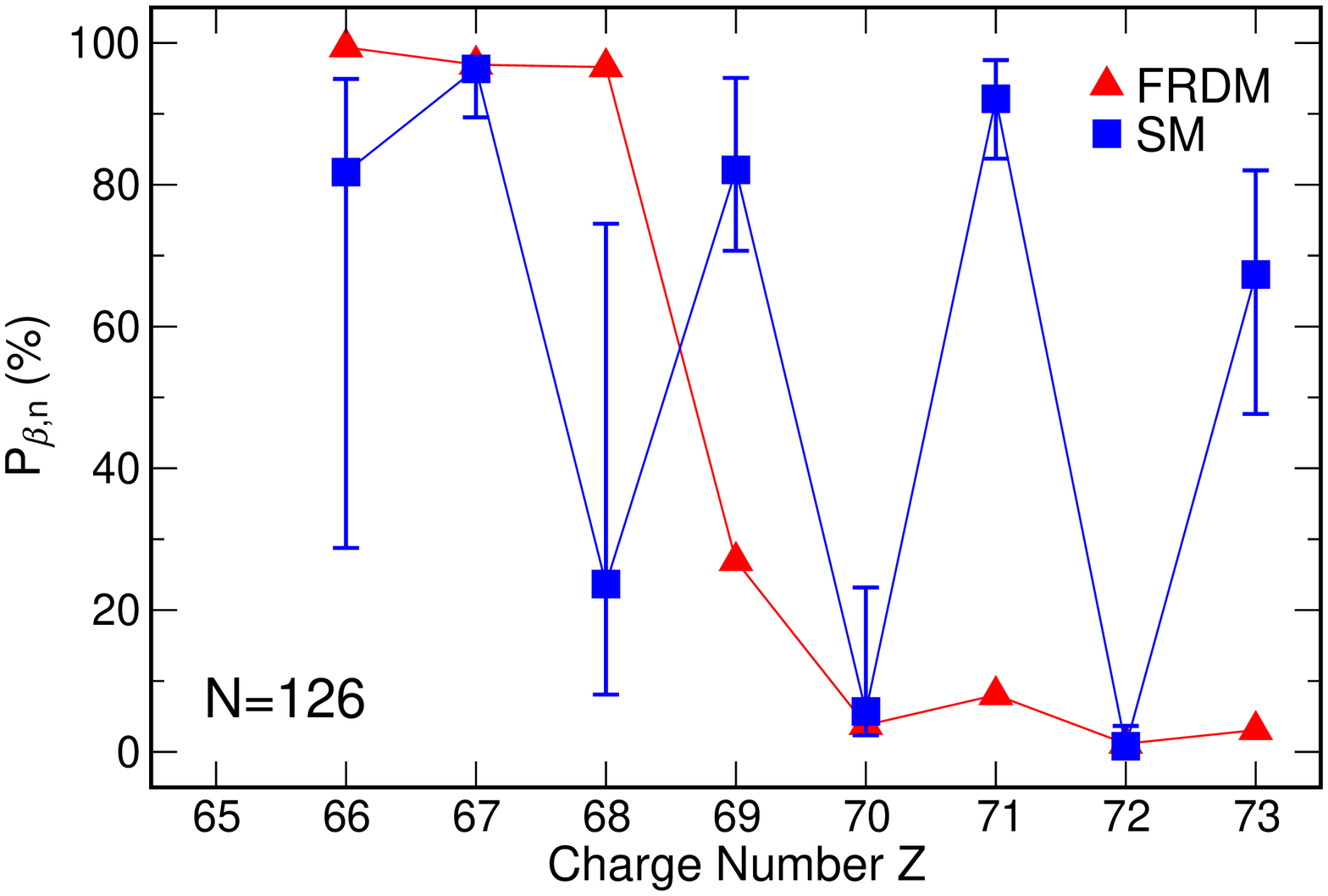}
  \caption{The left panels show the half-lives of the r-process waiting
    point nuclei with neutron numbers $N=82$ and 126, obtained within
    different theoretical models. In the case of $N=82$ the half-lives
    of $^{131}$In, $^{130}$Cd and $^{129}$Ag are 
    experimentally known \cite{Pfeiffer.Kratz.ea:2001}.
    The right panels compare the beta-delayed neutron emission
    probabilities for the $N=82$ (upper) and $N=126$ (lower)
    waiting-point nuclei as calculated in two different models. For
    the shell-model (SM) probabilities the error bars indicate the
    sensitivity of the calculations to a change of $\pm0.5$ MeV in the
    neutron separation energies. \label{fig:n82n126}}
\end{figure*}

Because of the huge number of nuclei relevant for the r-process, the
estimates of the half-lives are so far based on a combination of
global mass models and the quasiparticle random phase approximation
(QRPA), the latter to calculate the GT matrix elements.  Examples of
these models are the FRDM/QRPA \cite{Moeller.Nix.Kratz:1997} and the
ETFSI/QRPA \cite{Borzov.Goriely:2000}. Recently, calculations based on
the self-consistent Hartree-Fock-Bogoliubov plus QRPA model became
available for r-process waiting point nuclei with magic neutron
numbers $N=50$, 82 and 126 \cite{Engel.Bender.ea:1999}. The presence
of a closed neutron shell has allowed also for the study of these
nuclei by shell-model calculations
\cite{Martinez-Pinedo.Langanke:1999,Martinez-Pinedo:2001}, which is
the method of choice to determine the $\beta$-decay matrix elements.
However an adequate calculation of the nuclear masses for heavy nuclei
is yet prohibitive in the shell model.  In
\citet{Martinez-Pinedo.Langanke:1999}, and \citet{Martinez-Pinedo:2001}
the masses were adopted from the global model of
\citet{Duflo.Zuker:1995}.  Figure~\ref{fig:n82n126} compares the
half-lives predicted by the different approaches. For $N=82$, the
half-lives of $^{131}$In, $^{130}$Cd and $^{129}$Ag are known
experimentally \cite{Kratz.Gabelmann.ea:1986,Pfeiffer.Kratz.ea:2001}.
The comparison of the predictions of the different models with the few
experimental r-process benchmarks reveals some of their
insufficiencies. For example, the FRDM/QRPA half-lives show a
significant odd-even staggering which is not present in the data,
while the ETFSI/QRPA half-lives appear globally too long. The HFB/QRPA
and SM approaches obtain half-lives in reasonable agreement with the
data and predict shorter half-lives for the unmeasured waiting point
nuclei with $N=82$ than the global FRDM/QRPA and ETFSI/QRPA
approaches. For $N=126$ there is no experimental information so the
different models remain untested. While HFB calculations for the
half-lives of all the r-process nuclei are conceivable, similar
calculations within the shell-model approach are still not feasible
due to computer memory limitations.

While the $Q_\beta$ value for the decay of neutron-rich r-process
nuclei is large, the neutron separation energies are small. Hence
$\beta$ decay can lead to final states above neutron threshold and is
accompanied by neutron emission. If the r-process proceeds by an
$(n,\gamma)\rightleftarrows(\gamma,n)$ equilibrium the $\beta$-delayed
neutron emission probabilities, $P_{\beta,n}$, only play a role at the
end of the r-process when the neutron source has ceased and the
produced nuclei decay to stability. The calculated $P_{\beta,n}$
values are very sensitive to both the low-energy GT distribution and
the neutron threshold energies. No model describes currently both
quantities simultaneously with sufficient accuracy.
Figure~\ref{fig:n82n126} compares the $P_{\beta,n}$ computed in the
FRDM/QRPA and SM approaches for $N=82$ and 126.  The predictions
between different models can be quite different in experimentally
non-determined mass regions.  For the SM approach the error bars
indicate the sensitivity of the computed $P_{\beta,n}$ values to a
change of $\pm 0.5$ MeV in the neutron separation energies of the
daughter nucleus; the effect can be large.

It has been pointed out that first-forbidden transitions might have
important contributions in some nuclei close to magic numbers
\cite{Blomqvist.Kerek.Fogelberg:1983,Homma.Bender.ea:1996,Korgul.Mach.ea:2001}.
A systematic inclusion of first-forbidden transitions in the
calculation of r-process beta-decay half-lives in any of the many-body
methods used to describe the GT contributions has not been done.
However, a first attempt towards this goal
\cite{Moeller.Pfeiffer.Kratz:2002} has combined first-forbidden
transitions estimated in the Gross Theory
\cite{Takahashi.Yamada.Kondoh:1973} with GT results taken from QRPA
calculations. No significant changes compared to r-process studies,
which consider only the GT contributions to the halflives, have been
observed \cite{Kratz.pvt}.

The presence of low-lying isomeric states in r-process nuclei might
change the effective half-lives in the stellar environment. Currently
no estimates of these effect exists except for odd-A nuclei with
$N=82$. Here shell-model calculations predict half-lives for the
isomeric states very similar to the ground state half-lives
\cite{Martinez-Pinedo.Langanke:1999}. 
The halflives of the isomeric state in $^{129}$Ag has also been
estimated within the QRPA approach and by a second shell model
calculation finding values about a factor of 2 larger than 
the $^{129}$Ag ground state half-life \cite{Kratz:2001}. This reference
also reports about the first attempt to measure the half-life of the
isomeric state.

For heavy nuclei ($Z \ge 84$) some final states populated by
$\beta$-decay in the daughter nucleus can also decay by 
fission \cite{Cowan.Thielemann.Truran:1991}. The relevant beta-delayed
fission probabilities depend sensitively on the modelling of the fission
barriers \cite{Howard.Moeller:1980,Mamdouh.Pearson.ea:2001}.

\begin{figure}[htbp]
  \includegraphics[width=0.9\linewidth]{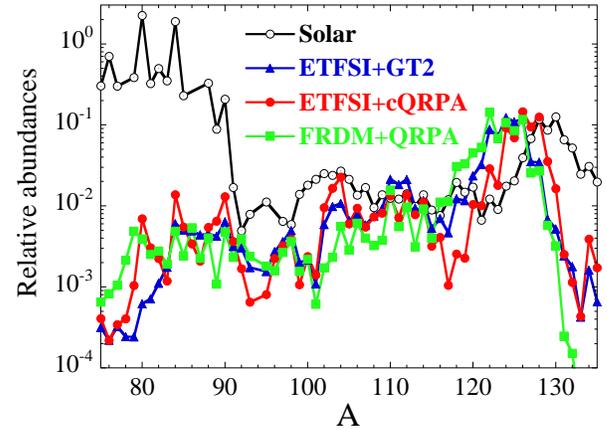}
  \caption{Abundances of r-process nuclides calculated in a dynamical
    r-process model and for different global sets of $\beta$-decay
    half-lives. In the dynamical r-process the matter flow timescale
    competes with the nuclear timescale, set by the $\beta$-decay
    half-lives.  As a consequence the magic neutron numbers (here
    $N=82$) are reached at different astrophysical conditions and
    hence at different proton numbers, which is reflected in the
    shifts of the abundance peaks \citep[from][]{Borzov.Goriely:2000}.
    \label{fig:gor2}}
\end{figure}

Borzov and Goriely have studied the influence of the $\beta$
half-lives on the r-process abundances within two distinct scenarios:
the canonical r-process picture with a exposure of the seed nucleus
$^{56}$Fe by a constant neutron density and temperature for a fixed
duration time (2.4~s) and the neutrino-driven wind model. In the
canonical model the location of the r-process abundance peaks depends on the
masses, but not on the $\beta$ half-lives, which act only as
bottlenecks for the matter-flow to more massive nuclei. In the
dynamical neutrino-driven wind model the half-lives affect the
abundance distribution. This comes about as at later times in this
model the environmental conditions shift the r-process to more
neutron-rich nuclei. Long half-lives then imply that the matter-flow
reaches the magic neutron numbers later, i.e.\ for more neutron-rich
nuclei.  Consequently the abundance peaks are shifted to smaller
$A$-values \citep[see figure~\ref{fig:gor2},][]{Borzov.Goriely:2000}.
Similar studies have been presented by
\citet{Kratz.Pfeiffer.Thielemann:1998}. 

\subsubsection{The possible role of neutrinos in the r-process}

Among the various possible astrophysical sites for the r-process, the
neutrino-driven wind model \cite{Witti.Janka.Takahashi:1994a,%
  Witti.Janka.Takahashi:1994b,Woosley.Wilson.ea:1994} has attracted
most attention in recent years.  Here it is assumed that the r-process
occurs in the layers heated by neutrino emission and evaporating from
the hot protoneutron star after core collapse in a type II supernova
\cite{Thompson.Burrows.Meyer:2001}.  Adopting the parameters of a
supernova simulation by Wilson \cite{Wilson:1985} Woosley and
collaborators obtained quite satisfying agreement between an r-process
simulation and observation \cite{Woosley.Wilson.ea:1994}. In the
classical picture the r-process nuclides are made by successive
capture of neutrons, starting from a seed nucleus with mass number
$A_{\text{seed}}$. Thus, to make the third r-process peak around $A \sim
200$ requires a large neutron-to-seed ratio of $n/s \sim 200 -
A_{\text{seed}}$. In the Wilson supernova models
\cite{Wilson:1985,Wilson:2001} this is achieved due to a high entropy
found in the neutrino wind at late times (a few seconds after the
bounce). However, other models with a different equation of state and
treatment of diffusion do not obtain such high entropies; in these
models the r-process fails to make the $A=200$ peak.  To explain the
strong sensitivity of the r-process nucleosynthesis on the entropy of
the environment \citet{Qian:1997} noted that the slowest reaction in
the nuclear network, which transforms protons, neutrons and $\alpha$
particles into r-process seed nuclei, is the 3-body $\alpha+\alpha+n
\rightarrow {}^9$Be reaction. Due to its low binding energy ($E_b =
1.57$ MeV), $^9$Be can be easily destroyed in a hot thermal
environment and thus the matter flow to nuclei heavier than $^9$Be
depends strongly on the entropy of the surrounding. The larger the
entropy, the smaller the abundance of surviving $^9$Be nuclei, which
are then transformed into seed nuclei, and the larger the
neutron-to-seed ratio.  \citet{Meyer:1995} pointed out that in a very
strong neutrino flux the slow 3-body $\alpha+\alpha+n \rightarrow
{^9}$Be reaction can be potentially bypassed by a sequence of two-body
reactions started by the neutrino-induced spallation of an $\alpha$
particle; e.g.  $^4$He($\nu,\nu^\prime
p)^3$H$({^4}$He,$\gamma)^7$Li($^4$He,$\gamma$)$^{11}$B and
$^4$He($\nu,\nu^\prime
n)^3$He$({^4}$He,$\gamma)^7$Be($^4$He,$\gamma$)$^{11}$C.  This would
speed up the mass flow to the seed nuclei and thus reduce the
neutron-to-seed ratio.

Systematic studies by \citet{Hoffman.Woosley.Qian:1997}
and \citet{Freiburghaus.Rembges.ea:1999} have shown that a successful
r-process requires either large entropies at the $Y_e$ values
currently obtained in supernova models, or smaller values for $Y_e$.

In the neutrino-driven wind model the extreme flux of $\nu_e$ and
$\bar{\nu}_e$ neutrinos from the protoneutron star interacts with the
free protons and neutrons in the shocked matter by charge-current
reactions, setting the proton-to-neutron ratio $n/p$ or equivalently
the $Y_e$ value of the r-process matter. As shown by Fuller and Qian
one has the simple relation \cite{Qian.Fuller:1995,Qian:1997}

\begin{equation}
\frac{n}{p} \approx \frac
{L_{\bar{\nu}_e} \langle E_{\bar{\nu}_e} \rangle}
{L_{\nu_e} \langle E_{\nu_e}
\rangle}
\end{equation}

As the neutrino energy luminosities are about equal for all species
($L_\nu \sim 10^{52}$ erg~s$^{-1}$) the n/p-ratio is set by the ratio of
average energies for the antineutrino and neutrino. As discussed
above, their different opacities in the protoneutron star ensure that
$\langle E_{\bar{\nu}_e} \rangle > \langle E_{\nu_e} \rangle$ and the
matter is neutron-rich, as is required for a successful r-process. When
the matter reaches cooler temperatures, nucleosynthesis starts and the
free protons are, in the first step, assemblied into
$\alpha$-particles, with some extra neutrons remaining. If these neutrons
are still exposed to a large neutrino flux, it will change some
of the neutrons into protons, which will then, together with
additional neutrons, also be bound very quickly into $\alpha$-particles.
Thus, this so-called $\alpha$ effect
\cite{Meyer.Mclaughlin.Fuller:1998} would severely reduce the final
neutron-to-seed ratio and is therefore very counter-productive to a
successful r-process. As mentioned above, the neutrino-nucleon cross
sections are only considered to lowest order in supernova simulations.
The correction, introduced by the weak magnetism, acts to reduce the
neutron-to-proton ratio in the neutrino-driven wind
\cite{Horowitz.Li:1999}.

There are possible ways out of this dilemma: One solution is to remove
the matter very quickly from the neutron star in order to reduce the
neutrino fluxes for the $\alpha$ effect.  Such a short dynamical
timescale of the material in the wind is found in neutrino-driven wind
models studied by Kajino and collaborators. These authors also observe
that relativistic effects as well as nuclear reaction paths through
neutron-rich light elements might  be helpful for a successful
r-process \cite{Otsuki.Tagoshi.ea:2000,Wanajo.Kajino.ea:2001,%
  Terasawa.Sumiyoshi.ea:2001a,Terasawa.Sumiyoshi.ea:2001b}. Another
intriguing cure is discussed by \citet{Mclaughlin.Fetter.ea:1999} and
\citet{Caldwell.Fuller.Qian:2000} invoking matter-enhanced
active-sterile neutrino oscillations to remove the $\nu_e$ from the
r-process site.

A simple estimate for the duration of the r-process can be had by
adding up the half-lives of the waiting point nuclei, which results in
about 1--2 seconds. However, in the neutrino-driven wind model it
appears that the ejected matter passes through the region with the
conditions suited for an r-process in shorter times ($\sim 0.5$ s),
implying that there might not be enough time for sufficient
matter-flow from the seed to nuclides in the $A\sim200$ mass region.
Such a `time problem' is avoided if, as indicated above, the
half-lives of the waiting point nuclei are shorter than conventionally
assumed, or if, in a dynamically changing environment, the matter,
which freezes out making the $A\sim200$ r-process peak, breaks through
the $N=50$ and 82 waiting points closer to the neutron dripline, i.e.\ 
through nuclei with shorter half-lives, than the matter which freezes
out at these lower magic neutron numbers.

\begin{figure}[htbp]
    \includegraphics[width=0.9\linewidth]{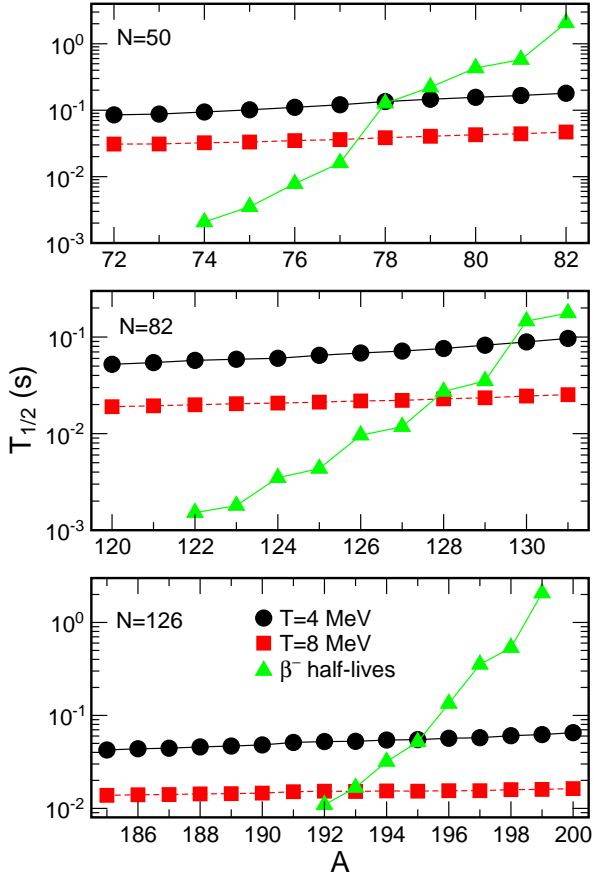}
    \caption{Half-Lives of r-process waiting point nuclei with $N=50$
      (top panel), 82 (middle panel) and 126 (bottom panel) against
      charged-current $(\nu_e,e^-)$ reactions. For the neutrinos a
      Fermi-Dirac distribution with $T=4$ MeV and zero chemical
      potential (circles) and a luminosity of ($L_\nu \sim
      10^{52}$~erg~s$^{-1}$) has been adopted. It is assumed that the
      reactions occur at a radius of 100 km, measured from the center
      of the neutron star. The half-lives can be significantly shorter
      if $\nu_e \rightleftarrows \nu_{\mu,\tau}$ oscillations occur.
      The squares show the half-lives for neutrinos with a Fermi-Dirac
      distribution with $T=8$ MeV and zero chemical potential, which
      corresponds to complete $\nu_e \rightleftarrows \nu_{\mu,\tau}$
      oscillations.\label{fig:cctau}}
\end{figure}

In an environment with large neutrino fluxes the matter-flow to
heavier nuclei can also be sped up by charged-current $(\nu_e,e^-)$
reactions \cite{Nadyozhin.Panov:1993,Qian.Haxton.ea:1997} which can
compete with $\beta$-decays. This is particularly important at the
waiting point nuclei associated with $N=50, 82$, and 126.
Figure~\ref{fig:cctau} shows the $(\nu_e,e^-)$ half-lives
\cite{Hektor.Kolbe.ea:2000,Langanke.Kolbe:2001} for these waiting
point nuclei, considering reasonable supernova neutrino
parametrizations and assuming that the ejected matter has reached a
radius of 100 km. Due to the dependence on $L_\nu$ the $(\nu_e,e^-)$
half-lives scale with $r^2$.  Indeed, a comparison with the $\beta$
half-lives (figure~\ref{fig:cctau}) shows that $(\nu_e,e^-)$ reactions
can be faster than the longest $\beta$-decays of the $N=50,82$, and
126 waiting point nuclei, thus speeding up the breakthrough of the
matter at the waiting points, if the r-process occurs rather close to
the surface of the neutron star.  This is even further enforced if
$\nu_e \rightleftarrows \nu_{\mu,\tau}$ oscillations occur due to the
higher energy spectrum of the supernova $\nu_{\mu,\tau}$ neutrinos.
However, the presence of charged-current reactions on nuclei in the
neutrino-driven wind model implies also neutrino reactions on nucleons
strengthening the $\alpha$ effect \cite{Meyer.Mclaughlin.Fuller:1998}.

\begin{figure}[htbp]
  \includegraphics[width=0.9\linewidth]{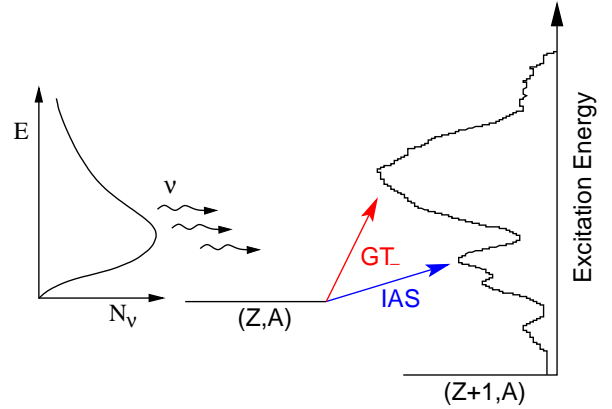}
  \caption{Schematic view of the $(\nu_e,e^-)$ reaction on r-process
    nuclei. Due to the high neutrino energies the cross sections are
    dominated by transitions to the Fermi (IAS) and Gamow-Teller
    resonances. \label{fig:nureso}} 
\end{figure}


Under a strong neutrino flux the weak flow is determined by an
effective weak rate given by the addition of the charged-current
$(\nu_e,e^-)$ and the nuclear beta-decay rates
\cite{Mclaughlin.Fuller:1997}. It has been argued that the solar
system r-process abundances provide evidence for the weak steady-flow
approximation, which implies that the observed abundances should be
proportional to the half-lives of their progenitor nuclei on the
r-process path \cite{Kratz.Thielemann.ea:1988,Kratz.Bitouzet.ea:1993}.
If this is so, the r-process freeze-out must occur at conditions
(i.e.\ radii) at which $\beta$-decay dominates over $(\nu_e,e^-)$
reactions, at least for late times.  The reason is that neutrino
capture on magic nuclei with $N=50,82,126$ is not reduced if compared
to the neighboring nuclides, as the capture occurs from a reservoir of
neutrinos with sufficiently high energies to allow for transitions to
the IAS and GT resonant states (see figure~\ref{fig:nureso}).
Consequently $(\nu_e,e^-)$ cross sections scale approximately like the
neutron excess $(N-Z)$, reflecting the Fermi and Ikeda sumrules (see
figure~\ref{fig:cctau}).  However, the observed solar abundances
around the $A=130$ ($N=82$) and 195 ($N=126$) peaks do not show such a
smooth dependence with $A$ as will be the case if the effective weak
rate is dominated by neutrino reactions.  If we require that the
$\beta$ decay half-lives dominate over the $(\nu_e,e^-)$ reactions we
can put constraints on the neutrino fluence in the neutrino-driven
wind scenario, which is particularly strict if neutrino oscillations
occur.  If, as an illustrative example, we apply the constraint to the
heavy waiting point nuclei with $N=126$ (e.g.\ the nuclei with $A \sim
199$) and adopt the neutrino and beta half-lives from
figure~\ref{fig:cctau}, $\beta$-decay is only faster if the
neutrino-nucleus reactions occur at distances larger than $\sim 500$
km.

\begin{figure}[htbp]
  \includegraphics[width=0.9\linewidth]{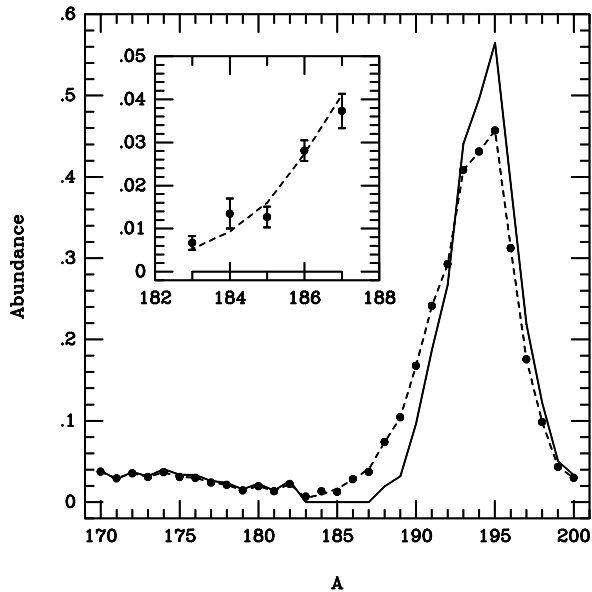}
  \caption{Effect of post-processing by neutrino-induced reactions on the
    r-process abundance. The unprocessed distribution (solid line) is
    compared with the distribution after post-processing (dashed
    line). A constant fluence of ${\cal F}=0.015$ has been assumed,
    which provides a best fit to the observed abundances for
    $A=183-87$ (see inset). The observed abundances are plotted as
    filled circles with error bars
    \citep[from][]{Qian.Haxton.ea:1997}). \label{fig:postpr}}
\end{figure}

As the $Q_\beta$ values in the very neutron-rich r-process nuclei are
large, the IAS and GT resonant states reside at rather high excitation
energies ($\sim 20$--30 MeV), in the daughter nuclei for $(\nu_e,e^-)$
reactions. This fact, combined with the small neutron separation
energies in these nuclei, ensures that $(\nu_e,e^-$) reactions, and
also neutral-current $(\nu,\nu^\prime$) reactions, spall neutrons out
of the target nuclei
\cite{Haxton.Langanke.ea:1997,Qian.Haxton.ea:1997} ($\sim 5$--7
neutrons for nuclei in the $A=195$ mass region
\cite{Haxton.Langanke.ea:1997,Hektor.Kolbe.ea:2000}). During the
r-process, i.e.\ as long as the neutron source is strong enough to
establish $(n,\gamma) \rightleftarrows (\gamma,n)$ equilibrium, the
neutrons will immediately be recaptured leading to no effect on the
abundance distribution.  However, once the neutron source has ceased,
e.g.\ after freeze-out, and if the r-process matter in the
neutrino-driven wind model is still subject to strong neutrino fluxes,
neutrons liberated during this phase by neutrino-induced reactions,
will not be recaptured and this post-processing
\cite{Qian.Haxton.ea:1997,Haxton.Langanke.ea:1997} leads to changes in
the r-process abundance distribution.  It is argued
\cite{Haxton.Langanke.ea:1997} that due to the smooth dependence of
the neutrino cross sections on the mass number, the post-processing,
in general, shifts abundances from the peaks to the wings at lower
$A$-values (figure~\ref{fig:postpr}). This shift depends on the
neutrino exposures and allows constraints to be put on the total
neutrino fluence in the neutrino-driven wind model
\cite{Qian.Haxton.ea:1997,Haxton.Langanke.ea:1997}.  The limits
obtained this way are compatible with the values predicted by
supernova models. Whether $\beta$-delayed neutron emission, which has
been neglected in \cite{Haxton.Langanke.ea:1997} might affect the
post-processing is an open question \cite{Kratz:2001,Kratz.pvt}.

Attempts to include neutrino-induced reactions in the r-process
network within the neutrino-driven wind model have been reported
in \cite{Fuller.Meyer:1995,Mclaughlin.Fuller:1995,%
Terasawa.Sumiyoshi.ea:2001a,Terasawa.Sumiyoshi.ea:2001b}.  In
particular \citet{Meyer.Mclaughlin.Fuller:1998} studied the
competition of the $\alpha$-effect with the possible speed-up of the
matter-flow by charged-current reactions on nuclei. These authors
estimated the respective charged-current cross sections for allowed
transitions on the basis of the independent particle model.  Improving
on this treatment, Borzov and Goriely calculated $(\nu_e,e^-)$ cross
sections for supernova neutrinos (Fermi-Dirac distribution with $T=4$
MeV) within the ETFSI method, consistently with their most recent
estimates for the $\beta$ half-lives \cite{Borzov.Goriely:2000}.
RPA-based neutrino-nucleus cross sections for selected nuclei have
been reported in \cite{Surman.Engel:1998,Hektor.Kolbe.ea:2000}.  Very
recently a tabulation with charged- and neutral-current total and
partial neutron spallation cross sections have become available for
the neutron-rich r-process nuclei
\cite{Langanke.Kolbe:2001,Langanke.Kolbe:2002}. This tabulation is
based on the RPA and considers allowed and forbidden transitions.
Furthermore, the cross sections are tabulated for various supernova
neutrino distributions, thus also allowing study of the influence of
complete neutrino oscillations on the r-process.

\begin{figure}[htbp]
  \includegraphics[width=0.9\linewidth]{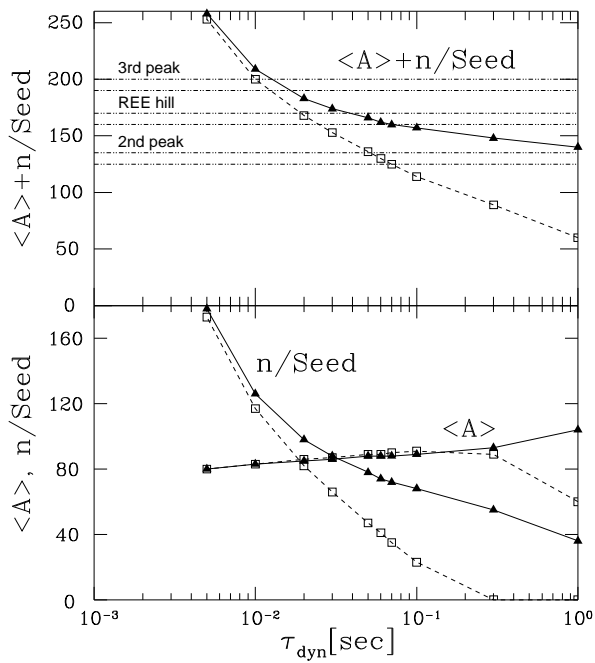}  
  \caption{The lower panel shows the average mass number of heavy seed
    nuclei $\langle A \rangle$ and the neutron-to-seed ratio (n/Seed)
    in a neutrino-driven wind simulation with an exponential time
    dependence of the matter flow, determined by the parameter
    $\tau_{\text{dyn}}$.  The open squares show the results for a
    simulation including neutrino-nucleus reactions, while the full
    triangles refer to a study in which the neutrino-reactions on
    nuclei have been switched off. The upper panel shows the sum of
    the mass number of the seed nucleus plus the neutron-to-seed
    ratio. This quantity shows up to which mass number the r-process
    can produce nuclides. The horizontal lines indicate the second
    ($A\sim 130$) and third ($A\sim 200$) r-process peaks as well as
    the position of the smaller r-process peak related to the deformed
    nuclei in the rare-earth region (REE peak), (courtesy of M.
    Terasawa). \label{fig:Terasawa}}
\end{figure}

\citet{Terasawa.Sumiyoshi.ea:2001a,Terasawa.Sumiyoshi.ea:2001b} have
performed studies similar to the pioneering work of
\citet{Meyer.Mclaughlin.Fuller:1998}, however, using their
neutrino-driven wind model and the complete set of RPA
neutrino-nucleus reaction rates
\cite{Langanke.Kolbe:2001,Langanke.Kolbe:2002}.  A typical result is
shown in figure~\ref{fig:Terasawa}, where, however, the simplifying, but
reasonable assumption of an exponential time-dependence of the
matter-flow, governed by the parameter $\tau_{\text{dyn}}$, away from
the neutron star has been assumed.  The quantity $\langle A \rangle$
defines the average mass of the heavy seed nuclei present at the beginning
of the r-process, defined at $T = 2.5 \times 10^9$ K. The
neutron-to-seed ratio $n/s$ is very sensitive to the dynamical
evolution time. This comes about as the shorter $\tau_{\text{dyn}}$,
the less time is available to assemble the seed nuclei from
$\alpha$-particles and neutrons. Consequently the abundance of seed
nuclei decreases for shorter $\tau_{\text{dyn}}$, increasing the $n/s$
ratio.  If the neutrino flux is artificially switched off, matter-flow
to the 3rd r-process peak at $A \sim 200$ (2nd r-process peak at $A\sim
130$) is achieved if $\tau_{\text{dyn}} \le 0.01$ s
($\tau_{\text{dyn}} \le 1$ s).  The 
consistent inclusion of neutrino reactions is counter-productive to a
successful r-process. This effect becomes more dramatic if the
matter-flow is slow as then the $\alpha$-effect strongly suppresses
the availability of free neutrons at the beginning of the r-process
\citep[see also][]{Meyer.Mclaughlin.Fuller:1998}. No r-process, i.e.\ no
production of nuclides in the second r-process peak at $A\sim130$, is
observed if $\tau_{\text{dyn}} \gtrsim 0.05$~s.

\begin{figure}[htbp]
  \includegraphics[width=0.9\linewidth]{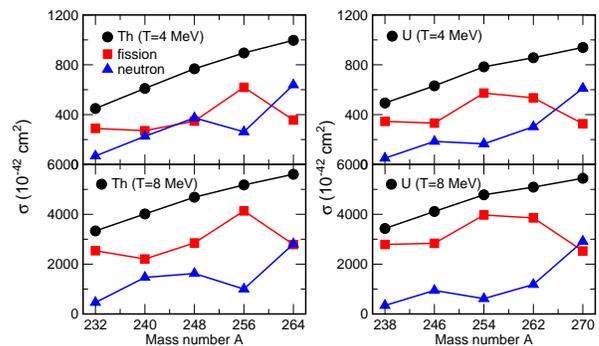}
  \caption{Total $(\nu_e,e^-)$ (circles) and partial $(\nu_e,e^- n)$
    (triangles) and neutrino-induced fission cross sections (squares)
    for selected thorium and uranium isotopes.  The calculations have
    been performed for Fermi-Dirac $\nu_e$ distributions with
    temperature $T=4$ MeV and 8 MeV. The first reflects a typical
    supernova $\nu_e$ spectrum, while the latter assumes complete
    $\nu_{\mu,\tau} \rightarrow \nu_e$ neutrino oscillations.}
  \label{fig:nufission}
\end{figure}

Recently \citet{Qian:2002} has suggested that, within the
neutrino-driven wind r-process scenario, charged-current neutrino
reactions can induce fission reactions on r-process progenitor nuclei
heavier than lead and that the fission products account for the
observed r-process abundance in metal-poor stars
\cite{Sneden.Cowan.ea:2000}. These observed abundances show a peak
around mass number $A \sim 195$, which follows the solar r-process
distribution, and enhanced structures at around $A \sim 90$ and $\sim
132$ which, due to the suggestion of \citet{Qian:2002}, are fission
products, which do not follow a solar r-process pattern.  First
calculations of neutrino-induced fission cross sections have been
performed by \cite{Kolbe.Fuller.Langanke:2002} using a combination of
RPA model, to calculate the $(\nu_e,e^-)$ excitation function, and
statistical model to determine the final branching probabilities.  The
neutrino-induced fission cross sections (see
figure~\ref{fig:nufission}) are quite large as the progenitor nuclei
are neutronrich, which increases the Fermi and Gamow-Teller
contributions to the total cross sections and places their strengths
at energies above the fission barrier in the daughter nucleus. The
calculations shown in figure~\ref{fig:nufission} use the fission
barriers derived by \citet{Howard.Moeller:1980}. Modern evaluations
predict larger fission barriers
\citep[e.g.][]{Mamdouh.Pearson.ea:2001} which would reduce the fission
cross section. The cross sections can be significantly enlarged if
$\nu_{\mu,\tau} \rightarrow \nu_e$ neutrino oscillations occur during
the neutrino-driven wind r-process scenario.

\section{The Neutrino Process}

When the flux of neutrinos generated by the cooling of the neutron
star passes through the overlying shells of heavy elements interesting
nuclear transmutations are induced, despite the small neutrino-nucleus
cross sections. Of particular interest here are neutral-current
reactions as they can be induced by $\nu_\mu,\nu_\tau$ neutrinos and
their antiparticles with the higher energy spectra ($\langle E_\nu
\rangle \sim 25$ MeV). These neutrinos are energetic enough to excite
the GT resonant state, and more importantly, also the giant dipole
resonant states. These states usually reside above particle thresholds
and hence decay mainly by proton or neutron emission, generating new
nuclides.  The neutrino reaction rates are too small to affect the
abundances of the parent nuclei, but they can noticeably contribute to
the production of the (sometimes much less abundant) daughter
nuclei. As a rule-of-thumb, the neutrino process, i.e.\ the
synthesis of nuclides by neutrino-induced spallation in a supernova,
can become a significant production process for the daughter nuclide
if one wants to explain abundance ratios of parent-to-daughter which
exceed about $10^3$ \cite{Woosley.pvt}.

The most interesting neutrino nucleosynthesis occurs in the outer
burning shells of the massive star which have not been affected by the
fatal core collapse in the center when the neutrinos pass through.
However, a little later the shock reaches these shells and the matter
will be subjected to rather high temperatures initiating fast nuclear
reactions involving also the nuclides just produced by the
neutrino-induced reactions. Hence studies of the neutrino process
depend on various neutrino-nucleus reaction rates and on the neutrino
spectra and fluxes, especially of the $\nu_\mu,\nu_\tau$ neutrinos,
and they require a rather moderate nuclear network to simulate the
effects of the re-processing by the shock.  The first investigation of
the neutrino process has been reported in
\cite{Woosley.Hartmann.ea:1990}. A more recent study
\cite{Woosley.Weaver:1995} confirmed the main result that a few
specific nuclei ($^7$Li, $^{11}$B, $^{19}$F) are being made by the
neutrino process in significant fractions. For example, $^{11}$B and
$^{19}$F are being made by $(\nu,\nu'p)$ and $(\nu,\nu'n)$ (followed
by a $\beta$-decay) reactions on the abundant $^{12}$C and $^{20}$Ne,
respectively. As noted by \citet{Woosley.Hartmann.ea:1990} neutrino
nucleosynthesis can also contribute to the production of $^{180}$Ta
(see above) and $^{138}$La. First calculations of the relevant total
and partial neutrino-induced cross sections have been reported in
\cite{Heger.Kolbe.ea:2002}, \citep[see also][]{Belic.Arlandini.ea:2002}.
The nucleus $^{138}$La is special as it is mainly made by the
charged-current reaction $^{138}$Ba$(\nu_e,e^-){}^{138}$La, while the
lighter nuclei ($^7$Li, $^{11}$B, $^{19}$F) and dominantly also
$^{180}$Ta, are being produced by neutral-current reactions induced by
$\nu_\mu,\nu_\tau$ neutrinos.  The neutrino process is therefore
sensitive to $\nu_e$ and ($\nu_\mu,\nu_\tau$) neutrinos, which are
expected to have different spectra in Type II supernovae (see section
V.C), and hence it can test this prediction. Moreover, the $^{138}$La
nucleosynthesis is sensitive to neutrino oscillations, as this nuclide
would be significantly overproduced if supernova $\nu_\mu,\nu_\tau$
neutrinos have a noticeably larger average energy and if they
oscillate into $\nu_e$ neutrinos before reaching the $^{138}$La
production site (helium shell) in massive stars.

Neutrino-induced nucleon spallation on $^{12}$C can also knock-out a
deuteron or a proton-neutron pair, in this way producing $^{10}$B. The
expected $^{10}$B/$^{11}$B abundance ratio in neutrino-nucleosynthesis
is $\sim 0.05$, which is significantly smaller than the observed
abundance ratio, 0.25 \cite{Haxton:2000}.  
Thus there must be a second process which
contributes to the production of $^{10,11}$B. These are reactions of
energetic protons on $^{12}$C in cosmic rays, which yield a ratio of
$^{10}$B/$^{11}$B of about 0.5, larger than the observed value. A
solution might be that the two nuclides are being produced by both
mechanisms, neutrino-nucleosynthesis and cosmic ray spallation. It is
interesting to note that the first process, being associated with
supernovae, is a primary process, while the latter is a secondary
process, as it requires the existence of protons and $^{12}$C in the
interstellar medium. As a consequence the $^{10}$B/$^{11}$B abundance
ratio should have changed during the history of the galaxy. This can
be tested once observers are able to distinguish between the
abundances of the two different boron nuclides in stellar
spectra \cite{Haxton:2000}. 

\section{Binary Systems}

Weak processes can also play interesting roles in the evolution and
nucleosynthesis processes in close binary systems where one component
is a compact object (white dwarf or neutron star) and the other a
massive star. If the latter expands during helium core burning, mass
flow from the hydrogen envelopment of the star onto the surface of the
compact object sets in. If the respective mass accretion rate is
rather low ($10^{-8}$--$10^{-10}$ $M_\odot$~y$^{-1}$), the hydrogen,
accreted on the surface of the compact object, burns explosively under
degenerate conditions leading to a nova (if the compact object is a
white dwarf) or an x-ray burst (neutron star).  This means that the
energy released by the nuclear reactions is used to heat the matter
rather than for expansion.  The rise in temperature increases the
nuclear reaction rates, triggering a thermonuclear runaway until the
degeneracy is finally lifted and an outer layer of matter is ejected.
In a type Ia supernova the faster mass accretion rate on to the
surface of a white dwarf (likely a carbon-oxygen white dwarf with
sub-Chandrasekhar mass $\sim 0.7\ M_\odot$) leads to steady hydrogen
and subsequently helium burning. If the growing mass of the white
dwarf exceeds the Chandrasekhar mass, contraction sets in and the
carbon in the center ignites by fusion reactions with screening
enhancement. As the environment is highly degenerate, a thermonuclear
runaway is triggered which eventually will explode the entire star.

\subsection{Novae}

The main energy source of a nova is the CNO cycle, with additional
burning from proton reactions on nuclei between neon and sulfur, if
the white dwarf also contained some $^{20}$Ne \cite{Truran:1982}. A
nova expels matter which is enriched in $\beta$-unstable $^{14,15}$O
for carbon-oxygen white dwarfs (leading to the production of stable
$^{14,15}$N nuclides) or can contain nuclides up to the sulfur mass
region for neon-oxygen novae \cite{Starrfield.Truran.ea:1998,%
  Starrfield.Sparks.ea:2000,Jose.Hernanz:1998}.  The most important
role of weak-interaction processes in novae is their limitation of the
energy generation during the thermonuclear runaway.  An interesting
branching occurs at $T \sim 8 \times 10^7$ K. For lower temperatures
the $\beta$-decay of $^{13}$N with a half-life of $\sim 10$ m
dominates over the $^{13}$N$(p,\gamma)$ reaction and sets the
timescale for the nuclear burning.  As charged-particle fusion
reactions are sensitively dependent on temperature, their reaction
rates strongly increase with rising temperature and for $T \ge 8
\times 10^7$ K and densities of order $10^4$~g~cm$^{-3}$ the proton
capture on $^{13}$N is faster than the $\beta$-decay. The CNO cycle
turns into the \emph{hot} CNO cycle and now the positron decay of
$^{15}$O with a half-life of 122 s is the slowest reaction occuring in
the CNO nova network.  It turns out that, once the degeneracy is
lifted, the dynamical expansion timescale is faster than the one for
nuclear energy generation, set by the $^{15}$O half-life. As a
consequence the runaway is quenched \cite{Truran:1982}.  We mention
that the determination of the dominant resonant contribution to the
$^{13}$N$(p,\gamma){}^{14}$O rate has been the first successful
application of the Coulomb dissociation technique in nuclear astrophysics
\cite{Motobayashi.Takei.ea:1991}.

Weak-interaction rates relevant for nova studies can be derived from
either experimental data or shell-model calculations.

\subsection{X-ray Bursts}

An x-ray burst is explained as a thermonuclear runaway in the
hydrogen-rich envelope of an accreting neutron star
\cite{Lewin.Paradijs.Taam:1993,Taam.Woosley.ea:1993}.  Due to the
higher gravitational potential of a neutron star, the accreted matter
on the surface reaches larger densities than in a nova (up to a few
$10^6$~g~cm$^{-3}$) and the temperature during the runaway can rise up
to $2 \times 10^9$ K \cite{Schatz.Aprahamian.ea:1998}.  Under these
conditions hydrogen burning is explosively fast. The trigger reactions
of the runaway are the triple-alpha reactions and the break-out from
the hot CNO cycle, mainly by $\alpha$-capture on $^{15}$O and
$^{18}$Ne. We note that these two reaction rates are yet
insufficiently known due to uncertain resonant contribution at low
energies
\cite{Mao.Fortune.Lacaze:1996,Goerres.Wiescher.Thielemann:1995}.  The
thermonuclear runaway is driven first by the $\alpha p$ process, a
sequence of $(\alpha,p)$ and $(p,\gamma)$ reactions which shifts the
ashes of the hot CNO cycle to the Ar and Ca mass region, and then by
the rp-process (short for rapid proton capture process). The
rp-process represents proton-capture reactions along the proton
dripline and subsequent $\beta$-decays of dripline nuclei processing
the material from the Ar-Ca region up to $^{56}$Ni
\cite{Schatz.Aprahamian.ea:1998}.  The $\beta$ half-lives on the
rp-process path up to $^{56}$Ni are known sufficiently well.

The runaway freezes out in thermal equilibrium at peak temperatures of
$(2\text{--}3)\times 10^9$ K with an abundance distribution rich in
$^{56}$Ni, forming Ni oceans on the surface of the neutron star.
Further matter flow at these temperatures is suppressed due to the low
proton separation energy of $^{57}$Cu and the long half-life of
$^{56}$Ni.  Re-ignition of the rp-process then takes place during the
cooling phase, starting with proton capture on $^{56}$Ni and
potentially shifting matter up to the $^{100}$Sn region where the
rp-process ends in a cycle in the Sn-Te-I range
\cite{Schatz.Aprahamian.ea:2001}. A matter flow to even heavier nuclei
is possible, if the rp-process operates in a repetitive mannor; i.e.\ 
a new rp-process is ignited after the ashes of the previous process
have decayed to stability and before these nuclei have sunk too deep
into the crust of the neutron star (see below). Such repetitive
rp-process models, shortly called $(\rm rp)^2$-process, have been
studied by \cite{Boyd.Hencheck.Meyer:2002}.
  
The reaction path beyond $^{56}$Ni runs through the even-even $N=Z$
nuclei which due to their known long half-lives ($^{64}$Ge has a
half-life of 63.7(25)~s) represent a strong impedance to the matter-flow.
This can also not be overcome always by proton captures as for some of
the $\alpha$-like nuclei the resulting odd-$A$ nucleus is
proton-unbound and exists only as a resonance. Such situations occur,
for example, for the $^{68}$Se$(p,\gamma){}^{69}$Br and
$^{72}$Kr$(p,\gamma){}^{73}$Rb reactions.  It has been
suggested \cite{Goerres.Wiescher.Thielemann:1995} that the gap in the
reaction path can be bridged by two-proton captures, with the
resonance serving as intermediate state (like in the triple-$\alpha$
reaction). The reaction rate for such a two-step process depends
crucially on the resonance energy, with some appropriate screening
corrections.

Most half-lives along the rp-process path up to the $^{80}$Zr region
are known experimentally. This region of the nuclear mass chart is
known for strong ground state deformations, caused by coupling of the
$pf$-shell orbitals to the $g_{9/2}$ and $d_{5/2}$ levels. The strong
deformation makes theoretical half-life predictions quite inaccurate,
mainly due to uncertainties in the $Q$-values stemming from
insufficiently wellknown mass differences.  We note that the effective
half-life of a nucleus along the rp-process path could be affected by
the feeding of isomeric states in the proton captures or by thermal
population of excited states in general. Again, deformation plays a
major role as then even in $\alpha$-like nuclei excited states are at
rather low energies (e.g.\ the first $2^+$ state in $^{80}$Zr is at 290
keV).  Apparently a measurement of the half-lives is indispensable for
rp-process studies beyond $A=80$.  An important step has recently been
taken by measuring the half-life of $^{80}$Zr at the Holifield Facility
in Oak Ridge \cite{Ressler.Piechaczek.ea:2000}. The experimental value
of $4.1^{+0.8}_{-0.6}$~s reduces the previous (theoretical)
uncertainty considerably and, in fact, it is shorter than the value
adopted previously in x-ray bursts simulations. Fast proton captures
on the daughter products $^{80}$Y and $^{80}$Sr allow matter-flow to
heavier nuclei, with the $\alpha$-nucleus $^{84}$Mo ($N=Z=42$) being
the next bottleneck. Experiments to measure this important half-life
are in progress.

The nucleosynthesis during the cooling phase in an x-ray burst alters
considerably the abundance distribution in the atmosphere, ocean and
crust of the neutron star. For example, the rp-process may be a
possible contributor to the presently unexplained relatively high
observed abundance of light p-nuclei like $^{92}$Mo and $^{96}$Ru
\cite{Schatz.Aprahamian.ea:1998}. This assumes that the matter
produced in the x-ray burst gets expelled out of the large
gravitational potential of the neutron star, which is still
questionable. Due to continuing accretion the rp-process ashes are
pressed into the ocean and crust of the neutron star, replacing there
the neutron star's original material. When the ashes sink into the
crust, they reach regions of higher densities, and relatedly, larger
electron chemical potentials. Thus, consecutive electron captures will
become energetically favorable and make the ashes more neutron-rich. At
densities beyond neutron drip ($\rho \sim (4\text{--}6) \times
10^{11}$~g~cm$^{-3}$) neutron emissions become possible and at even
higher densities pycnonuclear reactions can set in
\cite{Bisnovatyi-Kogan.Chechetkin:1979,Sato:1979,Haensel.Zdunik:1990}.
Importantly, these processes (electron capture, pycnonuclear
reactions) generate energies which can be locally stored in the
neutron star's ocean and crust and will affect their thermal
properties \cite{Brown.Bildsten:1998}. Previous studies of these
processes in accreting neutron stars have assumed that iron is the
endproduct of nuclear burning and the sole nucleus reaching the crust
of the neutron star \citep[e.g.][]{Haensel.Zdunik:1990}. But clearly
the rp-process produces a wide mixture of heavy elements
\cite{Schatz.Bildsten.ea:1999}, where the abundance distribution
depends on the accretion rate.

The ashes consist mainly of even-even $N=Z$ nuclei, for which electron
capture at neutron star conditions occur always in steps of two. At
first, the capture on the even-even nucleus sets in once sufficiently
high-energy electrons are available to effectively overcome the
$Q_{\text{EC}}$-value to the odd-odd daughter nucleus. Due to nuclear
pairing, which favors even-even nuclei, the $Q_{\text{EC}}$-value for
the produced daughter nucleus is noticeably lower so that electron
capture on the daughter readily follows at the same conditions. The
energy gain of the double-electron capture is of order the difference
of the two $Q_{\text{EC}}$-values; this gain is split between the
emitted neutrino and a local heating. Considering a blob of accreted
matter initially consisting solely of $^{56}$Ni and assuming
temperature $T=0$, the evolution of this blob was followed on the
neutron star surface until neutron-drip densities and beyond
\cite{Haensel.Zdunik:1990}. The $rp$-process simulations, however,
indicate a finite temperature of the ashes of a few $10^8$ K, allowing
electron capture already from the high-energy tail of the electron
distribution and significantly reducing the required densities.

\subsection{Type Ia Supernovae}

Type Ia supernovae at high redshifts serve currently as the standard
candles for the largest distances in the universe. Importantly recent
surveys of such distant supernovae provide evidence for an accelerating
expansion of the universe over the last several $10^9$ years 
\cite{Riess.Filippenko.ea:1998,Perlmutter.Aldering.ea:1999}.

Type Ia supernovae have been identified as thermonuclear explosions of
accreting white dwarfs with high accretion rates in a close binary
system.  While the general explosion mechanism is probably understood,
several issues are still open like the masses of the stars in the
binary or the carbon/oxygen ratio and distribution in the white dwarf.
The probably most important problem yet is the modelling of the matter
transport during the explosion and the velocity of the burning front,
both requiring multidimensional
simulations \citep[e.g.][]{Woosley.pvt,%
Reinecke.Hillebrandt.Niemeyer:1999,Hillebrandt.Niemeyer:2000}. 

\begin{figure}[htbp]
  \includegraphics[angle=90,width=0.9\linewidth]{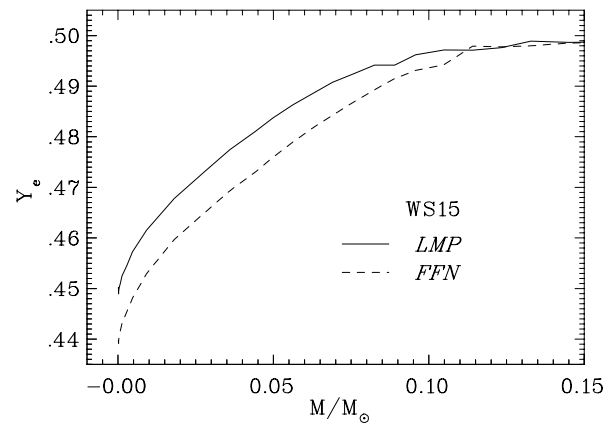}
  \caption{$Y_e$ profile as a function of radial mass for the 
    standard type Ia supernova model WS15
    \cite{Nomoto.Thielemann.Yokoi:1984} using the FFN and
    the shell-model weak-interaction rates (LMP), (courtesy of
    F. Brachwitz).\label{fig:yeprof}}
\end{figure}

\begin{figure*}[htbp]
  \hfill%
    \includegraphics[angle=90,width=0.45\textwidth]{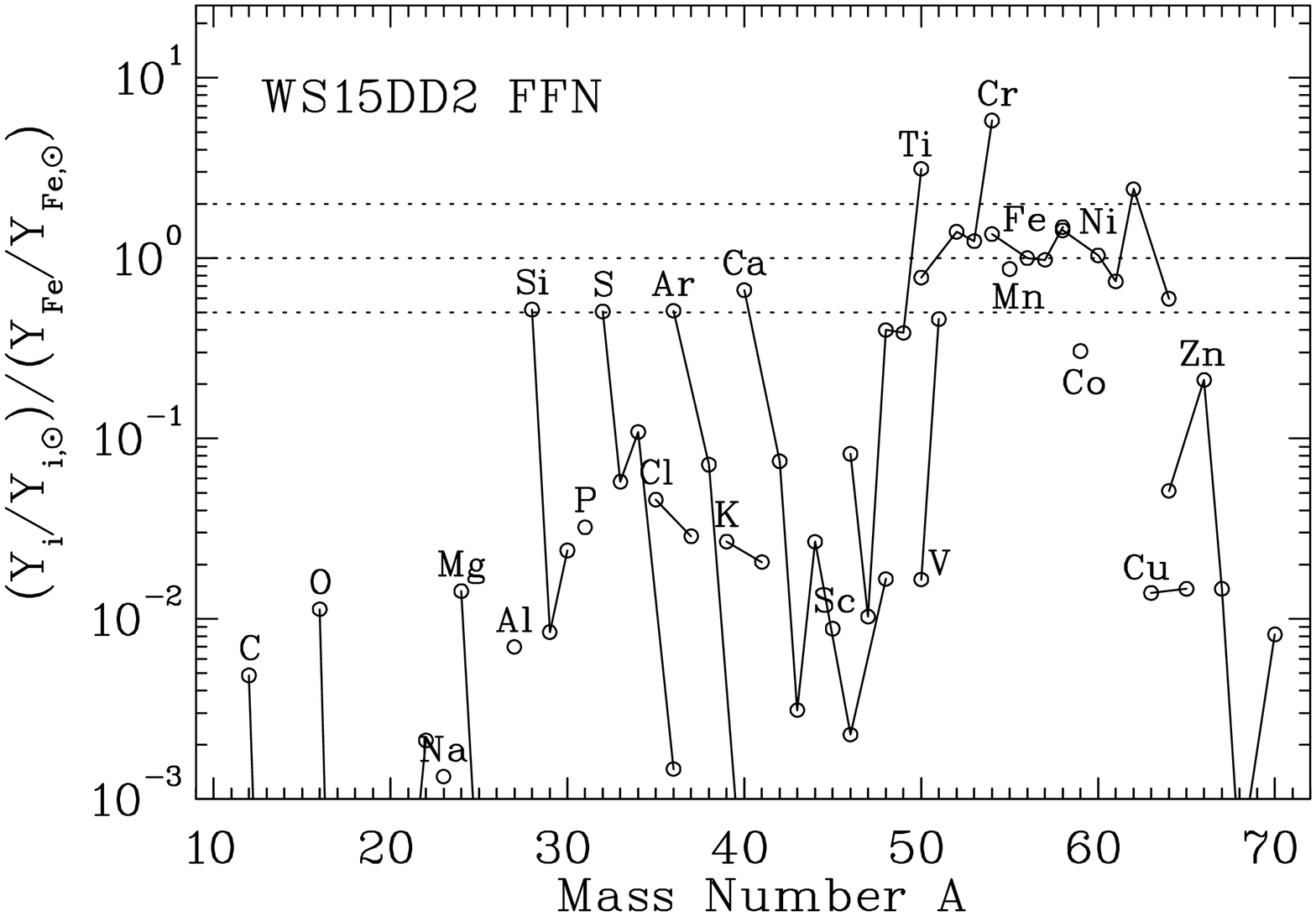}%
  \hfill%
    \includegraphics[angle=90,width=0.45\textwidth]{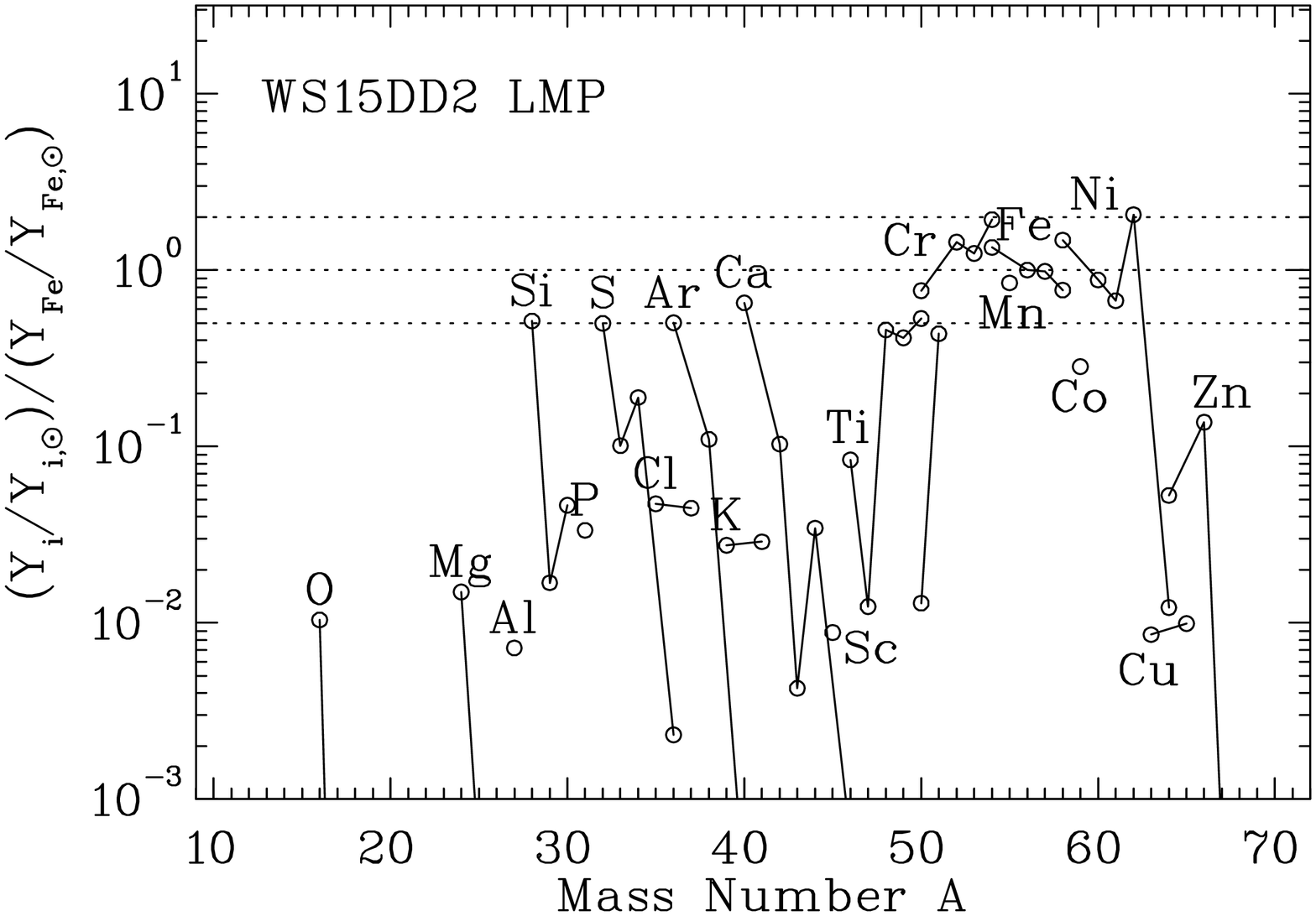}%
    \hspace*{\fill}
  \caption{Ratio of calculated to solar abundances predicted by the WS15 model
    \cite{Nomoto.Thielemann.Yokoi:1984} using the FFN and the
    shell-model rates (LMP). The ordinate is normalized to $^{56}$Fe.
    Intermediate-mass elements exist but are underproduced by a factor
    of 2--3 for SNe Ia models in comparison to Fe group elements.
    Using the FFN rates the Fe group does not show a composition close
    to solar. Especially $^{50}$Ti and $^{54}$Cr are strongly
    overproduced by more than a factor of 3. The change from FFN rates
    (left) to LMP (right) reduces the overproduction over solar close
    to the acceptable limit of a factor 2 (courtesy of F.
    Brachwitz).\label{fig:abundIa}}
\end{figure*}
   
Type Ia supernovae contribute about half to the abundance of Fe-group
nuclides in galactic evolution. Thus one can expect that type Ia
supernovae should not overproduce abundances of nuclides in the iron
group, such as $^{54}$Cr or $^{50}$Ti, relative to $^{56}$Fe by more
than a factor of two compared with the relative solar abundances.  This
requirement puts stringent constraints on models, in particular on the
central density of the progenitor white dwarf and the flame speed
\cite{Iwamoto.Brachwitz.ea:1999}. When the flame travels outwards, it
heats the matter to temperatures of a few $10^9$ K and brings its
composition close to nuclear statistical equilibrium (NSE).  As the
original matter ($^{16}$O, $^{12}$C) had an electron-to-baryon ratio
of $Y_e=0.5$, the NSE composition is dominated by $^{56}$Ni, which
after being expelled decays to $^{56}$Fe. However, behind the flame
front, which travels with a few percent of the local sound speed
\cite{Niemeyer.Hillebrandt:1995}, electron captures occur, which lower
$Y_e$ and drive the matter composition more neutron-rich. This effect
is larger the greater the central density of the white dwarf
(which increases the electron capture rates) and the slower the flame
speed (which allows more time for electron captures).
Figure~\ref{fig:yeprof} shows the $Y_e$ profile obtained in a standard
type Ia model WS15 \cite{Nomoto.Thielemann.Yokoi:1984}, with slow
deflagration flame speed (1.5\% of sound velocity), central ignition
density $\rho=2.1 \times 10^9$~g~cm$^{-3}$ and a transition from
deflagration to detonation at density $\rho = 2.1 \times
10^7$~g~cm$^{-3}$. The calculations have been performed by
\citet{Brachwitz:2001} with the FFN \cite{Fuller.Fowler.Newman:1982a}
and shell model \cite{Langanke.Martinez-Pinedo:2001} weak interaction
rate sets.  The differences are quite significant, even if one
considers that about $60\%$ of the captures occur on free protons,
which are unaffected by the differences in these rate sets.  Under
otherwise identical conditions the slower shell model rates yield a
central $Y_e$ value of 0.45, which is about 0.01 larger than for the
FFN rates. Consequently very neutron-rich nuclei with $Y_e \le 0.45$
are significantly suppressed (see figure~\ref{fig:abundIa}). In fact, no
nuclide is significantly overproduced in this model compared to the
solar abundance. The net effect of the new rates is that, for an
otherwise unchanged model, it increases the central density by about a
factor of 1.3 \cite{Brachwitz.Dean.ea:2000,Woosley.pvt}.  This can have
quite interesting consequences if one wants to use the nucleosynthesis
constraint to distinguish between two quite distinct type Ia models.
On the basis of recent models it has been concluded that the majority
of type Ia progenitors grow towards the Chandrasekhar mass through
steady hydrogen and helium burning \cite{Hachisu.Kato.Nomoto:1999}.
Such systems would lead to rather low central densities $\rho \le 2
\times 10^9$~g~cm$^{-3}$. In these models, only a small fraction
of progenitors would deviate from steady hydrogen burning at the end
of the accretion history experiencing weak hydrogen flashes; such
cases correspond to the W-model discussed above yielding higher
central densities. We stress, however, that changes in the
nucleosynthesis caused by differences in the central densities in the
models can be counterbalanced by changes of the flame speed.

\section{Conclusions and future perspectives}
\label{sec:concl-future-persp}

It has long been recognized that nuclear weak-interaction processes
play essential roles in many astrophysical scenarios. In a few cases
these are specific reactions, which are particularly important, and
these reactions have then been studied with increasingly refined
models. Examples are the various weak-interaction processes in the
solar hydrogen burning chains including the initial $p+p$ fusion
reaction \cite{Kamionkowski.Bahcall:1994,Park.Kubodera.ea:1998,%
Park.Marcucci.ea:2001a,Kong.Ravndal:2001} or the very
challenging $^3\text{He}+p$ reaction \cite{Marcucci.Schiavilla.ea:2001,%
Park.Marcucci.ea:2001b,Marcucci.Schiavilla.ea:2000}, which generates
the highest-energy neutrinos in the sun. Another typical example is
the solar electron capture rate on $^7$Be, where the nuclear matrix
element can be determined from the experimental lifetime of atomic
$^7$Be, while the proper description of the solar plasma effects on
the capture rate with the desired accuracy has been quite demanding.

However, most astrophysical applications require the knowledge of
weak-interaction rates for a huge body of nuclei. If, like for
s-process nucleosynthesis, the nuclei involved are close to the valley
of stability and hence quite long-lived, the needed rates (usually
half-lives) have been determined experimentally in decade-long
efforts.  For the s-process the challenge now focuses on the branching
point nuclei where the reaction flow branches into two (or more) paths
and where, in some cases, the observed relative abundances of nuclides
along the different paths depend on the stellar conditions
(temperature, neutron density) and hence allow determination of these
quantities inside the star. Again, the approach is to measure the
necessary data, e.g.\ half-lives of excited nuclear states.

Other astrophysical scenarios involve nuclei far-off stability, often
under extreme conditions (high density, neutron flux, temperature).
These astrophysical sites include core-collapse (type II) and
thermonuclear (type Ia) supernovae, r-process nucleosynthesis and
explosive hydrogen burning; and the interest in all of these has been
boosted recently by novel observations and data (supernova 1987A,
high-redshift supernova survey, Hubble Space Telescope, \ldots). A
direct experimental determination of the respective stellar
weak-interaction rates has been possible in a few cases, like the
half-lives of r-process and rp-process waiting point nuclei. However,
in nearly all cases the weak-interaction processes had to be
theoretically modeled so far -- a very demanding job if one considers
that often results for many hundreds of nuclei for a large range of
stellar conditions are needed. These data were then derived globally
based on  parametrized nuclear structure arguments, as an
appropriate treatment of the involved nuclear structure problem was
prohibited by both the available computational capabilities and the
lack of experimental guidance. Although evaluation of rate sets for
astrophysical purposes often appears to be a theoretical problem, the
second point -- experimental guidance -- is crucial and often
overlooked.  It is Willy Fowler's strategy and legacy that nuclear
models used to derive nuclear ingredients in astrophysical
applications, should be consistent, but more importantly they should
be accurate and, as a consequence, experimental data are to be used
whenever available. Therefore the role of experimental data in nuclear
astrophysics is twofold: If possible, they supply the needed
information directly, or equally important, they constitute
constraints and guidance for the nuclear models from which the needed
information is obtained. Thus, the renaissance of nuclear structure,
which we have witnessed in recent years, has two consequences in nuclear
astrophysics. The recent development of new facilities, techniques and
devices brought a large flood of new experimental information, in
particular for the proton- and neutron-rich nuclei away from
stability.  These data indicate that the nuclear structure models
adopted to derive the global data and rate sets were usually too
simple and improvements were warranted. The experimental renaissance
went hand-in-hand with decisive progress in nuclear structure theory,
made possible by the development of new models and better computer
hardware and software. Due to both experimental and theoretical
advances, it becomes now possible to calculate nuclear data sets for
astrophysical applications on the basis of realistic models rather
than on crude and often oversimplified parametrizations. This review
presents a summary of recent theoretical calculations.

The advances in nuclear structure modelling have also lead to progress
in astrophysically important nuclear input other than
weak-interaction processes. Typical examples are the equation of
state, derived on the basis of the relativistic mean-field model
guided by the relativistic Br\"uckner-Hartree-Fock theory
\cite{Shen.Toki.ea:1998a,Shen.Toki.ea:1998b}, which serves as an
alternative to the standard Lattimer-Swesty EOS
\cite{Lattimer.Swesty:1991}, or the nuclear mass table and level
density parametrizations determined within the framework of the
Hartree-Fock model with BCS pairing
\cite{Goriely.Tondeur.Pearson:2001,Demetriou.Goriely:2001b}. Such mass
tables, or equivalently neutron separation energies, play an essential
role in r-process nucleosynthesis and are conventionally derived by
parametrizations constrained to known masses.

Other astrophysical areas, loosely or indirectly related to the topic
of this review, have also benefitted from the progress in modelling
nuclear weak-interaction processes. A field with rapidly growing
importance is Gamma-ray Astronomy with beta-unstable nuclei. Due to
$\gamma$-ray observatories in space it has been possible in recent
years to search the sky for sources of known $\gamma$-rays which can
then be associated with recent nucleosynthesis activities. A highlight
has been the observation of the 1.157 MeV $\gamma$-line, produced in
the $\beta$-decay scheme of $^{44}$Ti, in the Cassiopeia supernova
remnant \cite{Iyudin.Diehl.ea:1994}. Knowing the date of the
supernova, the $^{44}$Ti half-life, the distance to the source and the
measured intensity of the $\gamma$-line allows determination of how
much $^{44}$Ti has been ejected into the interstellar medium by the
supernova event. Furthermore, as $\gamma$-rays can, in contrast to
optical wave lengths, escape from the galactic bulk, $\gamma$-ray
observation allows to detect historical supernovae which have not been
observed optically \cite{Iyudin.Schonfelder.ea:1998}, \citep[also
observed in x-rays by][]{Aschenbach:1998}. Such searches for
historical supernovae and an improved determination of the supernova
frequency in our galaxy will be one of the main missions of future
$\gamma$-ray observatories in space, like
INTEGRAL\footnote{http://astro.estec.esa.nl/SA-general/Projects/Integral/integral.html}.
A longer-lived radioactive nuclide produced in supernovae is
$^{60}$Fe. Investigations of rock samples taken from the ocean floor,
by precision accelerator mass spectroscopy, found a significant
increase of $^{60}$Fe abundance compared to other iron isotopes
pointing to a close-by supernova about 5 million years ago
\cite{Knie.Korschinek.ea:1999}.

Weak-processes on nuclei and electrons are the means to observe solar
and supernova neutrinos \cite{Balantekin.Haxton:1999}. Solar neutrinos
have rather low energies ($E_\nu \le 14$ MeV) and hence induce
specific low-lying transitions which are theoretically modelled best
by shell-model calculations.  Applications have been performed, for
example, for $^{37}$Cl \cite{Aufderheide.Bloom.ea:1994}, $^{40}$Ar
\cite{Ormand.Pizzochero.ea:1995} and $^{71}$Ga \cite{Haxton:1998}, the
detector material in the Homestake \cite{Cleveland.Daily.ea:1998},
ICARUS\footnote{http://www.aquila.infn.it/icarus}, and GALLEX/SAGE/GNO
detectors \cite{Hampel.Handt.ea:1999,Abdurashitov.Gavrin.ea:1999,%
  Altmann.Balata.ea:2000}, respectively. Supernova neutrinos have
higher energies; in particular, the energies of $\mu,\tau$ neutrinos
and antineutrinos are expected to be high enough to excite the giant
dipole resonances in nuclei.  Studies for detector materials like Na,
Fe, Pb have been performed in hybrid approaches combining shell model
calculations for allowed transitions with the RPA model for forbidden
transitions or treating the forbidden transitions on the basis of the
Goldhaber-Teller model
\cite{Fuller.Haxton.Mclaughlin:1999,Kolbe.Langanke:2001}.

Despite the experimental and theoretical progress, lack of knowledge
of relevant or accurate weak-interaction data still constitutes a
major obstacle in the simulation of some astrophysical scenarios
today. This refers mainly to type II supernovae and r-process
nucleosynthesis.

Core-collapse supernovae, as the breeding places of carbon and oxygen,
and hence life, in the universe, attract currently significant
attention and the quest to definitely identify the explosion mechanism
is on the agenda of several international and interdisciplinary
collaborations.  While most of the efforts concentrate on
computational developments towards a multidimensional treatment of the
hydrodynamics and the neutrino transport, some relevant and
potentially important nuclear problems remain. The improved
description of weak-interaction rates in the iron mass range has lead
to significant changes in the presupernova models. Evaluation of
electron capture rates for heavier nuclei will be available in the
near future. First results imply that capture on heavy nuclei, usually
ignored in collapse simulations, can compete with the capture on free
protons. Whether the inclusion of this process then leads to changes
in the collapse trajectory has yet to be seen. Further, finite
temperature neutrino-nucleus rates will also soon be available. This
will then test whether inelastic neutrino scattering on
nuclei, yet not modelled in the simulations, influences the
collapse dynamics or supports the revival of the shock wave by
preheating of the infalling matter after the bounce. The largest
nuclear uncertainty in collapse simulations are likely associated with
the description of nuclear matter at high density, extreme isospin and
finite temperature. In particular a reliable description of the
neutrino opacity in nuclear matter might very well be what is needed
for successful explosion simulations.  Nucleon-nucleon correlations
have been identified to strongly influence the neutrino opacities and
nuclear models like the RPA are quite useful to guide the way.
Ultimately one would like to see the many-body Monte Carlo
techniques, which have been so successfully applied to few-body
systems or the nuclear shell model, extended to the nuclear matter
problem. First steps along these lines have been reported by
\citet{Schmidt.Fantoni:1999} and \citet{Fantoni.Sarsa.Schmidt:2001},
who proposed a novel constrained-path Diffusion Monte Carlo model to
study nuclear matter at temperature $T=0$. The Shell Model Monte Carlo
model, formulated in momentum space and naturally at finite
temperature, constitutes an alternative approach. First attempts in
this direction have been taken by the Caltech group \cite{Zheng.pvt}
and by \citet{Rombouts.pvt}.  Whether the notorious sign-problem in
the SMMC approach can also be circumvented for nuclear matter has
still to be demonstrated.  \citet{Mueller.Koonin.ea:2000} have
investigated whether nuclear matter can be formulated on a spatial
lattice with nearest neighbor interactions, similar to the Hubbard
model for high-$T_c$ superconductors. Besides these theoretical
efforts it is equally important to improve our understanding of the
nuclear interaction in extremely neutron-rich nuclei (matter).

Despite four decades of intense research, the astrophysical site of
the r-process nucleosynthesis has yet not been identified. Recent
astronomical and meteoric evidence points now to more than one source
for the solar r-process nuclides, and it is clearly a major goal in
the astrophysics community to solve this cosmic riddle. However, the
puzzle will not be definitely solved if the nuclear uncertainties
involved are not removed.  This is even more necessary as recent
research shows that the r-process is a dynamical process under
changing astrophysical conditions. This implies dynamically changing
r-process paths. To determine the paths one needs to know the masses
of nuclei far-off stability accurately, besides the condition of the
astrophysical environment. In a dynamical r-process, the astrophysical
timescale will compete with the nuclear timescale, i.e.\ with the time
needed for the matter flow from the seed nuclei to the heavier
r-process nuclides. This nuclear timescale is set by the half-lives of
the nuclei along the paths, in particular by those of the longer-lived
waiting point nuclei associated with the magic neutron numbers. The
half-lives of waiting point nuclei are a very illustrative example for
the need of reliable experimental data: Modern global theoretical
models predict half-lives at the waiting points with a spread of
nearly one order of magnitude and only data can decide. For the $N=50$
and $N=82$ waiting points such data exist for a few key nuclides
\citep[e.g.][]{Pfeiffer.Kratz.ea:2001}, but not for the $N=126$
nuclei.

R-process nucleosynthesis as well as other astrophysical processes
will tremendously benefit from future experimental developments in
nuclear physics. Worldwide radioactive ion-beam facilities, with
important and dedicated programs in nuclear astrophysics, have just
started operation or are under construction or in the proposal stage.
These new facilities will boost our knowledge about nuclei far-off
stability, they will determine astrophysically relevant nuclear input
directly (e.g.\ masses and half-lives for the r-process, half-lives
and cross sections for the rp-process, etc). But equally important,
the radioactive ion-beam facilities will guide and constrain the
nuclear models, in this way indirectly contributing to the reduction
of the nuclear inaccuracies in astrophysical models. Ultimatively
nuclear physics can and will then become a stringent test and guidance
for astrophysical theories and ideas.

\acknowledgments

We like to thank many colleagues and friends for helpful
collaborations and discussions: D. Arnett, S.~M. Austin, J.~N.
Bahcall, R.~N. Boyd, F.  Brachwitz, R. Canal, E. Caurier, J.
Christensen-Dalsgaard, J.~J. Cowan, D.~J. Dean, R. Diehl, J.
Dobaczewski, J.~L. Fisker, G.~M. Fuller, E.  Garc\'{\i}a-Berro, S.
Goriely, J. G{\"o}rres, W.~C. Haxton, A. Heger, M. Hernanz, W.
Hillebrandt, R. Hix, J. Isern, H.-Th. Janka, J.  Jos\'e, T. Kajino, E.
Kolbe, S.~E. Koonin, K.-L. Kratz, F.  K{\"a}ppeler, M. Liebend\"orfer,
A. Mart\'{\i}nez-Andreu, G.~J. Mathews, A. Mezzacappa, Y.  Mochizuki,
E. M{\"u}ller, W.  Nazarewicz, F. Nowacki, I. Panov, 
B. Pfeiffer, A.  Poves, Y.-Z.
Qian, G. Raffelt, R.  Reiferth, A. Richter, C.~E. Rolfs, J.~M.
Sampaio, H. Schatz, M. Terasawa, F.-K. Thielemann, J.~W. Truran, P.
Vogel, M. Wiescher, S.~E. Woosley, and A.~P. Zuker. 
Special thanks are due to our two referees, R.N. Boyd and anonymous,
for very constructive and helpful comments on the manuscript. 
Our work has been
supported by the Danish Research Council and the Schweizerische
Nationalfonds.  Computational resources for our work were provided by
the Center for Advanced Computational Research at Caltech, by the
Danish Center for Scientific Computing and by the Center for
Computational Sciences at Oak Ridge National Laboratory.


\end{document}